\documentclass[preprint2]{aastex7}
\usepackage{color} 
\usepackage{tabularx} 
\usepackage{epsfig}
\usepackage{amsmath} 
\usepackage{amssymb} 
\usepackage{graphicx}
\usepackage{enumitem}
\usepackage{multirow}
\usepackage{cprotect}
\usepackage{natbib}
\setlist{nosep}
\usepackage{url}
\usepackage[export]{adjustbox}
\usepackage[column=0]{cellspace}
\definecolor{purp}{HTML}{8904B1}

\let\oldhat\hat
\renewcommand{\hat}[1]{\oldhat{\mathbf{#1}}}

\let\oldwidehat\widehat
\renewcommand{\widehat}[1]{\oldwidehat{\mathbf{#1}}}

\newcommand\fe[3]{$#1^{+#2}_{-#3}$}
\newcommand{\chandra}{\emph{Chandra }}

\shortauthors{Watson et al.}

\accepted{November 15, 2025}
\submitjournal{The Astrophysical Journal}

\begin{document}
\defcitealias{GM2019}{GM19}
\defcitealias{PM2013}{PM13}

\title{Deep Chandra X-ray Observations of Abell 2029: the Merger History of a Relaxed, Strong Cool Core Cluster}

\author[orcid=0000-0001-8456-4142]{Courtney B. Watson}
\affiliation{Institute for Astrophysical Research and Department of Astronomy, Boston University, Boston, MA 02215, USA}
\affiliation{Center for Astrophysics $|$ Harvard \& Smithsonian, 60 Garden Street, Cambridge, MA 02138, USA}
\email[show]{cbwatson@bu.edu}

\author[orcid=0000-0002-0485-6573]{Elizabeth L. Blanton}
\affiliation{Institute for Astrophysical Research and Department of Astronomy, Boston University, Boston, MA 02215, USA}
\email{eblanton@bu.edu}

\author[orcid=0000-0002-3984-4337]{Scott W. Randall}
\affiliation{Center for Astrophysics $|$ Harvard \& Smithsonian, 60 Garden Street, Cambridge, MA 02138, USA}
\email{scott.randall@cfa.harvard.edu}

\author[0000-0001-6812-7938]{Tracy E. Clarke}
\affiliation{Naval Research Laboratory, Code 7213, 4555 Overlook Avenue Southwest, Washington, DC 20375, USA}
\email{tracy.clarke@nrl.navy.mil}

\author[0000-0003-3175-2347]{John A. ZuHone}
\affiliation{Center for Astrophysics $|$ Harvard \& Smithsonian, 60 Garden Street, Cambridge, MA 02138, USA}
\email{john.zuhone@cfa.harvard.edu}

\begin{abstract}
We present results from very deep (485 ks) \chandra X-ray observations of the relaxed, cool core cluster Abell 2029 ($z = 0.0767$). A2029 hosts one of the longest, most continuous sloshing spirals ever observed, which we find extends nearly 600 kpc from the cluster core. In addition to providing detailed views of the sloshing spiral, imaging and spectroscopic analysis reveals ICM substructure related to the merger history including a broad ``splash'' of cooler gas and a potential merger shock. The radio lobes of the central WAT source show evidence of alignment with the sloshing motions, consistent with ICM bulk flow, rather than host-galaxy motion, being the primary driver of lobe bending. Comparison to a 1:10 mass-ratio off-axis merger simulation indicates that the observed ICM structures are relics of a second core passage of a subcluster $\sim$4 Gyr after the start of the merger, where the ``splash'' feature is revealed to be a wake of cool gas trailing behind the subcluster. Overall, our results suggest that A2029 is still settling from past interactions --- showing that even the initially most relaxed-looking clusters can be hiding a rich history of dynamical activity. 
\end{abstract}

\section{Introduction}

Galaxy clusters and their intracluster medium (ICM) -- the hot, ionized plasma that permeates throughout the cluster and represents almost 20\% of the total cluster mass --- play an integral role in our understanding of many layers of astrophysics. In their most relaxed state, clusters appear smooth, centrally concentrated, and symmetric in X-ray emission, making them attractive for cosmological studies under the assumption of hydrostatic equilibrium. However, increasingly deep, high-angular resolution observations have revealed that even apparently relaxed clusters can host  a surprising amount of substructure lurking beneath the surface, challenging these assumptions.

One long-standing issue in cluster astrophysics is the cooling flow problem: radiative cooling in the dense cores of relaxed clusters should, in principle, lead to large inflows of gas and high star formation rates in brightest cluster galaxies (BCGs). Yet, observed cooling rates are far lower than predicted \citep{Fabian2003,Peterson2006}, suggesting some heating mechanism must be involved to balance cooling. 

The most likely source of this heating comes from active galactic nuclei (AGN) feedback. The cooling flow of gas that would be deposited in the cluster center, should feed the cluster's central AGN, resulting in repeated outbursts that can inflate bubbles associated with the AGN's jets and lobes. These bubbles buoyantly rise to larger radii, distributing heat to the cooling ICM, though the precise mechanism of how this occurs is still debated \citep{McNamara2007, Blanton2011}. However, this is not the full solution. In some clusters, the central AGN cannot impart enough energy into the surrounding ICM and in these cases, additional processes, such as cluster-cluster interactions which give rise to gas sloshing and shock fronts, can contribute significant heating. While major mergers can permanently disrupt cool cores and raise central entropy beyond the reach of AGN feedback alone, even minor off-axis mergers may produce sloshing motions that help suppress runaway cooling in otherwise relaxed systems \citep{ZuHone2010, Chen2024}.

Prominent examples of such sloshing include the Perseus Cluster, which hosts a multi-scale spiral structure extending hundreds of kiloparsecs \citep{Churazov2001, Walker2017}; Abell 2052, which exhibits a sloshing spiral in addition to AGN-inflated bubbles \citep{Blanton2003, Blanton2009, Blanton2011}; and Abell 119, where deep Chandra imaging reveals a prominent sloshing spiral despite the absence of a traditional cool core \citep{Watson2023}. In systems such as these, sloshing of the cool, metal-rich ICM core led to contact discontinuities (a.k.a. cold fronts; observed as sharp edges in X-ray brightness) and sometimes merger shocks. At the interfaces of these contact discontinuities, shear flows can lead to the development of Kelvin-Helmholtz Instabilities (KHIs). The growth or suppression of KHI structures depend on the ICM magnetic field strength and viscosity, as shown in various numerical studies \citep{ZuHone2010, ZuHone2011, ZuHone2015, Roediger2012, Roediger2013a, Roediger2015}. 

Abell 2029 (A2029, hereafter) is a $z=0.0767$ \citep{Sohn2019} cool-core cluster, thought to be one of the most relaxed clusters in the universe \citep{Buote1996}. Even recently, two studies \citep{XRISMCollaboration2025,XRISMCollaboration2025a} using observations from X-ray Imaging and Spectroscopy Mission \citep[XRISM,][]{Tashiro2020} report negligible line-of-sight velocities in the ICM. \cite{XRISMCollaboration2025} find that within the central 180 kpc of the cluster core, the relative velocity of the ICM is at rest with respect to the central BCG, with a 3-sig upper limit of $|v| < 100$ km/s. Building on this, \cite{XRISMCollaboration2025a} extend the analysis out to 670 kpc from the cluster core, using three radial pointings covering just the NNW region of the cluster. Their findings also show low bulk motions of the ICM, however they note significant blueshifts in the northern regions suggesting increased bulk motions outside the cluster core. Both XRISM studies agree that the non-thermal pressure fraction is less than 3\% of the total pressure, consistent with a relaxed cluster. A third XRISM study \citep{Sarkar2025} using Resolve line-ratio diagnostics reports clear multi-temperature structure in A2029, with cooler ($\sim$3–4 keV) components superposed on the hot ($\sim$7–8 keV) ICM in the core and inner northern regions. These findings independently support the presence of cooling phases and azimuthal thermal complexity implicated by the sloshing features we map with Chandra.

Despite this, previous X-ray observations have revealed a large spiral in the ICM consistent with sloshing motions, suggesting recent or on-going merger activity \citep{Clarke2004, PM2013} occurring in A2029. In addition, \cite{Mirakhor2022} found that A2029 is possibly connected to a neighboring cluster in the north, Abell 2033 (z=0.0812 \citep{Sohn2019}; A2033, hereafter) by a bridge of X-ray gas (significant at $>6.5\sigma$) that is consistent with a $\sim 1$ Mpc wide filament. Further, \cite{Sohn2019} showed that A2033, in addition to a secondary infalling group which they label 'SIG' ($z= 0.0802$; located $\sim 27\arcmin$ south of A2029), are both gravitationally bound to A2029 and would be accreted within the next few Gyr. 

In this paper, we present the deepest Chandra observation of A2029 taken to-date, giving us an unprecedented look at the ICM structure. In comparing results of previous optical studies and simulations of cluster mergers, we build a picture of the merger history of A2029. 

Throughout this paper, we adopt a standard $\Lambda$CDM cosmology with $H_0 = 70$ km s$^{-1}$ Mpc$^{-1}$, $\Omega_M = 0.3$, and $\Omega_{\Lambda} = 0.7$. At the redshift of Abell 2029 (z=0.0767), the luminosity distance is $D_L = 347.5$ Mpc, the angular size distance is $D_A = 299.7$ Mpc, and $1 \arcsec = 1.453$ kpc.  Reported measurements include 1$\sigma$ confidence intervals unless otherwise noted. Unless specified in the text, reported angular values are measured towards north from west. 

\section{Chandra X-ray Observations of A2029}

A total of 24 Chandra pointings were used, totaling 515 ksec raw exposure, comprised of a set of two previous ACIS-S observations (ObsIDs 891 \& 4977) from Cycle 1, one previous ACIS-I observation (ObsID 6101) from Cycle 3, and 21 new ACIS-I observations (ObsIDs 2XXXX; see Tab. \ref{tab:obsinfo}) from Cycle 23. The data were processed using CIAO \citep[v4.17;][]{Fruscione2006} and CALDB (v4.12.0) provided by the Chandra X-ray Center (CXC). Level 2 event files were created using \verb|chandra_repro|, accounting for the given observations' data mode (FAINT or VFAINT). 

Background flares were removed by extracting light curves using \verb|dmextract|, and detecting and removing flares using \verb|deflare| --- a CIAO script that  iteratively removes periods where the count rate exceeds 1.2 times the mean background level. For back-illuminated (BI) chips we used a bin size of 1037.12 s while front-illuminated (FI) chips were filtered using a bin size of 259.28 s\footnote{\href{https://cxc.harvard.edu/contrib/maxim/acisbg/COOKBOOK}{https://cxc.harvard.edu/contrib/maxim/acisbg/COOKBOOK}}. After filtering for background flares, a total of 30 ksec was lost, for a final, cleaned exposure time of 485 ksec. 

Blank-sky background files were selected from CALDB, using \\ \verb|acis_background_lookup|, and reprojected to match the observations. Background images were exposure corrected and normalized such that the count rate in the hard band (10-12 keV) matched that of the observations. This was done on a chip-by-chip basis for each observation, while excluding point sources, to account for differences in exposure times between some of the ACIS-S blank-sky backgrounds as well as the different hard-band ratios due to the sensitivity of the FI vs BI chips. 

Aspect histograms, instrument maps, and exposure maps for each ObsID were created following the step-by-step guide outlined in the CIAO thread \emph{Multiple Chip ACIS Exposure Map}\footnote{\href {https://cxc.harvard.edu/ciao/threads/expmap_acis_multi/}{https://cxc.harvard.edu/ciao/threads/expmap\_acis\_multi/}}. Regions with exposure $<10\%$ of the maximum exposure time were excluded from further analysis, on an ObsID by ObsID basis. A combined exposure map was created by summing together the exposure maps of the individual ObsIDs. 

All 24 ObsIDs were stacked and reprojected to a common WCS tangent point, using \verb|reproject_events|, to create a final combined image, filtered to the 0.7-7 keV range, with a cleaned exposure time of 485 ksec. Table \ref{tab:obsinfo} lists the ObsID, observation date, data mode, mean focal plane temperature, and flare-cleaned exposure for each observation used in the combined mosaic and in the analysis presented here. The three sets of earlier observations consist of two pointings targeted on the ACIS-S array (ObsIDs 891 \& 4977), and another pointing targeted on the ACIS-I array (ObsID 6101) --- this group is often referred to in this paper as the ``pre-2005'' set. The set of 21 newer observations,  often referred to in this work as the ``new'' set, were taken using the ACIS-I configuration. 

\begin{deluxetable}{ccccc}
\tablecaption{Chandra observation information for Abell 2029 \label{tab:obsinfo}}
\tablehead{\colhead{ObsID} & \colhead{Date} & \colhead{Mode\tablenotemark{a}} & \colhead{FP Temp\tablenotemark{b} } & \colhead{Exp\tablenotemark{c}} \\
\colhead{} & \colhead{} & \colhead{} & \colhead{[K]} & \colhead{[ksec]} 
} 
\startdata
 891\tablenotemark{d} & 2000 Apr 12 & F & 154 & 19.82 \\
4977\tablenotemark{d} & 2004 Jan 9 & F & 154 & 76.88 \\ 
 6101 & 2004 Dec 17 & VF & 154 & 9.66 \\ 
25496 & 2022 Apr 7 & VF & 155 & 17.42 \\
26380 & 2022 Apr 10 & VF & 156 & 12.77 \\
25819 & 2022 Apr 15 & VF & 155 & 14.44 \\
26393 & 2022 Apr 16 & VF & 158 & 13.93 \\
25814 & 2022 May 22 & VF & 157 & 17.11 \\
26420 & 2022 May 23 & VF & 156 & 9.21 \\
25815 & 2022 May 28 & VF & 158 & 17.63 \\
25816 & 2022 May 29 & VF & 158 & 12.83 \\
26428 & 2022 May 29 & VF & 157 & 22.00 \\
25822 & 2023 Apr 8 & VF & 158 & 24.55 \\
25824 & 2023 Apr 11 & VF & 158 & 22.96 \\
25825 & 2023 Apr 12 & VF & 158 & 26.35 \\
25826 & 2023 Apr 22 & VF & 158 & 18.01 \\
27805 & 2023 Apr 23 & VF & 161 & 9.65 \\
25818 & 2023 May 19 & VF & 159 & 17.55 \\
25817 & 2023 May 21 & VF & 159 & 16.68 \\
25820 & 2023 May 22 & VF & 158 & 20.47 \\
27853 & 2023 May 23 & VF & 159 & 25.11 \\
25823 & 2023 May 27 & VF & 159 & 22.53 \\
25821 & 2023 Jun 3 & VF & 159 & 26.69 \\
27848 & 2023 Jun 6 & VF & 160 & 10.32 \\
\enddata
\tablenotetext{a}{Data mode: F=FAINT; VF=VFAINT}
\tablenotetext{b}{Mean focal-plane temperature during observation}
\tablenotetext{c}{Flare-cleaned exposure time}
\tablenotetext{d}{Taken with ACIS-S configuration, aimpoint on S3}
\end{deluxetable}

We found that the new data suffered from larger swings of the focal plane (FP) temperature during observation, than seen in the older, pre-2005 data. This resulted in a dependence on FP temperature of the best-fit ICM temperature during spectral fitting which led to a disagreement with the results of the pre-2005 data. We determined that this was the effect of a previously unidentified gain calibration issue that particularly affects high S/N ACIS-I observations of relatively hot clusters, as we have in A2029. %This resulted in the Chandra calibration team having to update calibration files appropriately, which is an ongoing effort. 
Until new calibration files are available, we determine that we are able to neglect the effects of the FP dependence by including gain parameters as well as constant scale factors to some of the fitted parameters of the new data. These interim measures significantly improve the spectral fits, bringing them into better alignment with those of the pre-2005 data. This is discussed in more detail in the Appendix \S\ref{sec:cal}.

\section{Imaging Analysis \label{sec:imaging}}

\begin{figure*}
    \center
    \begin{tabular}{rr}
  \includegraphics[width=0.47\textwidth]{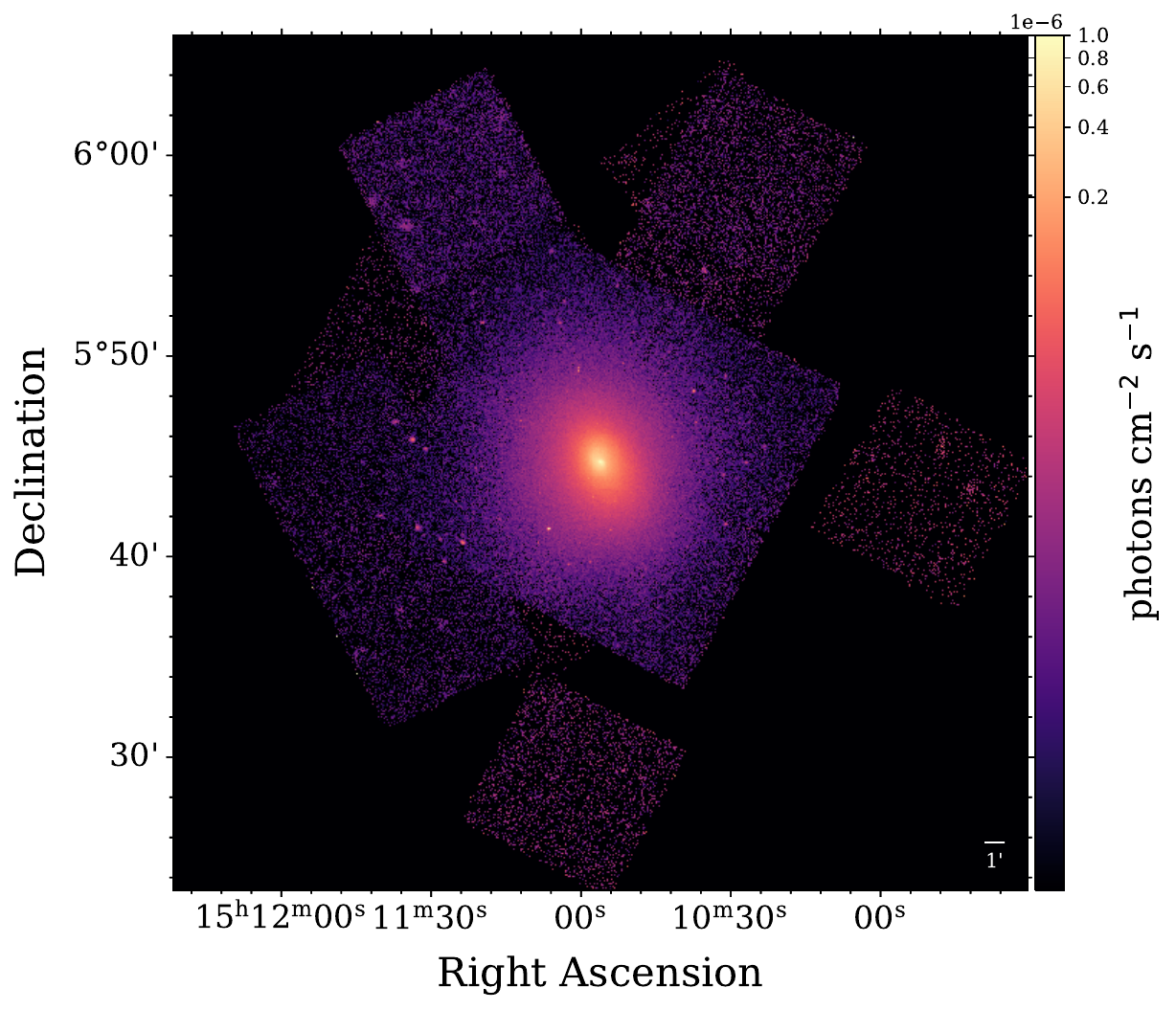} &\includegraphics[width=0.47\textwidth]{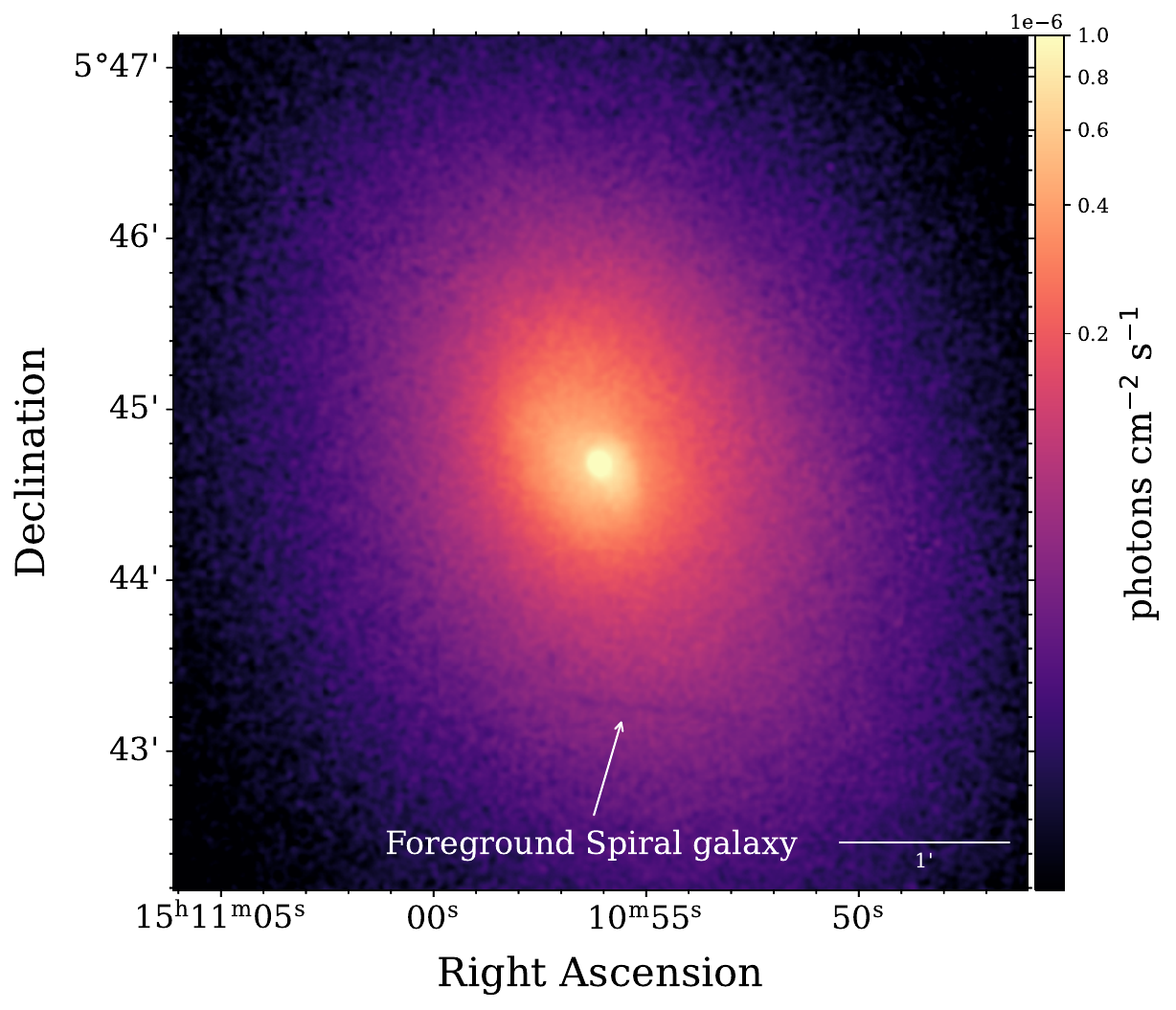} \\
\end{tabular}
    \caption{\emph{Left:} Background subtracted and exposure corrected X-ray image of A2029 in the 0.7-7 keV range, smoothed with a $\sigma = 1\farcs5$ Gaussian, with point sources included, covering a 40\arcmin$\times$ 40\arcmin\ area, the full FoV of the observations. \emph{Right:} Diffuse emission image of A2029, with point sources removed, in the same energy band, smoothed with a Gaussian of radius 1\farcs5, and showing the inner 5\arcmin$\times$ 5\arcmin\ region. A foreground, edge-on spiral galaxy, seen in absorption, is indicated with a white arrow. Note: $1\arcmin = 87$ kpc.}
    \label{fig:xray}
\end{figure*}

 A Gaussian smoothed (using a radius of $\sigma = 1\farcs5$) background-subtracted and exposure-corrected X-ray image of A2029 in the 0.7-7 keV range is shown in the left panel of Fig.~\ref{fig:xray}. The overall emission is elliptical, with major axis along the NE-SW direction. 

 We use the CIAO \verb|wavdetect| routine, with wavelet scales of 2,4,6,and 8 pixels (1 pix = 0.492\arcsec), to detect point sources in the unsmoothed X-ray image, with corrections for off-axis PSF. The detected regions were visually examined to exclude regions that do not have observed counterparts, suggesting they could be just gas structure erroneously detected as point sources. The CIAO routine \verb|dmfilth| is used to remove the point sources from the observed image and replace the removed point sources' pixels with interpolated pixel values from the surrounding area. The resulting product is an image of the isolated diffuse emission from the cluster ICM. 
 
 The right panel of Fig.~\ref{fig:xray} shows the diffuse emission in the inner $5\arcmin \times 5\arcmin$ area. A foreground, edge-on spiral galaxy, seen in absorption \citep{Clarke2004b}, is highlighted with a white arrow and this region is excluded from all detailed analyses presented in this work. 
 
\begin{figure*}
    \centering
    \plotone{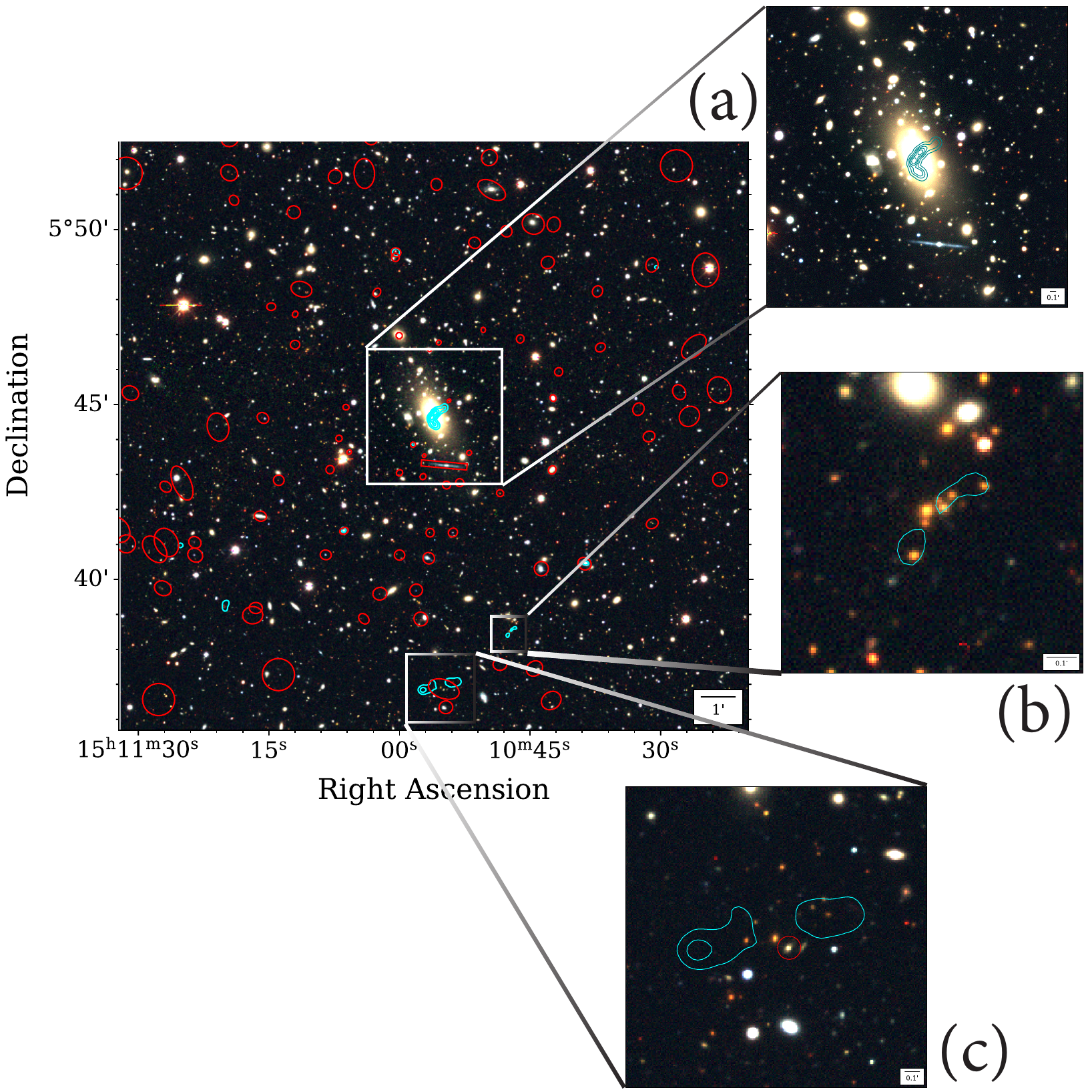}
    \caption{False color optical image of A2029 taken from the Legacy Survey Sky Browser, comprised of SDSS g-, r-, and z-band images. The image has a scaling of 0\farcs5/px. Red ellipses mark X-ray point sources detected by \texttt{wavdetect}. The 1.4 GHz VLA FIRST radio emission is shown with cyan contours which are drawn at levels of 0.0005, 0.0036, 0.0129, 0.0283, and 0.05 Jy beam$^{-1}$ and smoothed with a Gaussian kernel of width 4 pixels. The central red rectangle denotes the foreground spiral galaxy that is seen in absorption in the X-ray imaging (see Fig.\ref{fig:xray}). Inset are thumbnails showing zoom-ins of some radio sources in the field. The central WAT source of A2029 is shown in a $5\arcmin\times 5\arcmin$ inset thumbnail (a; upper right). The smaller radio source to the SW, is shown in a 1\arcmin$\times$ 1\arcmin\ thumbnail (b; middle right). The southernmost source is shown in a 2\arcmin$\times$2\arcmin\ thumbnail (c; lower right), with the detected X-ray point source indicated by a red circle. The two southern sources are discussed in more detail in \S\ref{sec:optical}. Image credits: Legacy Surveys / D. Lang (Perimeter Institute). }
    \label{fig:optical}
\end{figure*}

 Figure~\ref{fig:optical} shows an optical RGB image of A2029 obtained from the Legacy Survey Sky Browser\footnote{\href{https://www.legacysurvey.org/viewer?ra=227.7315&dec=5.7431&layer=ls-dr67&zoom=10}{www.legacysurvey.com/viewer}}, comprised of SDSS g-, r-, and z-band images. Overlaid are the X-ray point sources identified via \verb|wavdetect| (red ellipses). Also overlaid are cyan contours of 1.4 GHz radio emission, taken from the Very Large Array (VLA) Faint Images of the Radio Sky at Twenty-cm (FIRST) survey, with a beam which has a FWHM size of $6\farcs4\times 5\farcs4$ \citep{Becker1995}. Contours are shown at levels of 0.0005, 0.0036, 0.0129, 0.0283, and 0.05 Jy beam$^{-1}$, smoothed with a Gaussian kernel of width 4 pixels. Two additional radio sources south of the central bent-lobe AGN are shown in inset thumbnails (panels (b)-(c) in Fig.~\ref{fig:optical}). We discuss these sources and their relation to the observed ICM substructure in more detail in \S\ref{sec:azprofs} and \S\ref{sec:optical}.
 
\subsection{ICM Substructure \label{sec:substr}}

To highlight substructure that may not be readily seen in the source image, we employ the techniques of beta-model subtraction and Gaussian Gradient Magnitude (GGM) filtering, each of which is described in detail below. Unless otherwise noted, we use an unsmoothed, merged, background and exposure corrected 0.7-7 keV image of the diffuse cluster emission. 

\subsubsection{2D Beta-Model \label{sec:beta2d}}

To highlight substructure that may not be readily seen in the source image, we fit a 2D $\beta$-model (\verb|beta2d| model in \verb|Sherpa| \citep[v4.16.0;][]{Siemiginowska2024}) to the unsmoothed diffuse emission image in the 0.7-7 keV range using a circular region with radius 12\farcm3. Corrections for background emission and instrumental exposure were made using the merged blank-sky background image and combined exposure map. The center position, core radius, ellipticity, position angle, amplitude, and power law index of the beta-model were allowed to vary while the amplitudes of the supplied background image and exposure map were frozen. Errors were computed with Cash statistics.

The best-fit 2D $\beta$-model parameters (with 1-$\sigma$ errors) are found to be $r_{core} = 48.2 \pm 0.10$ kpc, ellipticity $e = 0.23 \pm 0.0007$, position angle $PA = 20^{\circ}\pm 0.^{\circ}4$ (measured from north towards east), and index $\beta = 0.53 \pm 0.0002$. The peak of the X-ray emission is not found to be significantly offset from the central BCG.

\begin{figure*}
\centering
\begin{tabular}{cc}
\multicolumn{2}{c}{\includegraphics[width=0.7\linewidth]{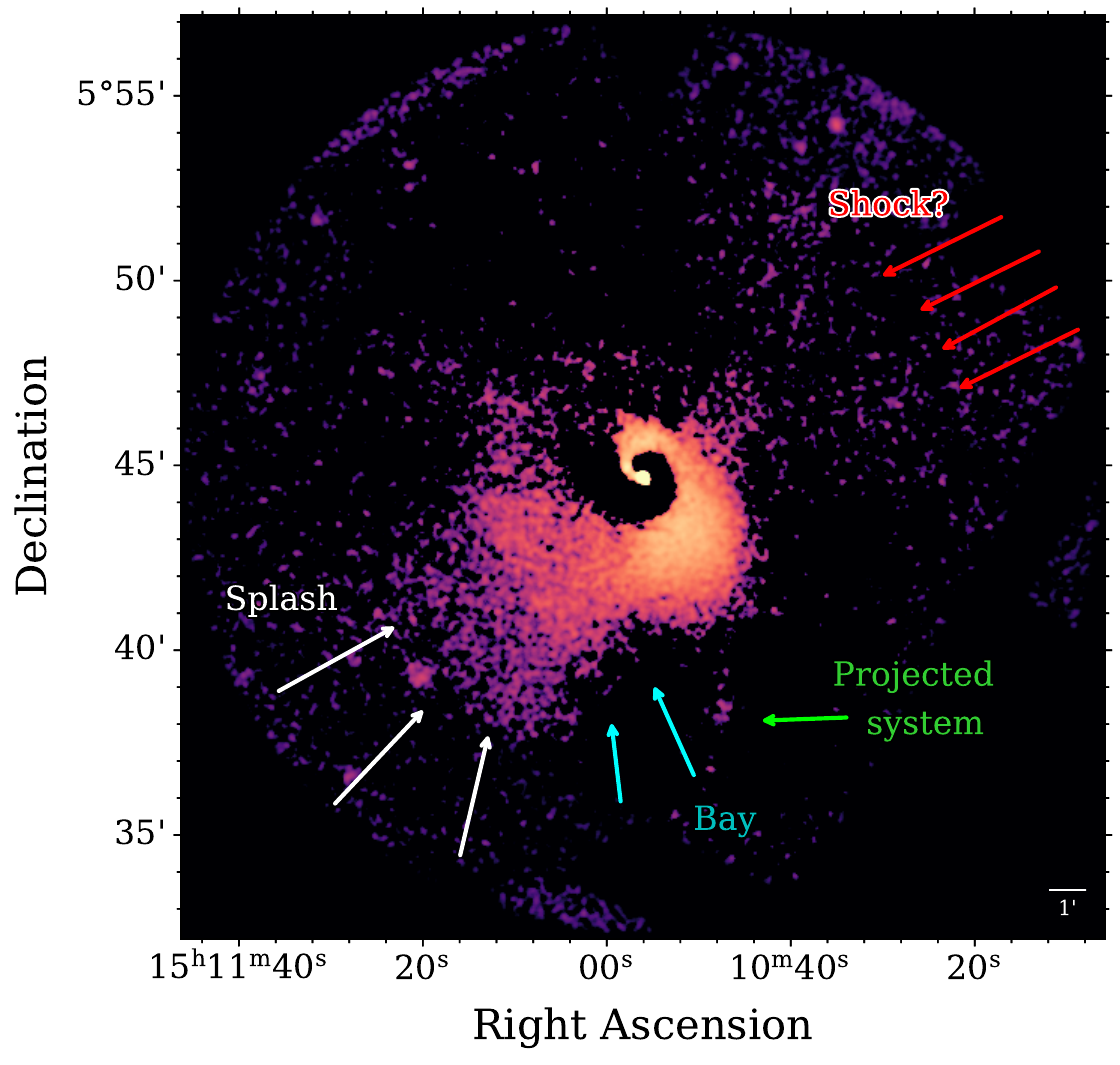}} \\
\includegraphics[width=0.5\linewidth]{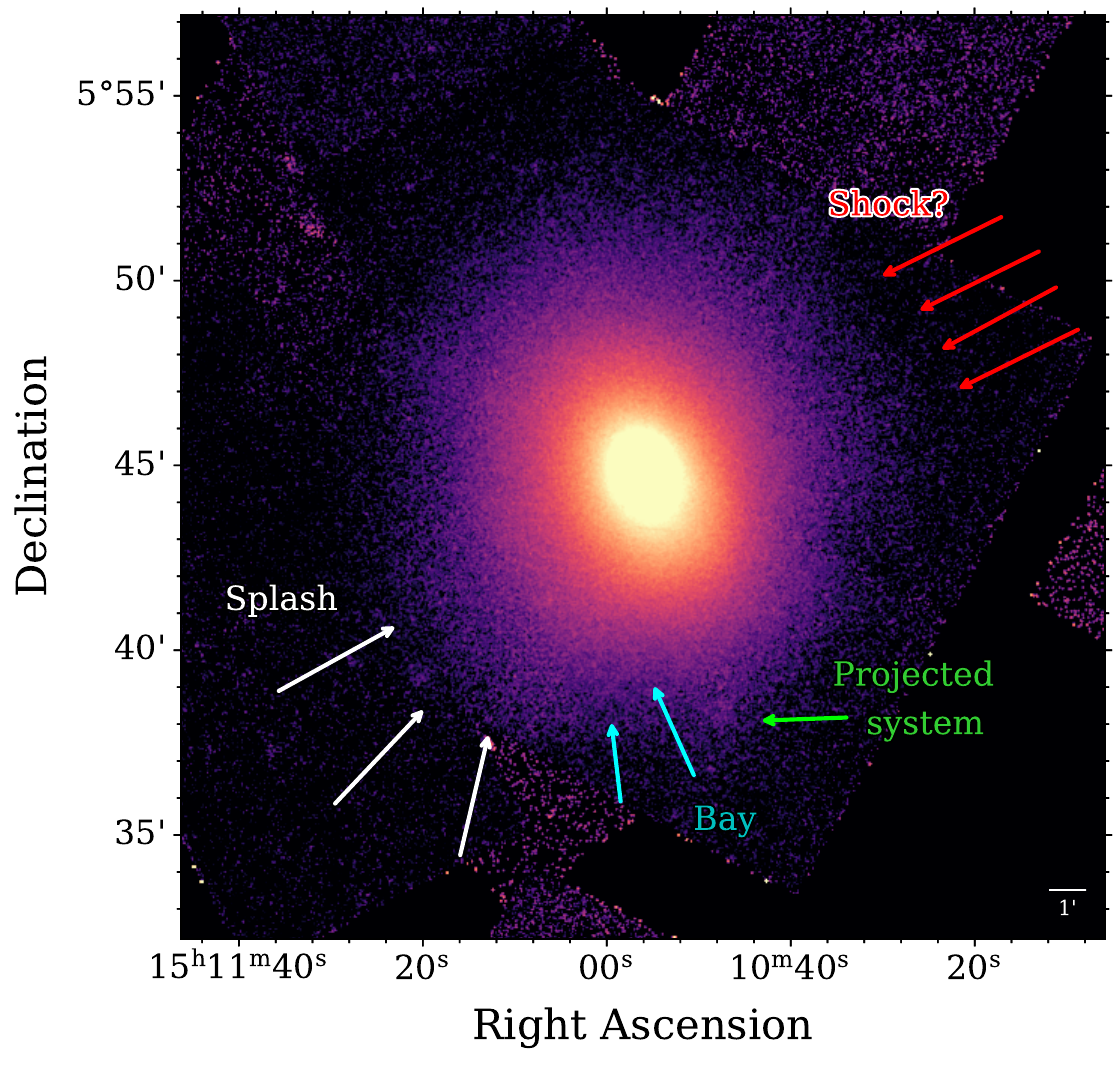} & \includegraphics[width=0.4\linewidth]{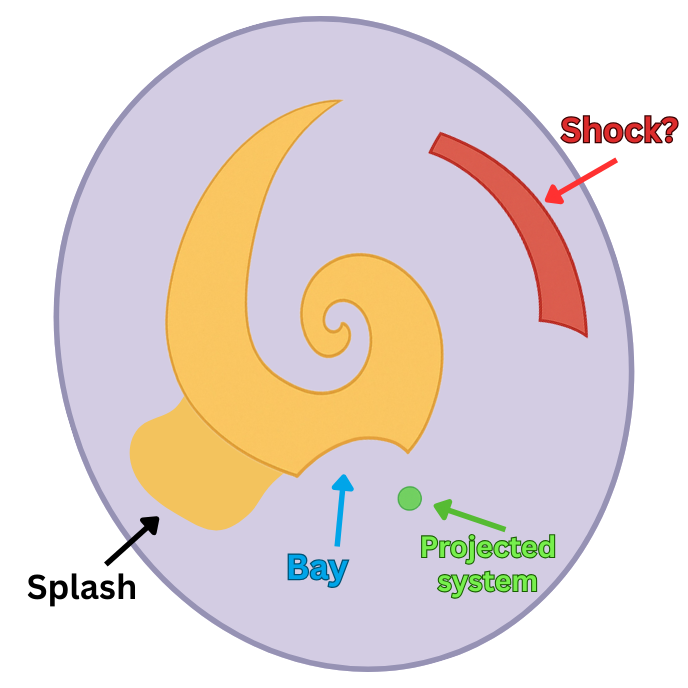}
\end{tabular}
    \caption{Residual emission (\emph{top center}) in the 0.7-7 band, created by subtracting a 2D $\beta$-model that was fit to the diffuse source image (\emph{bottom left}). The residual image was smoothed using a Gaussian of $\sigma=$7\farcs5, while the diffuse image was smoothed with $\sigma=3\arcsec$, and both panels cover a 25\arcmin$\times$25\arcmin\ area. Arrows indicate notable features, which are illustrated schematically in the cartoon (bottom right) and examined in more detail in \S\ref{sec:azprofs}. Note: $1\arcmin = 87$ kpc.}
    \label{fig:beta2d}
\end{figure*}

The fitted 2D $\beta$-model is subtracted from the unsmoothed X-ray image, and the resulting residual image is shown in Fig.~\ref{fig:beta2d}. The residual emission image was smoothed with a 7\farcs5 Gaussian, covers an area of 25\arcmin$\times$25\arcmin, and includes arrows pointing to some notable features. The sloshing ICM is easily identified with its spiral-like morphology, spiraling anti-clockwise towards the cluster center (or clockwise outwards). Notably, the spiral excess appears to wrap back to the N, a feature not observed in earlier, shallower Chandra studies \citep{Clarke2004, PM2013}. The spiral excess extends outwards to $\sim 430$ kpc (excluding the southeastern splash feature), while the SE ``splash'' (white arrows in Fig.~\ref{fig:beta2d}) extends $\sim 660$ kpc beyond the cluster core. 

A small clump of excess emission is seen in the SW (green arrow in Fig.~\ref{fig:beta2d}) that could be due to ICM emission from a fore- or background group or cluster (see discussion in \S\ref{sec:optical}), or the structure could be connected to the bay feature that is seen due east (cyan arrows). These X-ray features are explored further in \S\ref{sec:azprofs}. The red arrows in Fig.~\ref{fig:beta2d} indicate the region of a potential merger shock identified NNW of the cluster which is discussed in more detail in \S\ref{sec:shock}. A cartoon illustrating these features (the splash, bay, shock, and projected system) is shown in the bottom right panel of Fig.~\ref{fig:beta2d}.

\subsubsection{Gaussian Gradient Magnitude Filter}

A Gaussian Gradient Magnitude (GGM) filter was applied to the X-ray images using the \verb|gaussian_gradient_magnitude| function from \texttt{SCIPY} \citep{scipy2020}. This filtering method computes the gradients in the two-dimensional data under the assumption of Gaussian derivatives, thereby enhancing the visibility of surface brightness edges \citep{Sanders2016}. The key parameter in this technique is $\sigma$, which defines the width of the Gaussian kernel; smaller $\sigma$ values emphasize fine-scale structures, whereas larger $\sigma$ values reveal more extended features. 

\begin{figure}
\centering
\includegraphics[width=\linewidth]{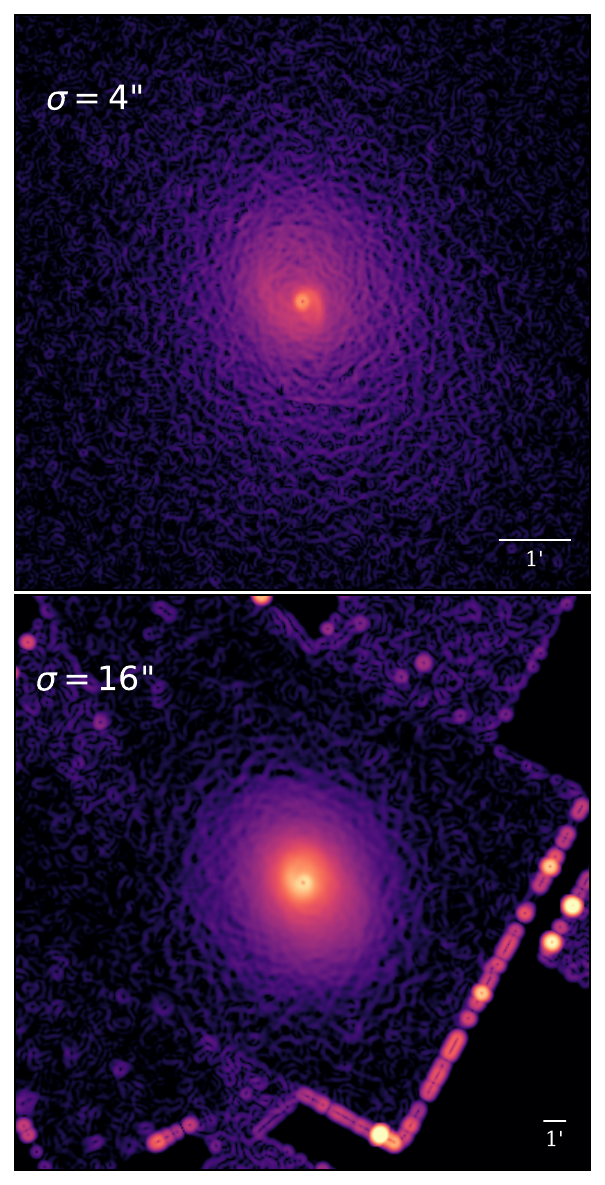}
    \caption{Gaussian gradient magnitude (GGM) filtered, with $\sigma = 4\arcsec$ (\emph{top}) and $16\arcsec$ (\emph{bottom}), images of A2029 in the 0.7-7 keV band covering a 25\arcmin$\times$ 25\arcmin\ area. The brightest regions correspond to the steepest changes in X-ray surface brightness. }
    \label{fig:ggm1}
\end{figure}

Fig.~\ref{fig:ggm1} shows the GGM-filtered images with $\sigma = 4\arcsec$ and $16\arcsec$ to show both small- and large-scale brightness variations over a 25\arcmin$\times$25\arcmin area. The finer-scale $\sigma$ is more sensitive to fine gradients near the core, whereas the larger $\sigma$ filtered image highlights broader, large-scale variations in surface brightness. In the finer scaled ($\sigma = 4\arcsec$) GGM filtered image (Fig.~\ref{fig:ggm1}, top panel), the sloshing cold front is easily seen towards the cluster core.  In the larger-scaled ($\sigma = 16\arcsec$) GGM filtered image (Fig.~\ref{fig:ggm1}, bottom panel), broader edges emerge that likely correspond to the outer boundaries of the sloshing spiral structure.  Taken together, they reinforce the presence of a sloshing spiral in A2029 without revealing additional major edges that would correspond to cavities or shocks. However, a more detailed analysis of the surface brightness edges in relation to AGN bubbles and potential shocks is presented in \S\ref{sec:sbps}.

\begin{figure}
\centering
\includegraphics[width=\linewidth]{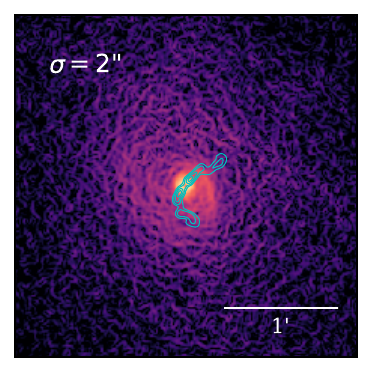}
\caption{Gaussian gradient magnitude (GGM) filtered, with $\sigma$ = 2\arcsec, image of A2029 in the 0.7-7 keV range and covering a 3\arcmin$\times$ 3\arcmin area. The brightest regions correspond to the steepest changes in X-ray surface brightness. The inner region of the sloshing spiral can be seen, starting in the cluster core and passing between the radio lobes, shown by the 1.4 GHz radio emission in cyan.}
\label{fig:ggm2}
\end{figure}

Fig.~\ref{fig:ggm2} shows a $\sigma=2\arcsec$ GGM filtered image, zoomed into the inner $3\arcmin\times 3\arcmin$ area, with the 1.4 GHz radio emission contours overlaid in cyan. The WAT lobes show some evidence of being swept back, following the sloshing motions towards the cluster core. This is consistent with what was previously seen in \cite{Clarke2004} and \cite{PM2013}.

\section{Spectral Analysis \label{sec:spec}}

We determine the thermodynamic properties of the ICM in A2029 by extracting X-ray spectra from our \chandra observations. The CIAO tool \verb|specextract| was used to extract the point-source free source spectra, binned such that each energy bin contained a minimum of 1 count per bin, with weighted RMFs and ARFs. Background spectra were taken from the reprojected blank-sky background observations; however, no background response files were created since the background was subtracted using the scaled blank-sky background observations. Unless otherwise noted, all source spectra, background spectra, and response files were extracted separately for each of our 24 ObsIDs.  Data from BI and FI CCDs were treated as separate data groups for each observation.

Extracted spectra are fit in the 0.7-7 keV range using \verb|XSPEC| (v12.14.1) \citep{Arnaud1996}. A corrective factor that depends on the observation specific ratio of exposure times and count rates in the hard-band (10-12 keV) range was applied to the observation specific background spectrum before subtracting it from the corresponding source spectrum. We adopted the \cite{Grevesse1998} (\verb|grsa|) solar abundance table, unless otherwise noted. Since spectra are binned to 1 count per bin, all fits are done using C-statistics. However, to report an estimate of the goodness of fit, we take the best-fit models and apply them to spectra that were re-binned to 40 counts/bin and do not re-fit. The resulting $\chi^2/DoF$ is what is reported here. For each spectral fit, the neutral hydrogen column density was fixed at the Galactic value, $N_H = 3.15\times 10^{20}$ cm$^{-2}$ \citep{Dickey1990}, and the redshift was fixed to $z=0.0767$ \citep{Sohn2019}.

We fitted the pre-2005 and new observations simultaneously, applying the appropriate scale factors to temperatures and abundances to the new set and including gain corrections on the 2023 group to account for systematic offsets related to evolving ACIS calibration. The scale factors and gain correction, determined by direct comparison with earlier data, correct for discrepancies that arise from time-dependent changes in detector response (see Appendix \ref{sec:cal}).

\subsection{Global Spectral Properties}

To compare with previous studies, we extract and fit spectra, with weighted responses from the inner 169 kpc, corresponding to the 116\arcsec region of \cite{Sarazin1992}, adjusted for our cosmology. The temperature within this region has been measured, using the A2029 ACIS-S observations, by \cite{Clarke2004} as $kT = $ \fe{7.27}{0.19}{0.26} keV and by \cite{PM2013} as $kT = $ \fe{7.54}{0.16}{0.16} keV, where the errors reported by both studies are 90\% confidence ranges. Using the same region, we obtain a single-temperature \texttt{APEC} fit of $kT =$ \fe{7.14}{0.02}{0.03} keV and $Z =$ \fe{0.75}{0.01}{0.01} $Z_{\odot}$ with $\chi^2_{red} =9926/8517 = 1.17$, consistent with \cite{Clarke2004} and \cite{Vikhlinin2005} and slightly below \cite{PM2013}. Differences in and abundance tables (they used \cite{Anders1989}) likely contribute. We emphasize that the statistical uncertainties on the projected temperature and abundance are significantly smaller than the systematic uncertainties introduced by our scaling factors (Appendix \ref{sec:cal}). Thus, small differences in the assumed scaling factor could produce larger shifts than the quoted statistical errors.

\begin{figure}
    \centering
    \includegraphics[width=\linewidth]{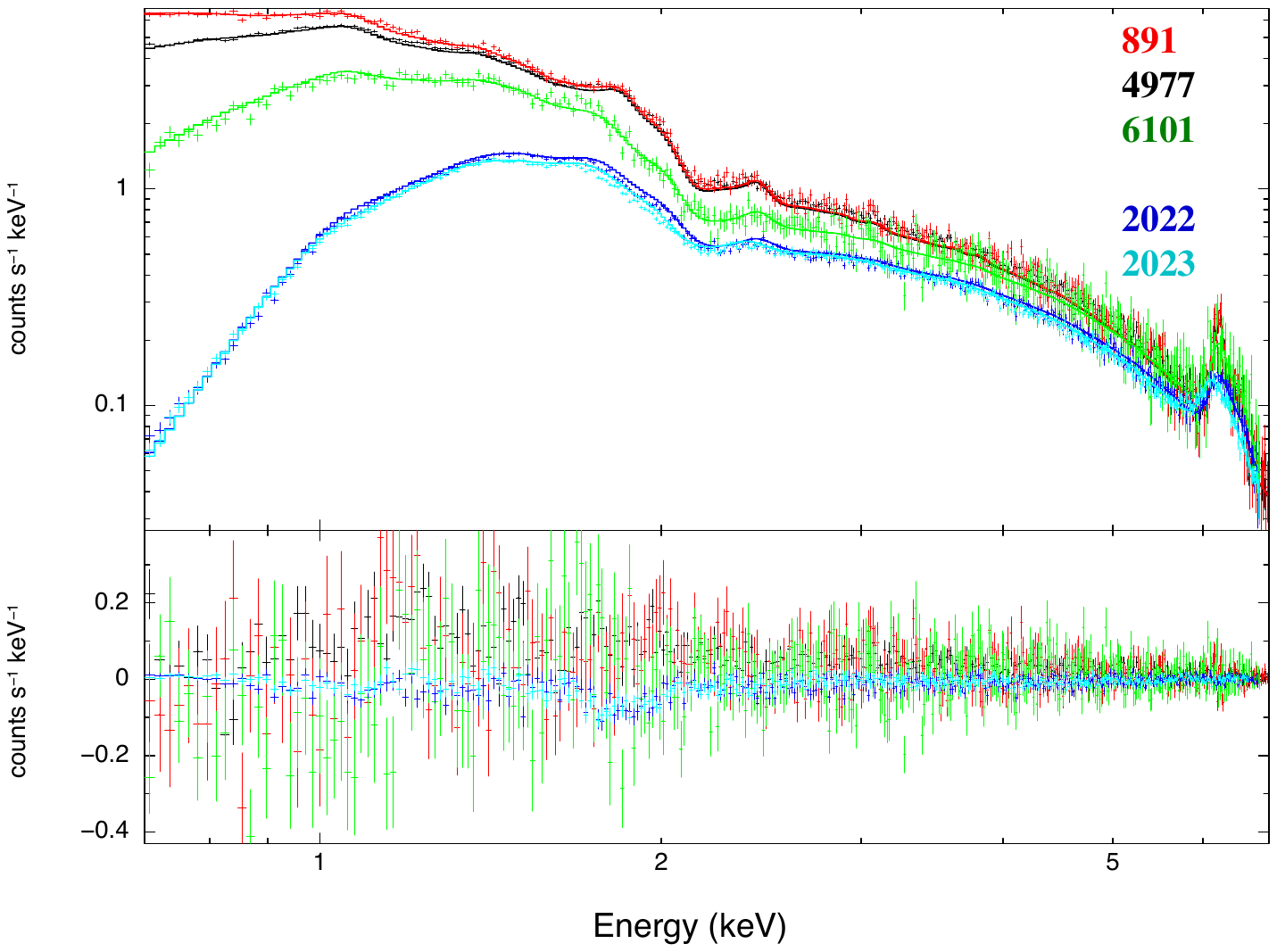}
    \caption{Chandra spectrum for A2029 in the 0.7-7 keV range for the central 116\arcsec region. The best-fitting cooling flow model is overlaid. Bottom panel shows the fit residuals.}
    \label{fig:spec}
\end{figure}
 
We have already noted a complex sloshing dynamic in the ICM; thus, there could be more than one temperature component of the gas in the fitted region. Fitting a double \texttt{APEC} model gives best fit parameters of $kT_{low} =$ \fe{5.63}{0.10}{0.11} keV, $kT_{high} =$ \fe{12.90}{0.18}{0.33} keV, $Z = $ \fe{0.82}{0.01}{0.01} $Z_{\odot}$, and $\chi^2_{red} = 9770/8516 = 1.15$. The appropriate scale factors were applied to the upper and lower temperatures and the abundance parameter of the new observations when tied to the fit parameters of the pre-2005 observations, following Appendix \ref{sec:cal}. Running an F-test to compare the single and double \texttt{APEC} fits shows significant improvement with the second \texttt{APEC} component, with an F-statistic of 45.3 and probability $4.8\times 10^{-29}$. In comparison, with the older, pre-2005 dataset, \cite{PM2013} find best fit values of $kT_{low} = $ \fe{5.97}{0.56}{0.68} keV and $kT_{high} =$ \fe{11.16}{33.7}{1.63} keV and abundance $Z = $ \fe{0.52}{0.03}{0.02} $Z_{\odot}$, where they give the 90\% confidence ranges. Thus, we are able to obtain much tighter constraints on the upper and lower confidence of the $kT_{high}$ component. 

In the case of A2029, the mass deposition rate, $\dot{M}$, is still debated. \cite{Sarazin1992} and \cite{Peres1998} used ROSAT imaging to estimate a mass deposition rate within the central 116\arcsec\ and found that rates of $>100$ M$_{\odot}$ yr$^{-1}$ (scaled to our cosmology) were needed to match observations. From spectra extracted in the same region, \cite{PM2013} used the old Chandra ACIS-S observations (ObsID 891 \& 4977) and measured an even higher rate of $\dot{M} =$ \fe{783}{371}{268} M$_{\odot}$ yr$^{-1}$, cooling over a range of \fe{10.08}{2.30}{1.12} to \fe{4.24}{0.80}{0.67} keV. In contrast, other studies find much lower values --- e.g., $\dot{M} \sim 56$ M$_\odot$ yr$^{-1}$ \citep{Clarke2004};  however, this value is for ICM cooling to very low temperatures, 0.08 keV. Our analysis methods are most comparable to \citet{PM2013} and \citet{Clarke2004}, who both fit spectroscopic cooling flow models, allowing for a more direct comparison of results.

 In the same 116\arcsec\ region, we fit the pre-2005 and new observations to an \texttt{APEC}+\texttt{MKCFLOW} cooling flow model, including Galactic absorption, applying the appropriate scale factors to temperatures and abundance to the new set, and including gain corrections on the 2023 group (see Appendix \ref{sec:cal}). The upper temperature of the \texttt{MKCFLOW} component is tied to the \texttt{APEC} temperature. However, we also explore keeping the low temperature, $kT_{low}$, fixed to the \texttt{XSPEC} lower minimum of 0.0808 keV. All these results are summarized in Tab.~\ref{tab:fittests}. 
 
When allowing both the upper and lower temperatures of the model to vary, we find $kT_{low} =$ \fe{3.71}{0.12}{0.12} keV and $kT_{high} =$ \fe{11.82}{0.30}{0.31} keV, $Z =$\fe{0.77}{0.01}{0.01} $Z_\odot$, and $\dot{M} =$ \fe{784}{40}{36} M$_\odot$ yr$^{-1}$ with $\chi^2_{red} = 11554/8516 = 1.36$. We are consistent with what was seen in \cite{PM2013} where we see a high $\dot{M}$ with cooling over a limited temperature range, down to a lower temperature of 3.7 keV. When $kT_{low}$ is fixed at 0.0808 keV, we find  $kT_{high} =$ \fe{7.18}{0.04}{0.04} keV,  $Z =$ \fe{0.83}{0.01}{0.01} $Z_\odot$, and $\dot{M} =$\fe{51}{3}{3} M$_\odot$ yr$^{-1}$ with $\chi^2_{red} =13859/8517 = 1.63$. This scenario is more consistent with what was seen in \cite{Clarke2004} with a much lower $\dot{M}$ of $\sim 50$ M$_\odot$ yr$^{-1}$, but cooling all the way down to low temperatures (0.08 keV). The fit with the higher $\dot{M}$ and cutoff in the low temperature of $\sim$ 3.7 keV, however, is more preferred based on the fit statistics.

Additionally, we fit the pre-2005 and new observations within just the central 40\arcsec, encompassing the central AGN and radio emission since this would provide a mass deposition rate that is better to compare directly with the star formation rate. This region is contained within the cD galaxy. We again fit all observations to an \texttt{APEC}+\texttt{MKCFLOW} model, with Galactic absorption, applying the appropriate scale factors to temperatures and abundance to the new set, and including gain corrections on the 2023 group (see Appendix \ref{sec:cal}). Again, the upper temperature of the \texttt{MKCFLOW} component is tied to the \texttt{APEC} temperature. These results are listed in Table \ref{tab:fittests}. Letting both temperature components vary, we find best-fit parameters of $kT_{low} =$ \fe{3.31}{0.12}{0.16} keV, $kT_{high} =$ \fe{10.26}{0.40}{0.29} keV, $Z = $ \fe{0.96}{0.02}{0.02} $Z_\odot$, and $\dot{M} =$\fe{348}{20}{24} M$_\odot$ yr$^{-1}$ with $\chi^2_{red} = 7393/6110 = 1.21$. Keeping $kT_{low}$ fixed at 0.0808 keV yielded $kT_{high} =$ \fe{6.21}{0.05}{0.04}, $Z = $ \fe{1.01}{0.02}{0.02} $Z_\odot$, and $\dot{M} =$ \fe{17}{2}{2} M$_\odot$ yr$^{-1}$ with $\chi^2_{red} = 8216/6111 = 1.34$. This latter $\dot{M}$ value is best used for comparing directly with the star formation rate in the cD galaxy. This $\dot{M}$ is still higher than the measured star formation rate in the cD, as discussed in \S~\ref{sec:bubbles}.

Previous high–resolution grating work with XMM-Newton/RGS \citep{Liu2019} placed constraints on gas cooling below $\leq$1 keV in A2029, with upper limits on mass-deposition rates far below the classical cooling-flow prediction. Our Chandra data reveal significant emission from gas at intermediate temperatures of $\sim$3–4 keV, implying significant gas cooling down to these levels. Thus our results extend the earlier RGS constraints by demonstrating that while cooling to sub-keV gas remains strongly suppressed, substantial amounts of gas may be cooling down to a few keV within the core. Note that cooling to 1/2 to 1/3 of the starting temperature is seen very often in cooling flows \citep[e.g., A2052;][]{Blanton2001,Blanton2003,Blanton2009,Blanton2011}.

\begin{deluxetable}{clccccr}
\tablecaption{\texttt{XSPEC} fits to inner regions of A2029. Uncertainties are 1$\sigma$ confidence levels.\label{tab:fittests}}
\tablehead{
    \colhead{} & 
    \colhead{Model} &
    \colhead{$kT_{low}$} & 
    \colhead{$kT_{high}$} & 
    \colhead{Abundance} & 
    \colhead{$\dot{M}$} &\colhead{$\chi^2$/DoF\tablenotemark{*}} \\
    \colhead{} & \colhead{} & \colhead{[keV]} & \colhead{[keV]} & \colhead{$Z_\odot$} & \colhead{[M$_\odot$ yr$^{-1}$]} & \colhead{} 
} 
\startdata
\multicolumn{7}{l}{$r= 116\arcsec$ (169 kpc)  region\tablenotemark{a}} \\\hline
& \texttt{APEC} & \nodata & \fe{7.14}{0.02}{0.03} & \fe{0.75}{0.01}{0.01} & \nodata & $9926/8517 = 1.17$\\
& \texttt{APEC}+\texttt{APEC} & \fe{5.63}{0.10}{0.11} & \fe{12.90}{0.18}{0.33} & \fe{0.82}{0.01}{0.01} & \nodata & $9770/8516 = 1.15$ \\
& \texttt{APEC}+\texttt{MKCFLOW} & \fe{3.71}{0.12}{0.12} & \fe{11.82}{0.30}{0.31} & \fe{0.77}{0.01}{0.01} & \fe{784}{40}{36} & $11554/8516 = 1.36$ \\
& \texttt{APEC}+\texttt{MKCFLOW} & [0.0808] & \fe{7.18}{0.04}{0.04} & \fe{0.83}{0.01}{0.01} & \fe{51}{3}{3} & $13859/8517 = 1.63$ \\ \hline \hline
 \multicolumn{7}{l}{$r= 40\arcsec$ (58 kpc) region\tablenotemark{b}} \\\hline
& \texttt{APEC}+\texttt{MKCFLOW} & \fe{3.31}{0.12}{0.16} & \fe{10.26}{0.40}{0.29} & \fe{0.96}{0.02}{0.02} & \fe{348}{20}{24} & $7393/6110 = 1.21$ \\
& \texttt{APEC}+\texttt{MKCFLOW} & [0.0808] & \fe{6.21}{0.05}{0.04} & \fe{1.01}{0.02}{0.02} & \fe{17}{2}{2} & $8216/6111 = 1.34$
\enddata
\tablenotetext{*}{From applying best-fit model to re-binned spectra (see \S\ref{sec:spec})}
\tablenotetext{a}{Following \cite{Sarazin1992, Clarke2004, PM2013}}
\tablenotetext{b}{Innermost cD region which is better proxy for SFR}
\end{deluxetable}

\subsection{Temperature and Abundance Maps}

Spectral maps of A2029 were created following the methods of \cite{Randall2008, Randall2009}. For each spectral map pixel (9\farcs8 pixel$^{-1}$), spectra were extracted from circular regions whose radius was allowed to increase until a minimum of 20,000 background-subtracted counts were contained. The radius of the extraction region varies from $\sim$10\arcsec\ ($\sim$ 14 kpc) in bright regions near the core to $\sim350\arcsec$\ ($\sim$ 509 kpc) in faint regions near the edges of the map. Within this region, the spectrum of each ObsID is extracted and then fit simultaneously using an absorbed \texttt{APEC} model, with only the hydrogen column density --- $N_H$ --- and redshift --- $z$ --- frozen during fitting. 

\begin{figure*}
    \centering
  \includegraphics[width=\textwidth]{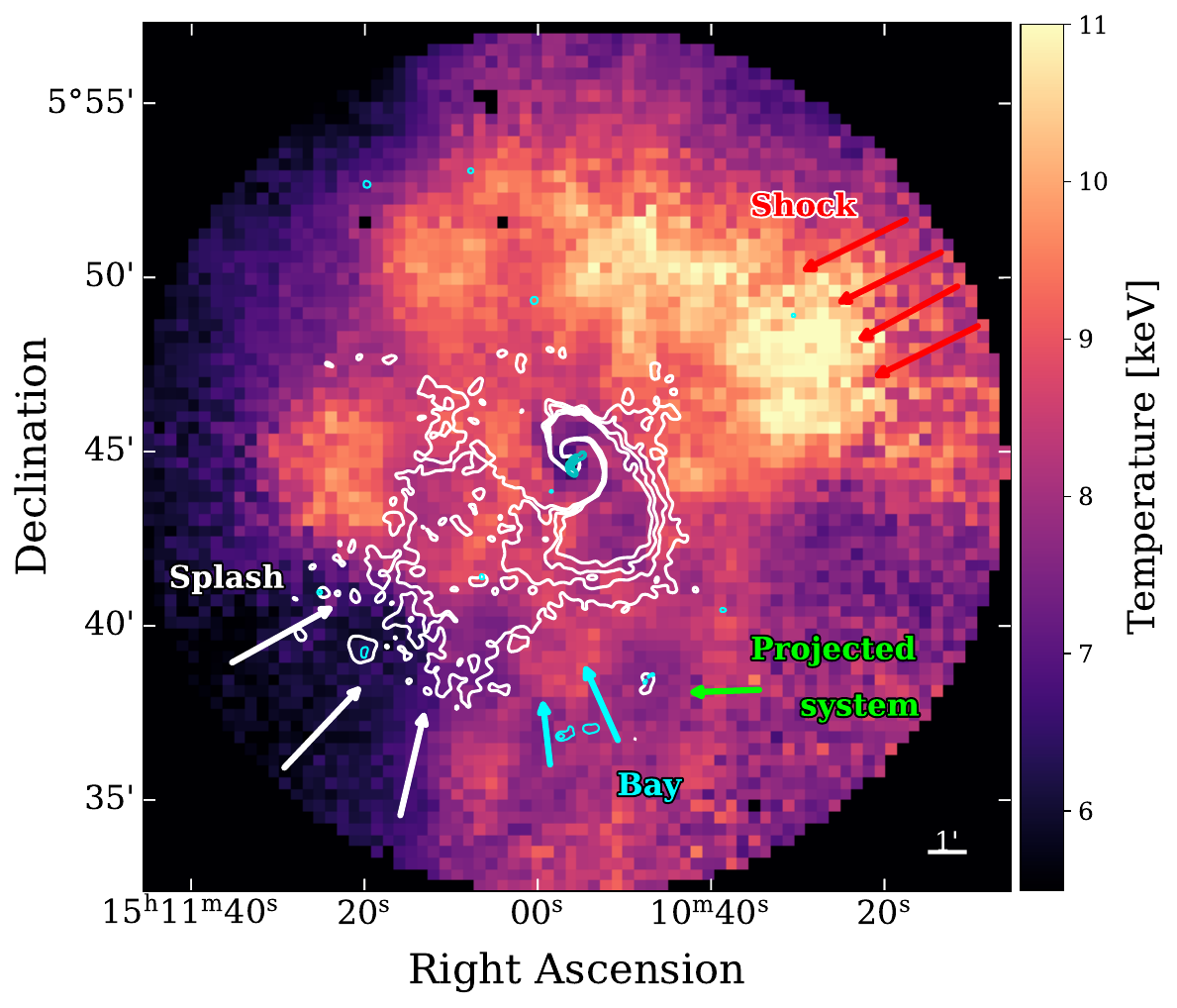} 
    \caption{Projected temperature map of A2029. Overlaid are contours of residual X-ray emission (white) and 1.4 GHz radio emission (blue). Colored arrows are the same as those shown in Fig.~\ref{fig:beta2d}. The appropriate scale factor was applied to the fitted temperatures of the new data (see Appendix \ref{sec:cal}) }
    \label{fig:tmap}
\end{figure*}

\begin{figure*}
    \centering
  \includegraphics[width=\textwidth]{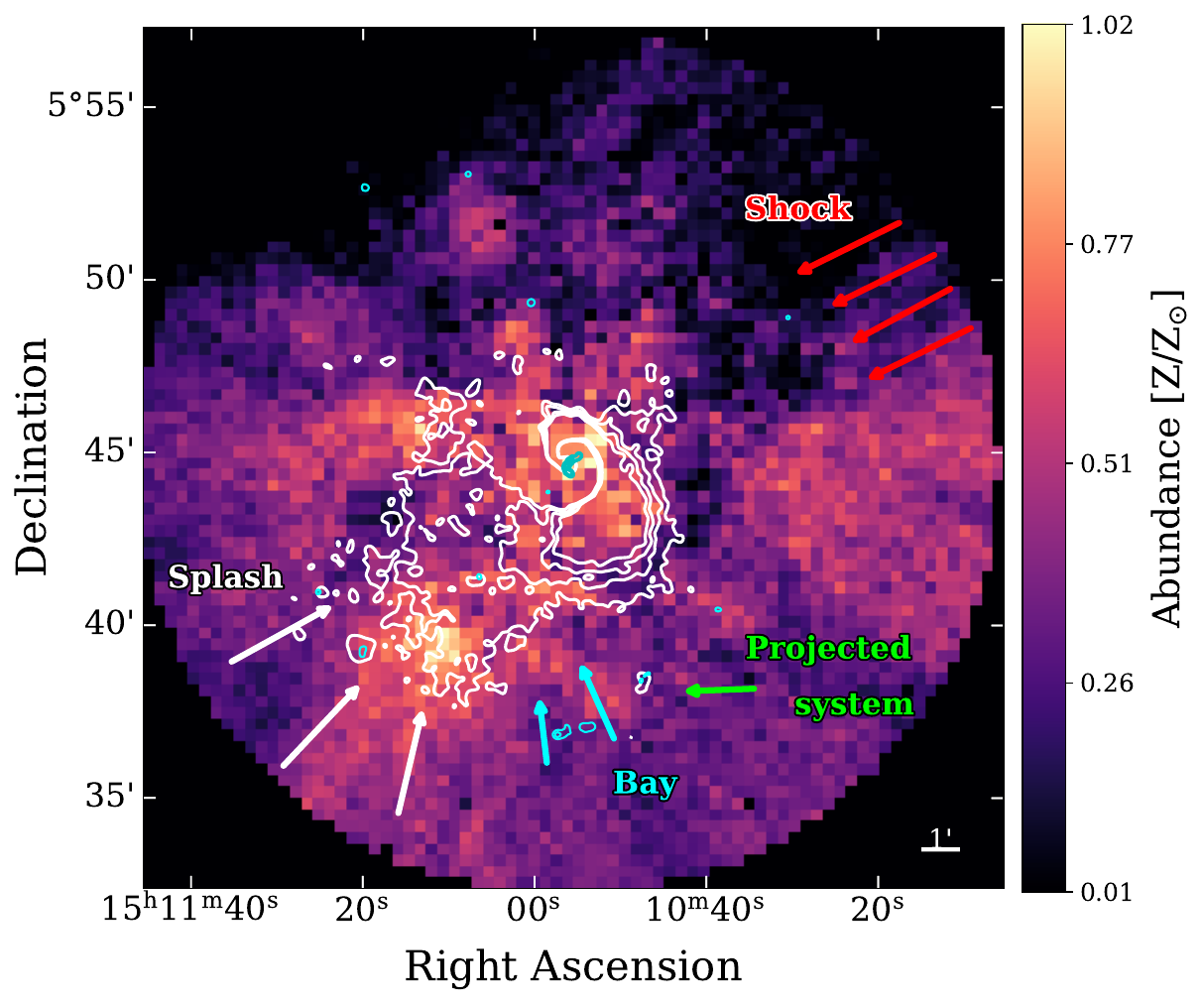} 
    \caption{Projected abundance map of A2029. Overlaid are contours of residual X-ray emission (white) and 1.4 GHz radio emission (blue). Colored arrows are the same as those shown in Fig.~\ref{fig:beta2d}. The appropriate scale factor was applied to the fitted temperatures of the new data (see Appendix \ref{sec:cal}) }
    \label{fig:abund}
\end{figure*}

Figures~\ref{fig:tmap} and \ref{fig:abund} show the projected temperature and abundance maps of A2029, respectively, with contours of excess emission (white) and 1.4 GHz radio emission (blue) overlaid. The appropriate scale factors on temperature and abundance were applied, following Appendix \ref{sec:cal}. Errors on the temperature map range from  $\sim 3\%$ in the cluster core to $\sim6\%$ in the outskirts. Errors on the abundance map range from $\sim15\%$ in the core to $\sim40$\% in the outskirts. Note that in low surface brightness regions, the extraction apertures become larger than the map pixel size, effectively smoothing the maps toward the cluster outskirts. Thus, small-scale features near the outskirts may be less sharply resolved than those near the core.

The innermost spiral excess (highlighted by the white contours) traces $\sim 7.5$ keV gas temperatures and higher metallicities, as expected from sloshing ICM. The SW ``splash'' feature (white arrows in Fig.~\ref{fig:tmap}) traces regions of $\sim 7-8$ keV gas and leads into the coldest regions, with temperatures around $5-6$ keV. This cold region was previously seen in XMM and Chandra observations of A2029 \citep{Bourdin2008, Walker2012}. The splash also leads into a region of higher ($\sim0.9$ Z$_{\odot}$) metallicity (Fig.~\ref{fig:abund}). The region of isolated excess emission to the SW of the cluster (green arrow in Fig.~\ref{fig:tmap}) corresponds to a region of colder ($\sim7$ keV) gas, which may be associated with a background group or cluster (discussed further in \S\ref{sec:optical}). A large ($\sim$ 0.9 Mpc) hot ($\sim 11$ keV) region to the NW of the cluster center may indicate shock-heated gas in the region. 

While the spectral maps aid in visualizing the thermodynamic structure of the ICM, they are only meant to guide the eye, and more quantitative analyses must be performed on identified features to determine their robustness and potential origins. In order to determine the significance of the features seen in the spectral maps, follow-up detailed analysis is performed on a number of regions of interest and is discussed further in \S\ref{sec:sbps} and \S\ref{sec:azprofs}.

\section{Results and Discussion}

\subsection{Mapping the Sloshing Spiral\label{sec:sbps}}

Substructure within the ICM can lead to anisotropies of the X-ray surface brightness. Cluster-cluster mergers can result in enormous amounts of energy being poured into the ICM, often redistributed through shock fronts, turbulence, and sloshing motions that mix hot and cold gas. AGN feedback can displace the surrounding ICM and previous bursts of AGN activity can result in ``bubbles'' which buoyantly rise to larger radii. Whether driven by mergers or AGN feedback, substructure in the ICM can be observed through the mark these processes leave behind on the X-ray brightness. 

\begin{figure}
    \centering
    \includegraphics[width=\linewidth]{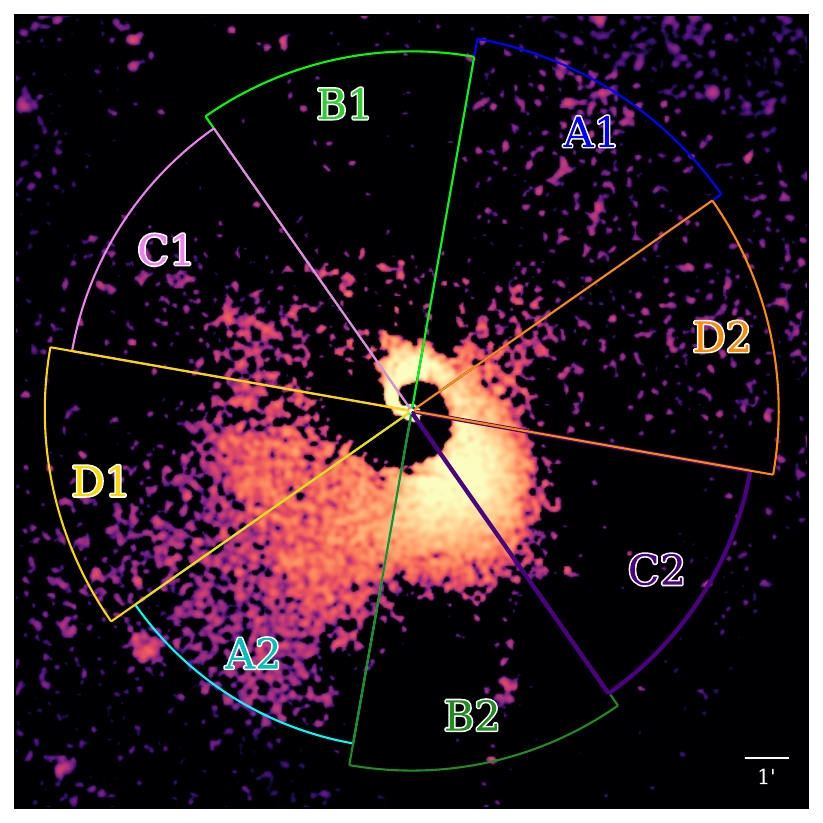}
    \caption{Residual emission image, smoothed with $\sigma = 7\farcs5$ Gaussian, in the 0.7-7 keV range. Overlaid are the four sets of diametrically opposed wedge regions, labeled A1/A2 (blue/cyan), B1/B2 (light/dark green), C1/C2 (magenta/purple), and D1/D2 (yellow/ orange).}
    \label{fig:profileregs}
\end{figure}

\begin{figure*}
\centering
\begin{tabular}{cc}
  \includegraphics[width=0.46\linewidth]{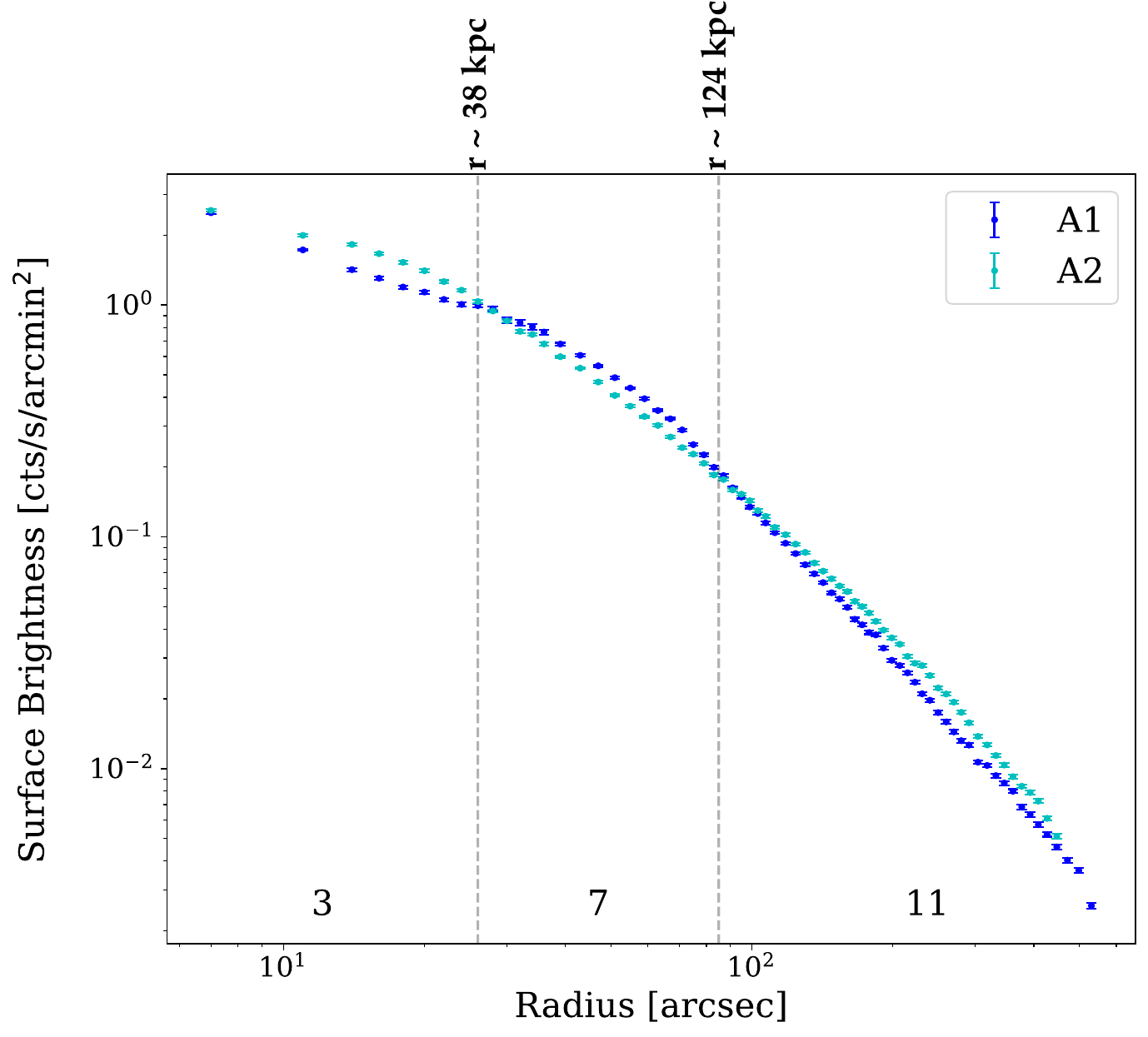} &\includegraphics[width=0.46\linewidth]{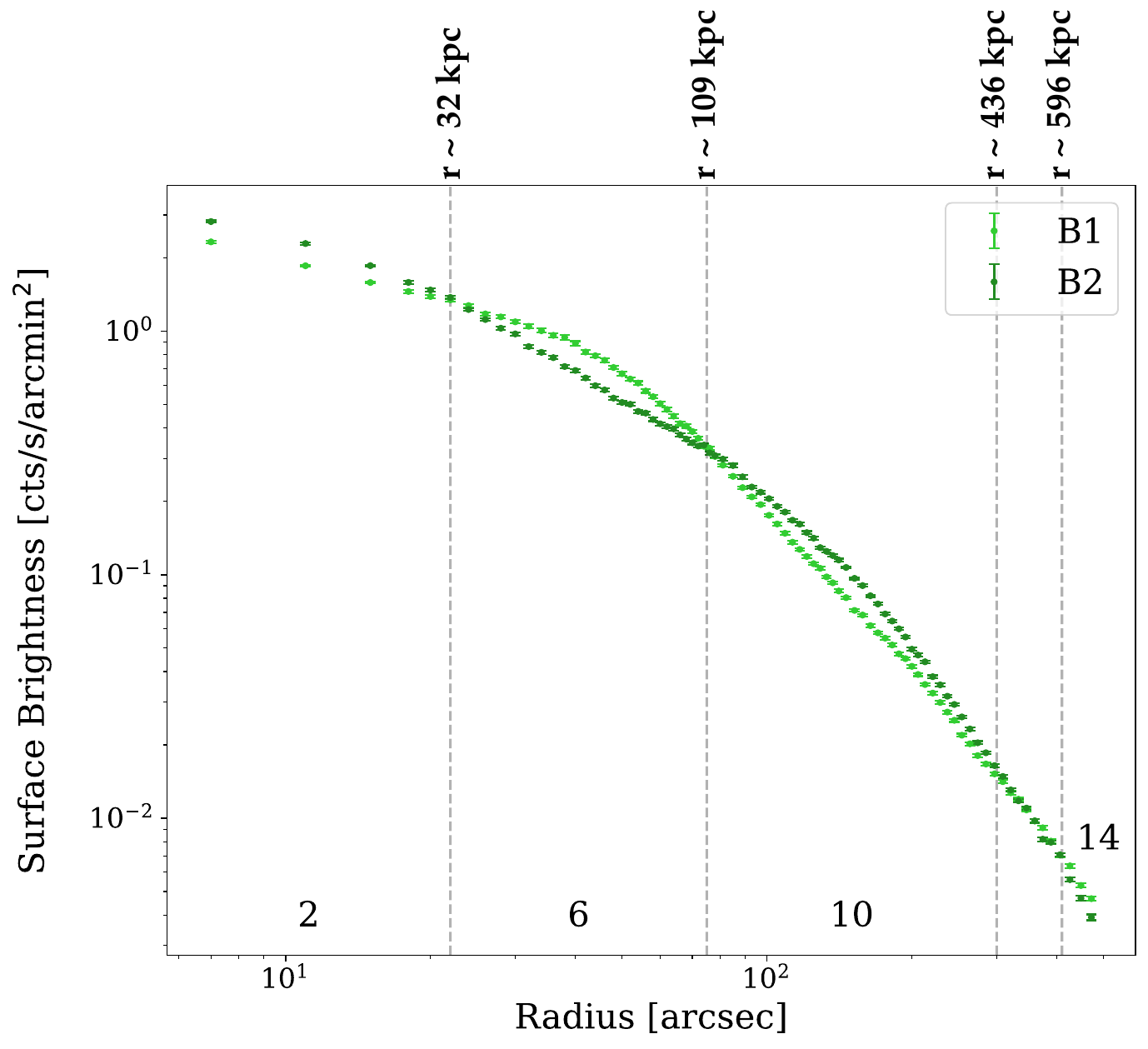} \\
   \includegraphics[width=0.46\linewidth]{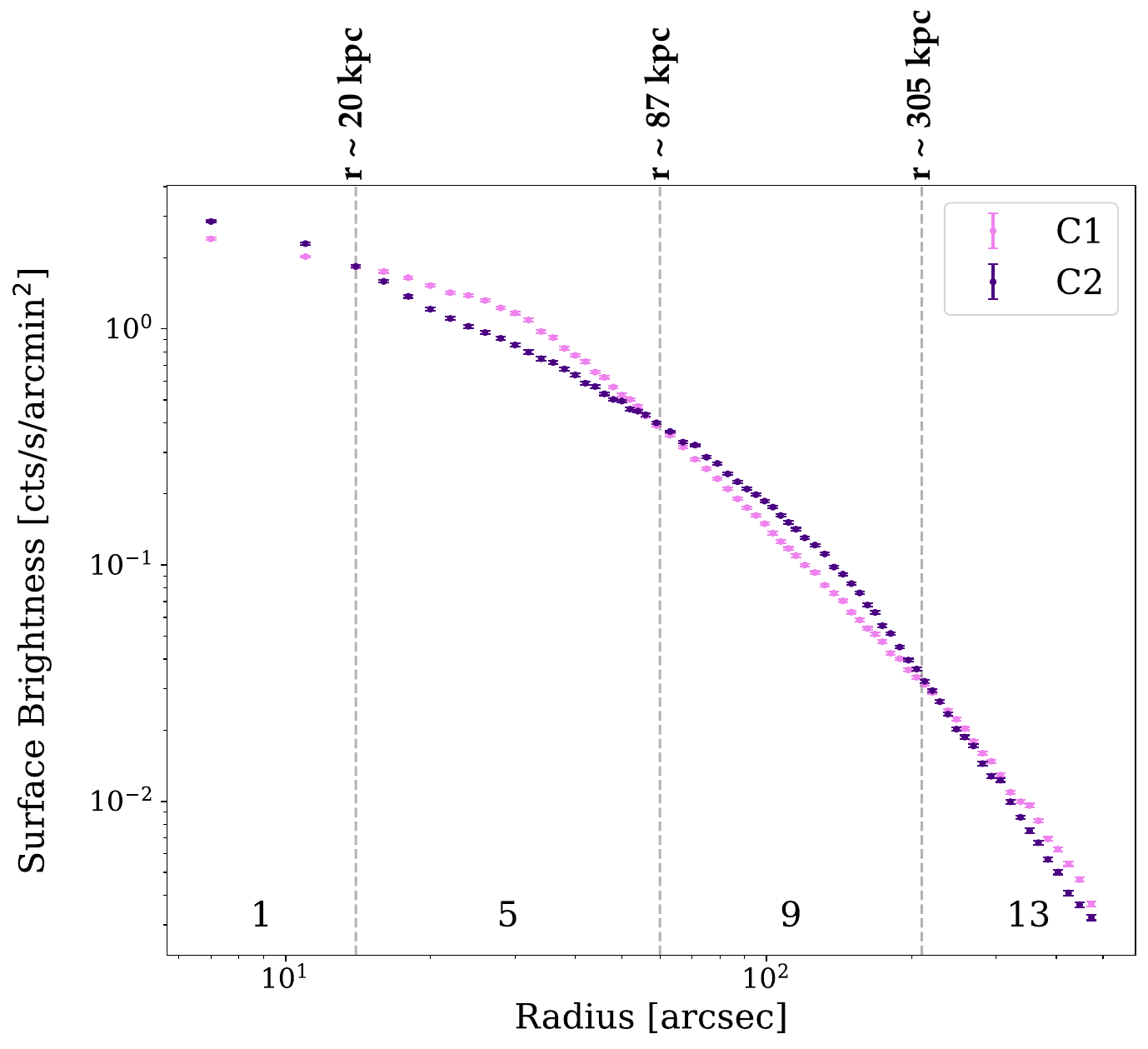} &\includegraphics[width=0.46\linewidth]{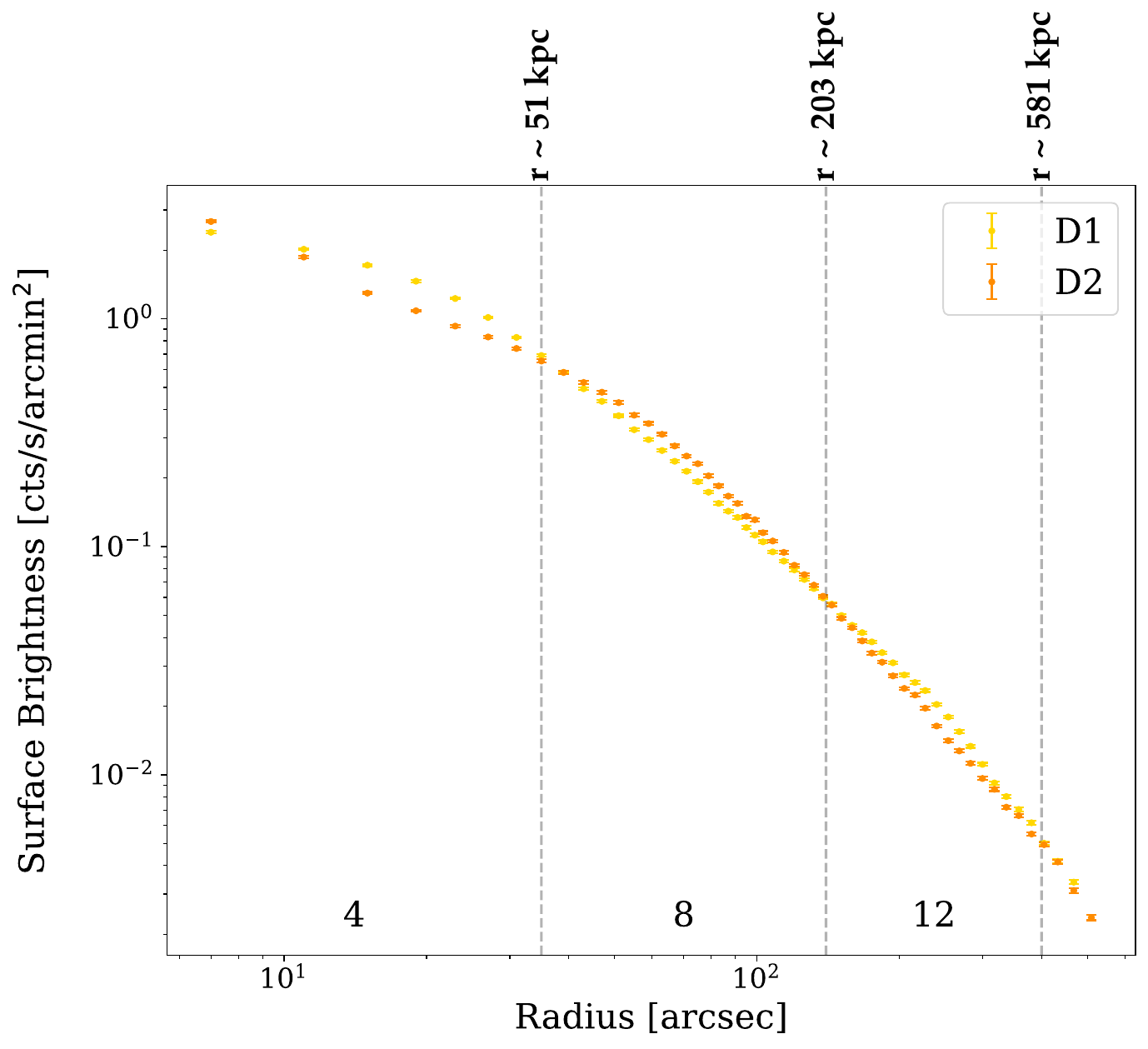} \\
\end{tabular}
\caption{Surface brightness profiles in the 0.7-7 keV energy range extracted from diametrically opposed wedge regions, shown in Fig.~\ref{fig:profileregs}. The legend labels and point colors correspond to the labels and colors shown in  Fig.~\ref{fig:profileregs}. Dashed lines indicate the radius where the opposing profiles ``criss-cross'', denoting the regions where one side of the ICM shows excess emission compared to the other at the same radius. The numerical labels at the bottom of each plot are used to denote different regions of excess emission, which are highlighted in Fig.~\ref{fig:spiralexcess}. \label{fig:profiles}}
\end{figure*}

 By comparing the surface brightness in wedge regions of opposing angular extent but at the same radius from the cluster center\footnote{Defined at the peak of the X-ray surface brightness, found during the $\beta$-model fitting in \S\ref{sec:beta2d}}, we can map out the regions of excess emission, corresponding to areas where the cooler, brighter core gas has been sloshed out to larger radii, through the ``criss-crossing'' of the radial profiles. To do this, we define four sets of diametrically opposed wedge-shaped regions, centered on the peak of the X-ray emission as found in the 2D-$\beta$ fitting of \S\ref{sec:beta2d}, which are shown in Fig.~\ref{fig:profileregs} and labeled A1/A2 (blue/cyan), B1/B2 (light/dark green), C1/C2 (magenta/purple), and D1/D2 (yellow/ orange). We extract the surface brightness profile in each of these regions and show opposing profiles in Fig.~\ref{fig:profiles}. Error bars are 1$\sigma$ confidence levels, but are often smaller than the plotted point. Dashed vertical lines in Fig.~\ref{fig:profiles} mark the locations of the excesses in each region, which is labeled by a number at the bottom of each plot. 

Relative to region A1, region A2 shows excess emission (\#3) out to a radius of $\sim 38$ kpc [26\arcsec], where the profiles cross, and then region A1 shows excess emission. Region A1 remains in excess (\#7) of the brightness in the opposing region A2 until a radius of $\sim 124$ kpc [85\arcsec] from the core, where the profiles cross yet again, and region A2 shows a brightness excess (\#11) over A1. Beyond $\sim124$ kpc [85\arcsec], region A2 remains in excess of the opposing A1. 

Regions B1 and B2 most closely match the direction of ellipticity for the overall X-ray emission (see \S\ref{sec:beta2d}). The southern B2 region shows excess emission (\#2) until $\sim 32$ kpc [22\arcsec] and then again between $\sim 109$ kpc [75\arcsec] to about $\sim 436$ kpc [300\arcsec] (\#10). The northern B1 region shows excess emission between $\sim 32$ kpc [22\arcsec] and $\sim 109$ kpc [75\arcsec] (\#6) and then again beyond $\sim596$ kpc [410\arcsec] (\#14). Between $\sim436$ kpc [300\arcsec] and $\sim596$ kpc [410\arcsec], the regions remain at roughly the same brightness, as would be expected for ICM gas at similar radii in opposite directions for a fully relaxed cluster.

In the opposing regions C1 and C2, the southern region is in excess out to a radius of $\sim 20$ kpc [14\arcsec] (\#1) and again between $\sim 87$ kpc [60\arcsec] to $\sim 305$ kpc [210\arcsec] (\#9). The northern region shows excesses between $\sim 20$ kpc [14\arcsec] and $\sim 87$ kpc [60\arcsec] (\#5) and then remains in excess beyond $\sim305$ kpc [210\arcsec] (\#13). 

Within regions D1 and D2, the northern region shows excess emission (\#4) out to $\sim 51$ kpc [35\arcsec] and again between $\sim 203$ kpc [140\arcsec] and $\sim 581$ kpc [400\arcsec] (\#12). The southern D2 region shows excess emission between $\sim51$ kpc [35\arcsec] and $\sim203$ kpc [140\arcsec] (\#8). 

\begin{figure*}
\centering
\begin{tabular}{cc}
   \includegraphics[width=0.5\linewidth]{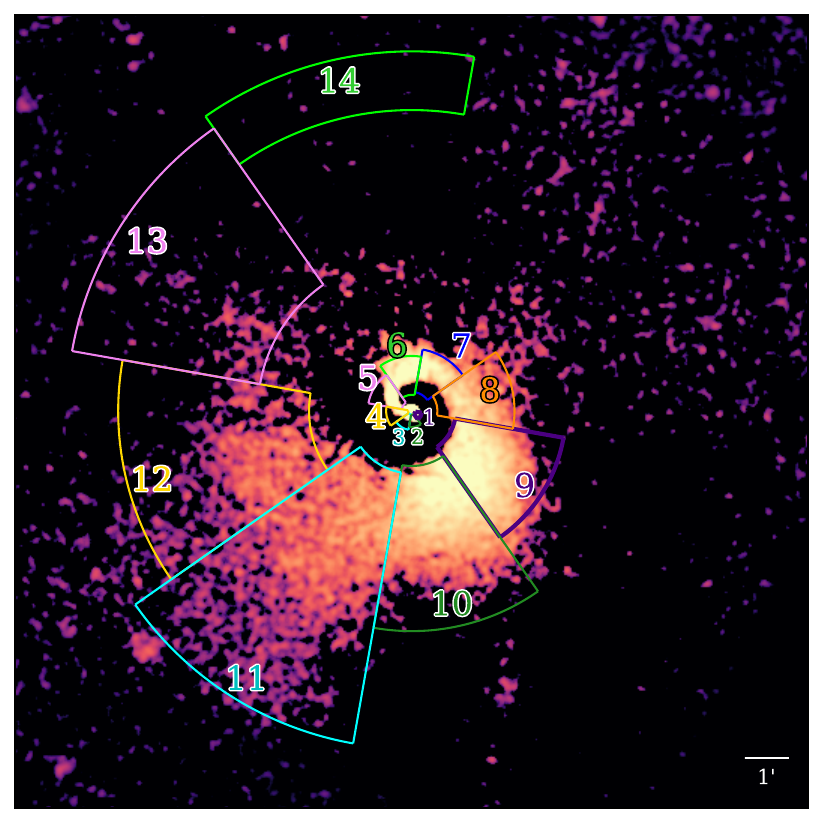} &
   \includegraphics[width=0.5\linewidth]{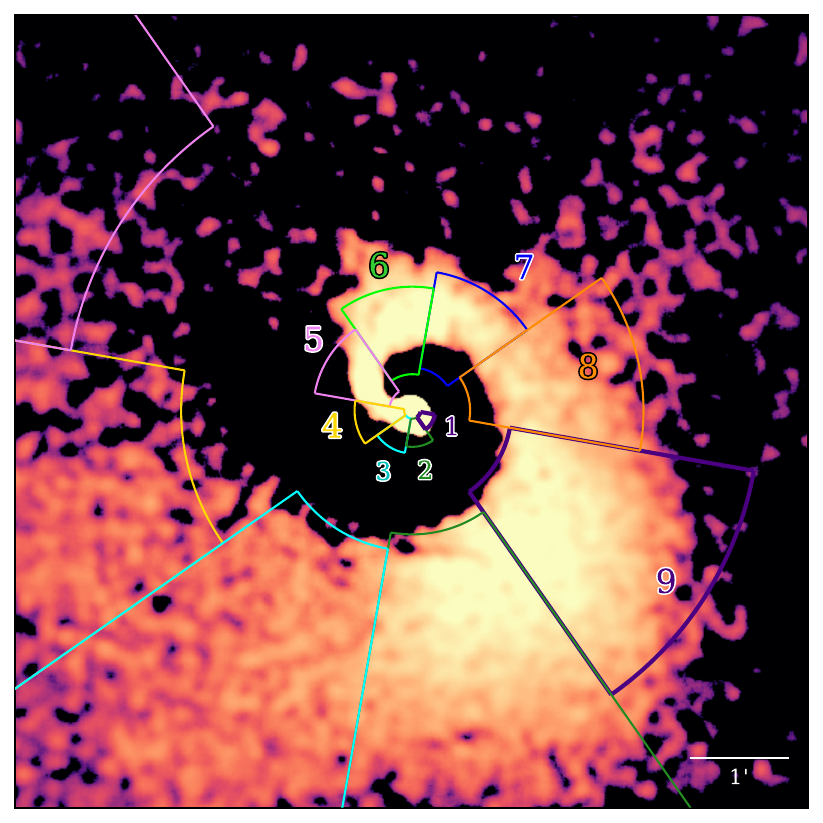}
   \end{tabular}
    \caption{2D-beta model-subtracted residual image of A2029, smoothed with $\sigma = 7\farcs5$, in the 0.7-7 keV range. Overlaid are the sectors containing excess surface brightness in comparison to a sector at a diametrically opposed angle. The numerical labels and colors correspond to those shown in each panel of Fig.~\ref{fig:profiles}. The spiral pattern is immediately evident and extends $\sim600$ kpc from the cluster core. The right panel shows a zoom-in on the inner 8\arcmin $\times$ 8\arcmin area to better show innermost region.}
    \label{fig:spiralexcess}
\end{figure*}

For each of the opposing wedge regions, we highlight the radius in which a given side was found to be in excess, based on the radial profiles of Fig.~\ref{fig:profiles}. These are shown in Fig.~\ref{fig:spiralexcess} and numbered according to the numbers denoting each excess region in Fig.~\ref{fig:profiles}. The spiral pattern is clearly visible. Notably, the radial profiles indicate that the spiral pattern extends far more to the North than what is apparent by eye in the X-ray and residual images.

\subsection{Potential Merger Shock \label{sec:shock}}

We used the python package \texttt{pyproffit} \citep[v0.7.0;][]{Eckert2020} to plot the surface brightness profile in a region over the large hot region observed to the NW in the temperature map (Fig.~\ref{fig:tmap}). We used a background subtracted image with point sources removed, an exposure map in units of seconds, and a user-generated error map to extract the surface brightness profile of the shock region using \texttt{pyproffit}. The surface brightness profile was extracted in a circular sector spanning 1.5$^\circ$ to $60^\circ$, centered on the peak of the X-ray emission, with a maximum radial extent of 630\arcsec, and a bin size of 10\arcsec\ in order to ensure a minimum of 5,000 net counts in the outermost bin.

\begin{figure}
    \centering
    \includegraphics[width=\linewidth]{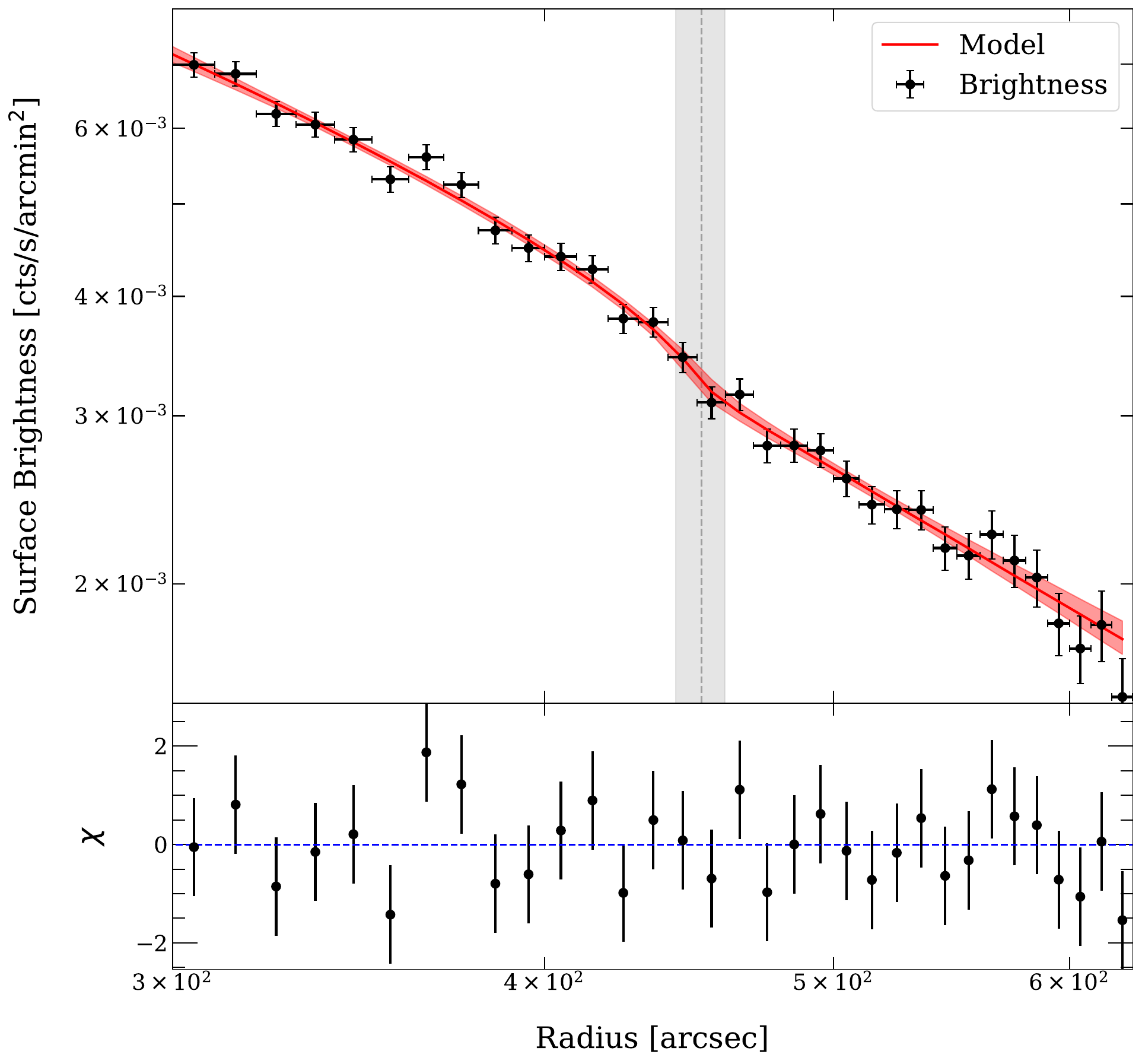}
    \caption{Surface brightness profile and best fitting broken power-law density model (red) of the shock region (see blue region in Fig.~\ref{fig:shock}). The edge is fit at $r_{jump} =$ \fe{451}{8}{9} arcsec = \fe{656}{12}{13} kpc (denoted by the vertical dashed line with the 1$\sigma$ confidence range shaded gray) from the cluster core and has a corresponding density jump $\frac{\rho_2}{\rho_1} = $ \fe{1.17}{0.06}{0.05}.}
    \label{fig:shockfit}
\end{figure}

We fit the 2D surface brightness within the $\sim436$ kpc [300\arcsec] to $\sim915$ kpc [630\arcsec] range using a projected broken power-law 3D density model \citep{Markevitch2007, Owers2009, Tiwari2022, Rajpurohit2024} of the form:
\begin{equation} 
n(r) \propto {\left\{\begin{matrix} n_0 \left(\frac{r}{r_{\text{edge}}}\right)^{-\alpha_1} \quad \text{if } r < r_{\text{edge}} \\ 
\frac{n_0}{C} \left(\frac{r}{r_{\text{edge}}}\right)^{-\alpha_2} \quad \text{if } r \geq r_{\text{edge}} \end{matrix}\right.} 
\end{equation} 
where $r_{\text{edge}}$ is the distance of the putative edge from the cluster center, $n_0$ is a normalization factor, $C = \frac{\rho_2}{\rho_1}$ is the density jump factor, and $\alpha_1$ and $\alpha_2$ are the power-law slopes inside and outside of the edge. The best-fit model is shown in Fig.~\ref{fig:shockfit} as a red line, with 1$\sigma$ confidence range shaded red. We found best-fit values of the power-law indices of $\alpha_1 =$ \fe{1.01}{0.12}{0.11} and $\alpha_2 =$ \fe{1.42}{0.09}{0.10}. The brightness edge was found to lie at a radius of $r_{jump} =$ \fe{451}{8}{9} arcsec = \fe{656}{12}{13} kpc from the core and has a gas density jump of $C = \frac{\rho_2}{\rho_1} =$ \fe{1.17}{0.05}{0.06}. 

If the front is a shock, the Mach number can be calculated from the fitted density jump as
\begin{equation}
    \mathcal{M} = \sqrt{\frac{2\frac{\rho_2}{\rho_1}}{\gamma + 1 - \frac{\rho_2}{\rho_1}(\gamma-1)}},
    \label{eq:mach}
\end{equation}
where $\gamma = \frac{5}{3}$ is the assumed adiabatic index. From the best-fit density jump factor, $\frac{\rho_2}{\rho_1} =$ \fe{1.17}{0.06}{0.05}, we found a Mach number, $\mathcal{M} =$ \fe{1.12}{0.04}{0.04}. The density discontinuity is significant at $3.1\sigma$, relative to a smooth ICM profile (i.e., $\frac{\rho_2}{\rho_1} = 1$), providing strong evidence of the presence of a weak shock.

\begin{figure*}
    \centering 
        \begin{tabular}{cc}
      \includegraphics[width=0.5\linewidth]{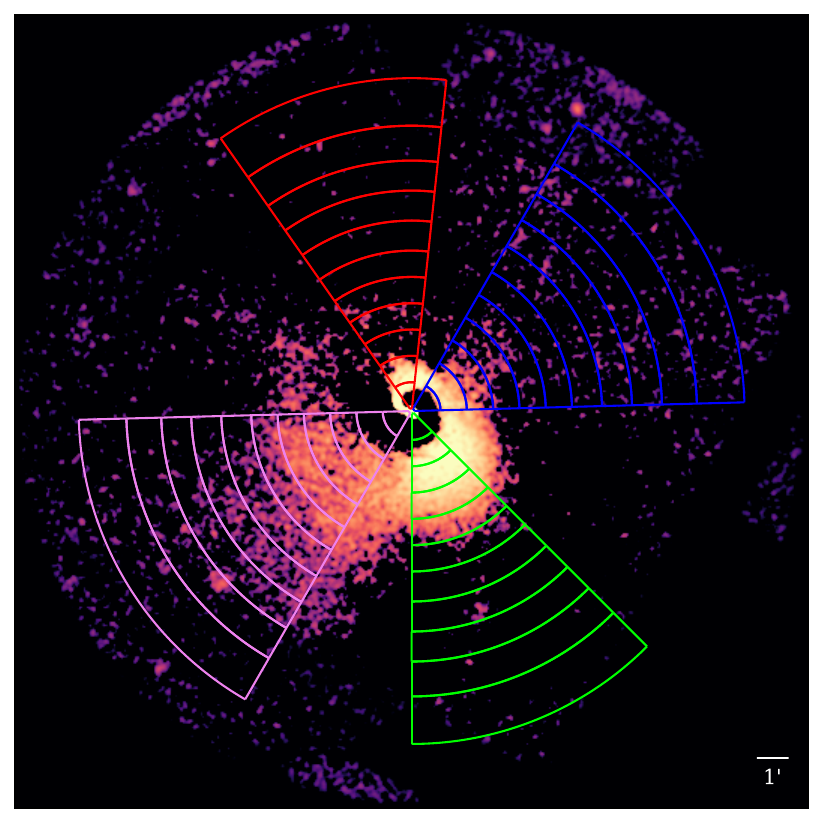} &
      \includegraphics[width=0.5\linewidth]{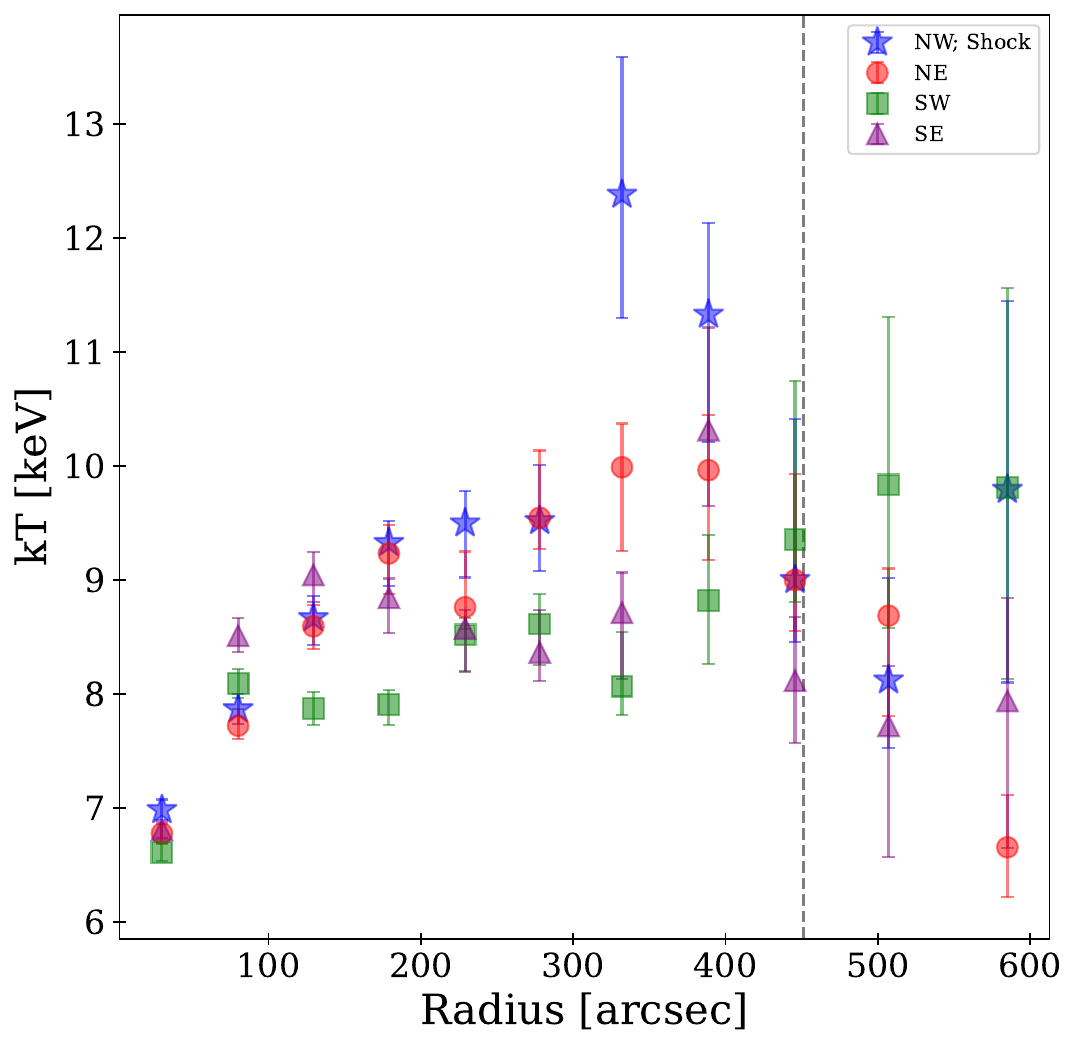} \\
\end{tabular}
    \caption{Residual emission image (\emph{left}) showing the regions used to extract the projected temperature profiles (\emph{right}). The dashed black line denotes the location of the fitted density jump at $r_{jump} =$ \fe{451}{8}{9} arcsec = \fe{656}{12}{13} kpc.}
    \label{fig:shock}
\end{figure*}

As another test of this potential shock region, we extracted temperature profiles in regions of opposing angular extents using a method similar to the one used to map the sloshing spiral. Projected temperatures were obtained from fitting spectra extracted from the annular regions shown in Fig.~\ref{fig:shock}, following the methods of \S\ref{sec:spec}. Constant scale factors for temperature and abundance were applied to the 2022 and 2023 fitted spectra, following Appendix \ref{sec:cal}. The resulting projected temperature profiles, as a function of radius from the cluster core, are shown plotted together in the right panel of Fig.~\ref{fig:shock}. The NE and SW sectors were defined to minimize contributions from the sloshing cold front, to aid in comparison against the NW-shock region, while the SE sector was defined to maximize coverage of the cold ``splash'' area to see how the temperature varies in this region. 

The NW sector (blue) shows a clear temperature jump across the shock region in the projected temperature profile shown in Fig.~\ref{fig:shock}. The temperature increases from \fe{8.12}{0.89}{0.60} keV in the radial bin just outside the fitted shock edge, to a peak value of \fe{12.38}{1.21}{1.08} keV within the edge. This temperature rise is consistent with shock heating due to a merger-driven front. In comparison, the projected temperature profiles in other directions at similar radii show smooth and more gradual changes, with temperatures remaining $\lesssim 10$ keV across all radii. Interestingly, the temperature peak is located $\sim$140\arcsec\ inside the fitted edge found above using \texttt{pyproffit}, which is denoted in Fig.~\ref{fig:shock} with a dashed black line. This could be due to projection effects and the line-of-sight depth through the Mach cone. 

We can determine the Mach number of the potential shock front using a similar equation as Eq.~\ref{eq:mach}, but using the temperature jump, expressed as a function of the density jump \citep{Sanders2016a}:
\begin{equation}
    \mathcal{M}_T = \sqrt{\frac{\bigg(8\frac{T_2}{T_1} - 7\bigg) + \bigg[\bigg(8\frac{T_2}{T_1}-7\bigg)^2+15\bigg]^{1/2}}{5}},
\end{equation}
where $T_1$ and $T_2$ are the upstream and downstream gas temperatures, respectively. For regions where $\frac{\rho_2}{\rho_1} > 1$, when the temperature ratio, $\frac{T_2}{T_1}$, is greater than unity it suggests the presence of a shock front, while ratios of less than unity indicate a potential cold front \citep{Rajpurohit2024}. 

Given the projection effects discussed previously, the observed temperature peak appears at a smaller radius than the fitted density jump (see Fig.~\ref{fig:shock}). To better reflect the physical temperature discontinuity across the shock, we adopted a pre-shock temperature of $T_1 = $\fe{8.12}{0.89}{0.60} keV, taken from the radial bin immediately outside the radius of the best-fit density jump, $r_{jump}$. The post-shock temperature was taken as $T_2 = $ \fe{12.38}{1.21}{1.08} keV, corresponding to the hottest temperature inside the edge. With these, we found a Mach number of $\mathcal{M}_T = $\fe{1.53}{0.25}{0.27}. The temperature discontinuity is detected at 2.8$\sigma$, providing additional support for the presence of a shock front. This Mach number is slightly higher than calculated using the density jump, corresponding to a $\sim1.6\sigma$ difference between the two measurements.

\subsection{Azimuthal profiles of ICM substructure \label{sec:azprofs}}

\subsubsection{Bay Feature and SW Excess \label{sec:khi}}

As noted in \S\ref{sec:imaging}, there appears to be a concave bay-like structure located  $\sim350$ kpc south of the cluster core that could be attributed to a Kelvin-Helmholtz instability (KHI). Similar structures have been reported in previous clusters \citep[e.g., Perseus, Centaurus, and Abell 1795;][]{Sanders2016, Walker2014, Walker2017} and are thought to be associated with large-scale KHI bays. 

\begin{figure*}
    \centering
    \begin{tabular}{cc}
      \includegraphics[width=0.5\linewidth]{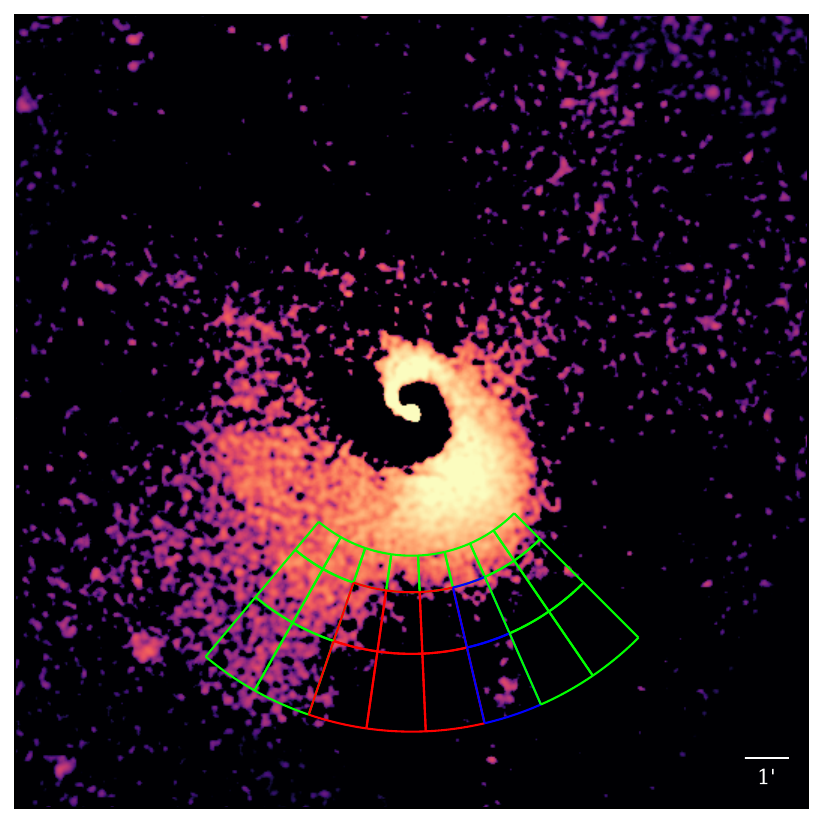} &
      \includegraphics[width=0.5\linewidth]{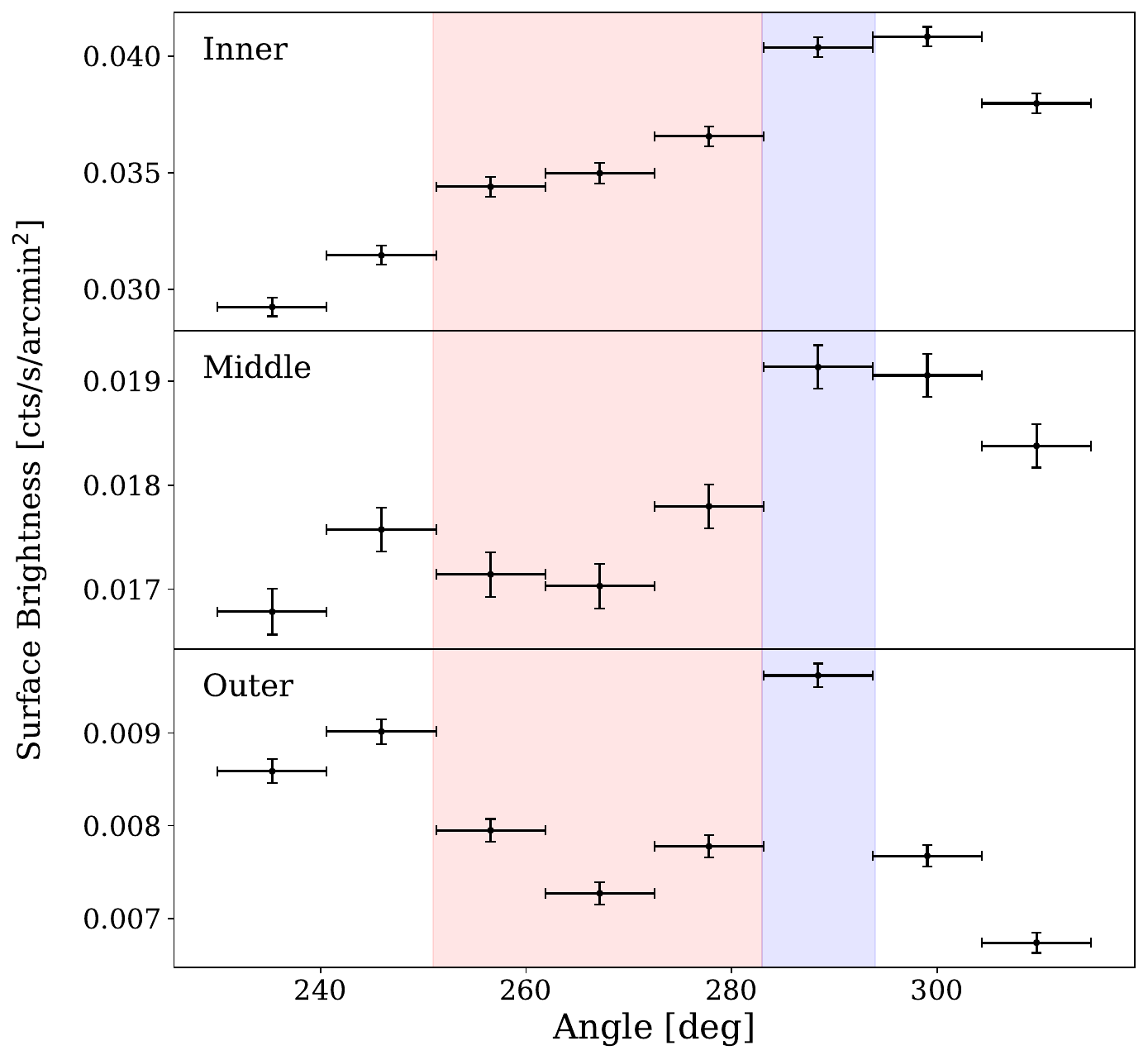} \\
\end{tabular}
    \caption{Residual emission image (\emph{left}) showing the regions used to extract the azimuthal surface brightness profiles (\emph{right}) of the inner (top panel), middle (middle panel), and outer (bottom panel) annuli. The KHI bay-like feature spans 251$^\circ$ to 283$^\circ$ in angular bins, which is highlighted red in all panels. The SW region of excess likely attributed to ICM emission from a separate cluster or group covers 283$^\circ$ to 294$^\circ$ and is highlighted by blue in all panels. }
    \label{fig:bayprofs}
\end{figure*}

To examine this structure further, we extracted the azimuthal surface brightness from 3 annular regions, each of which spans the angular extent between 230$^\circ$ to 315$^\circ$, measured from the RA axis towards the west. The regions were defined to isolate the sloshing front from the surrounding ICM and enhance the contrast of the potential KHI feature. The inner annulus spans radii of $r_i = 286$ kpc to $r_o = 359$ kpc and represents the reference, as it covers the region just inside the southern edge of the sloshing spiral. The middle annulus has inner and outer radii of $r_i = 359$ kpc and $r_o = 481$ kpc, respectively, while the outer annulus spans $r_i = 481$ kpc to $r_o = 635$ kpc. These three regions are shown overlaid on the residual emission image in the left panel of Fig.~\ref{fig:bayprofs}. The resulting azimuthal brightness profiles are presented on the right side of Fig.~\ref{fig:bayprofs}, for the inner (top panel), middle (middle panel), and outer (bottom panel) annuli regions. 

The region of decreased X-ray emission that was identified as a potential KHI bay feature in the X-ray and residual emission images (see cyan arrows in Figs.~\ref{fig:xray} \&~\ref{fig:beta2d}) corresponds to the angular bins spanning 251$^\circ$ to 283$^\circ$ (highlighted by red regions in Fig.~\ref{fig:bayprofs}) in the middle and outer annuli regions. Indeed, we observe a dip in the surface brightness observed in the middle and outer annuli within these angular bins, in comparison to the brightness observed in the same bins of the inner annulus. The red-shaded region highlights this in the brightness profiles of Fig.~\ref{fig:bayprofs}. The potential detection of a KHI bay not only supports the presence of sloshing motions in A2029 but also provides constraints on the effective viscosity and magnetic field configuration of the ICM. Lower viscosity and a weaker magnetic field tend to promote the growth of KHIs, while higher viscosity or strong magnetic tension can suppress their development.

Looking at the X-ray and residual emission images, it was unclear whether the region of excess emission observed to the SW (green arrows in Fig.~\ref{fig:beta2d}) is connected to the sloshing structure or represents separate ICM belonging to a background cluster or group. However, the azimuthal brightness profiles in the right side of Fig.~\ref{fig:bayprofs} favor the latter scenario. The region of excess emission corresponds to the bin spanning 283$^\circ$ to 294$^\circ$ (highlighted blue in Fig.~\ref{fig:bayprofs}) and shows a clear excess in only the outer annulus region. 

Thus, while the X-ray depression (highlighted in red) could be associated with a KHI bay, the region of excess emission (highlighted in blue) appears to be more likely associated with ICM emission from a projected group or cluster. This is explored further in \S\ref{sec:optical}.

\subsubsection{SE `Splash'}

\begin{figure*}[h!]
    \centering
    \begin{tabular}{cc}
      \includegraphics[width=0.5\linewidth]{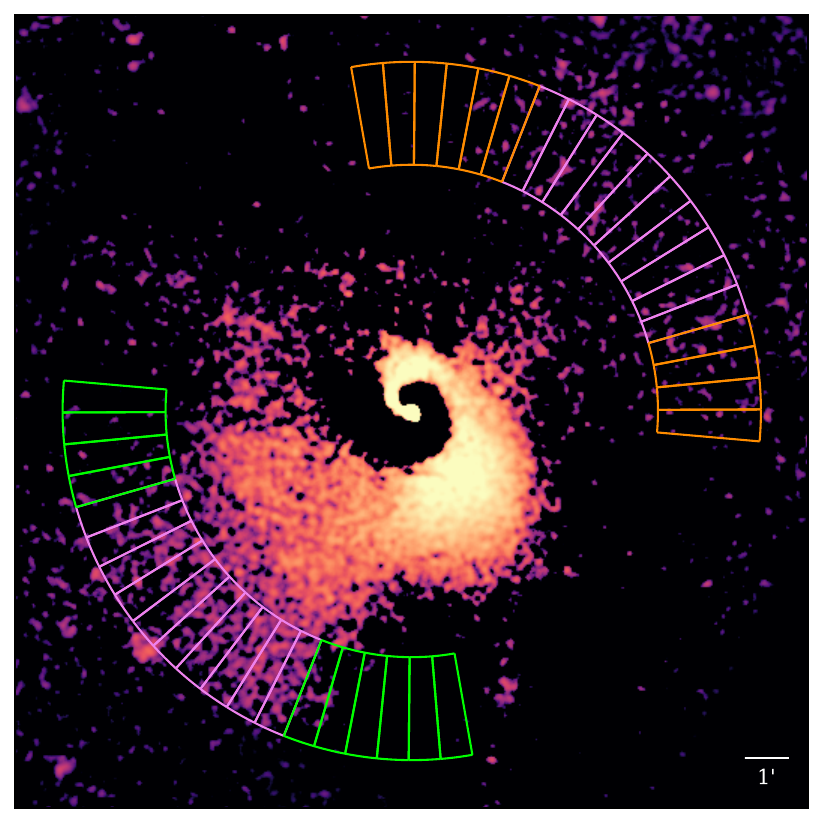} &
      \includegraphics[width=0.5\linewidth]{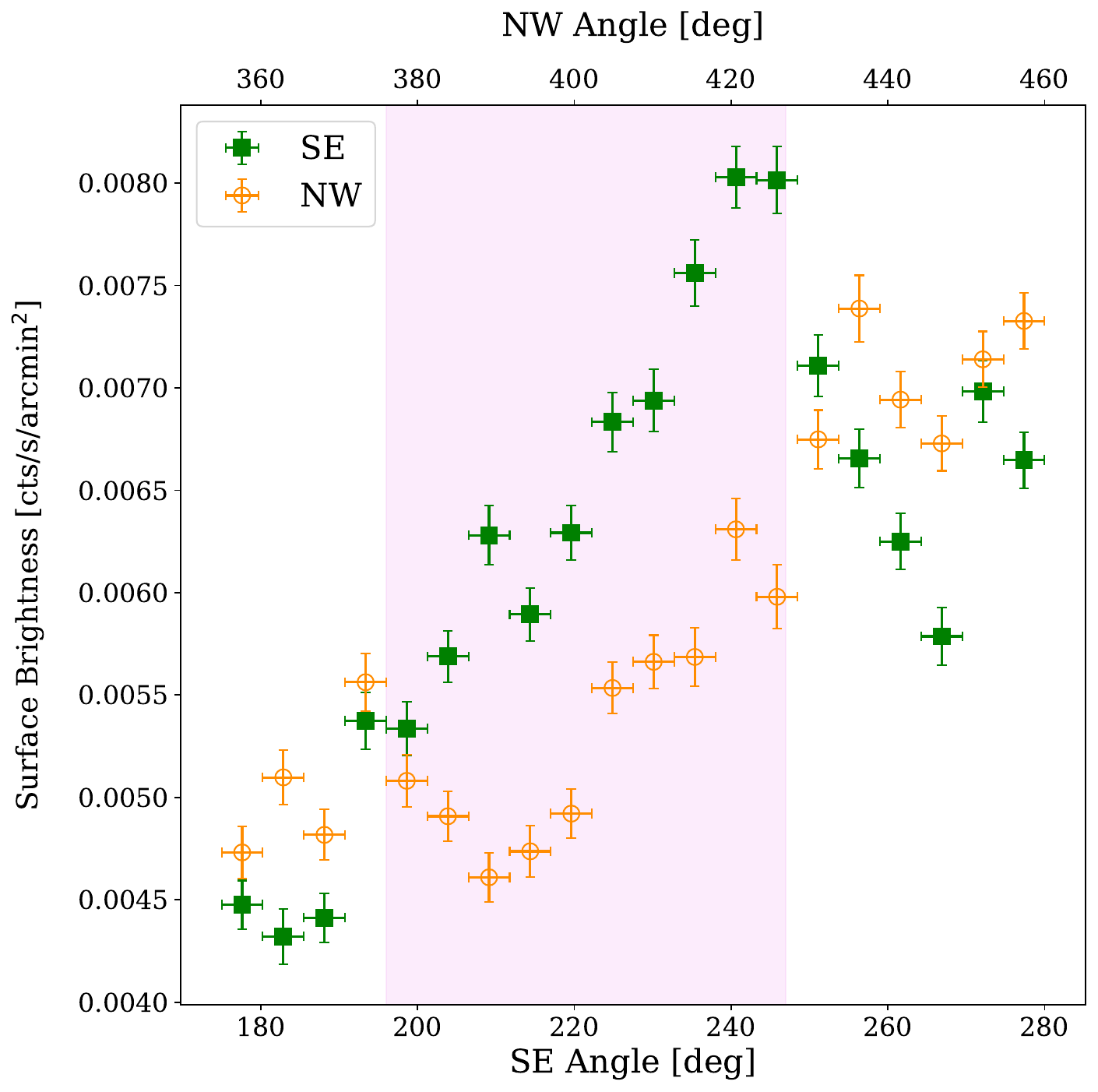} \\
\end{tabular}
    \caption{Residual emission image (\emph{left}) showing the regions used to extract the azimuthal surface brightness profiles (\emph{right}) for the SE (solid green squares) and NW (open orange circles) directions. The ``splash'' feature spans 196$^\circ$ to 247$^\circ$ (indicated by purple shaded regions in each panel), extending beyond 690 kpc from the cluster center.}
    \label{fig:az_splash}
\end{figure*}

In \S\ref{sec:imaging}, we noted a region of excess emission SE of the cluster core, which we dubbed the ``splash'' feature, noted in Fig.~\ref{fig:beta2d} with white arrows. Figure~\ref{fig:az_splash} illustrates residual emission (left panel) overlaid with the annular sectors used to extract the azimuthal surface brightness profiles (right panel). We compared the SE direction (solid squares) with the northwestern (NW) direction (open circles) in the right panel. The SW splash excess is observed between 196$^\circ$ to 247$^\circ$, highlighted in purple in Fig.~\ref{fig:az_splash}. The source of this extended feature could be due to the second, perturbing cluster system. Some simulations \citep{ZuHone2010, ZuHone2016} have shown similar ``splash''-like features, in conjunction with sloshing spiral structures, that results after the second core passage of the perturbing body.

\subsubsection{AGN Bubbles and Feedback\label{sec:bubbles}}

At the core of A2029 lies the AGN radio source PKS1508+059. AGN are known to inflate bubbles in the ICM that can offset radiative cooling. Previous studies of A2029 have been unable to detect any significant\footnote{\cite{Rafferty2006} report the detection of one cavity, but at low significance.} evidence of such bubbles in X-ray imaging. While the observations here similarly reveal no clear evidence for such bubbles, we also cannot rule out the possibility that the southern bay feature could be associated with the inner edge of an outer ghost cavity that could have been inflated from the southern lobe. A similar situation arose in the Ophiuchus cluster, where a concave X-ray edge was initially deemed unrelated to the radio emission, but much deeper low-frequency radio data later revealed a matching ghost cavity \citep{Giacintucci2020}.

\begin{figure*}
\centering
    \begin{tabular}{cc}
      \includegraphics[width=0.46\linewidth]{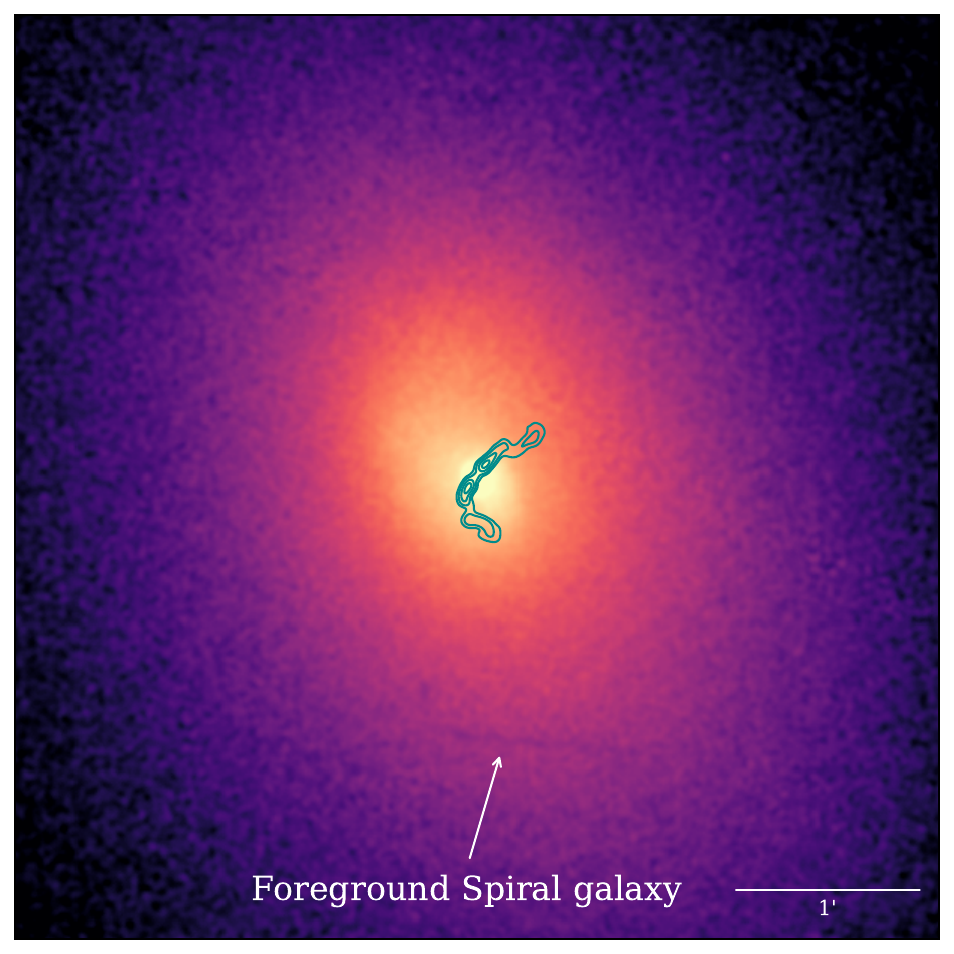} &
      \includegraphics[width=0.5\textwidth]{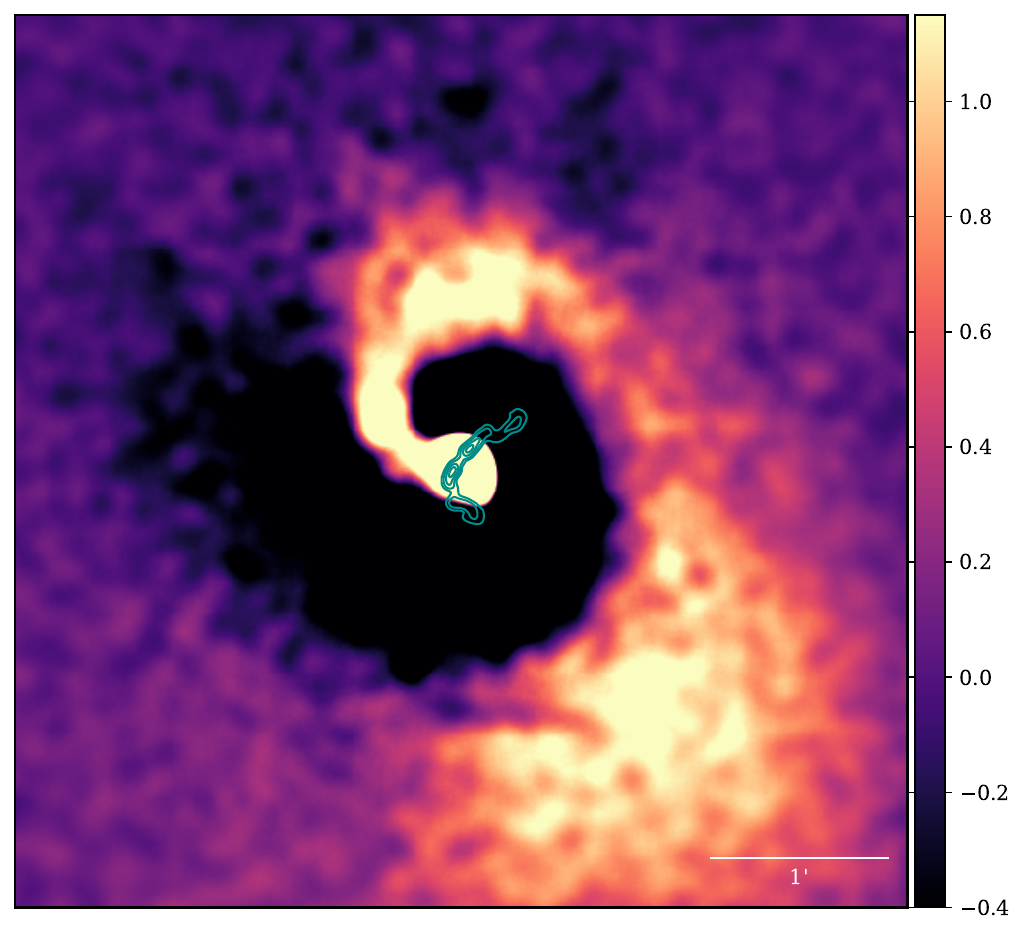} \\
\end{tabular}
\caption{Gaussian smoothed X-ray (\emph{left}; $\sigma = $1\farcs5) and residual emission after $\beta$-model subtraction (\emph{right}; $\sigma =$ 7\farcs5) images of A2029 in the 0.7-7 keV range, showing the 1490 MHz radio emission (cyan contours) of the central AGN. Both panels show a $5\arcmin\times5\arcmin$ FoV. Contours are drawn at levels corresponding to 0.0006, 0.0024, 0.0079, and 0.017 Jy beam$^{-1}$.}
\label{fig:wat}
\end{figure*}

Figure~\ref{fig:wat} shows the X-ray and residual emission images in the inner 5\arcmin$\times$ 5\arcmin\ cluster core to highlight the central WAT source. The overlaid cyan contours are the 1490 MHz radio emission taken from VLA observations presented in \cite{Taylor1994}. Contours are drawn at levels corresponding to 0.0006, 0.0024, 0.0079, and 0.017 Jy beam$^{-1}$. The X-ray images show no obvious signs of isolated regions of decreased emission near the radio lobes, nor anywhere in the ICM at larger radii. In the $\beta$-model subtracted residual emission image, though showing decreased emission in the regions of both lobes, the observed depressions appear to be more closely associated with the cooling spiral structure rather than with distinct cavities. However, cavities may be particularly hard to detect in these regions because of confusion with the sloshing features.

\begin{figure*}
    \centering
\begin{tabular}{cc}
      \includegraphics[width=0.46\linewidth]{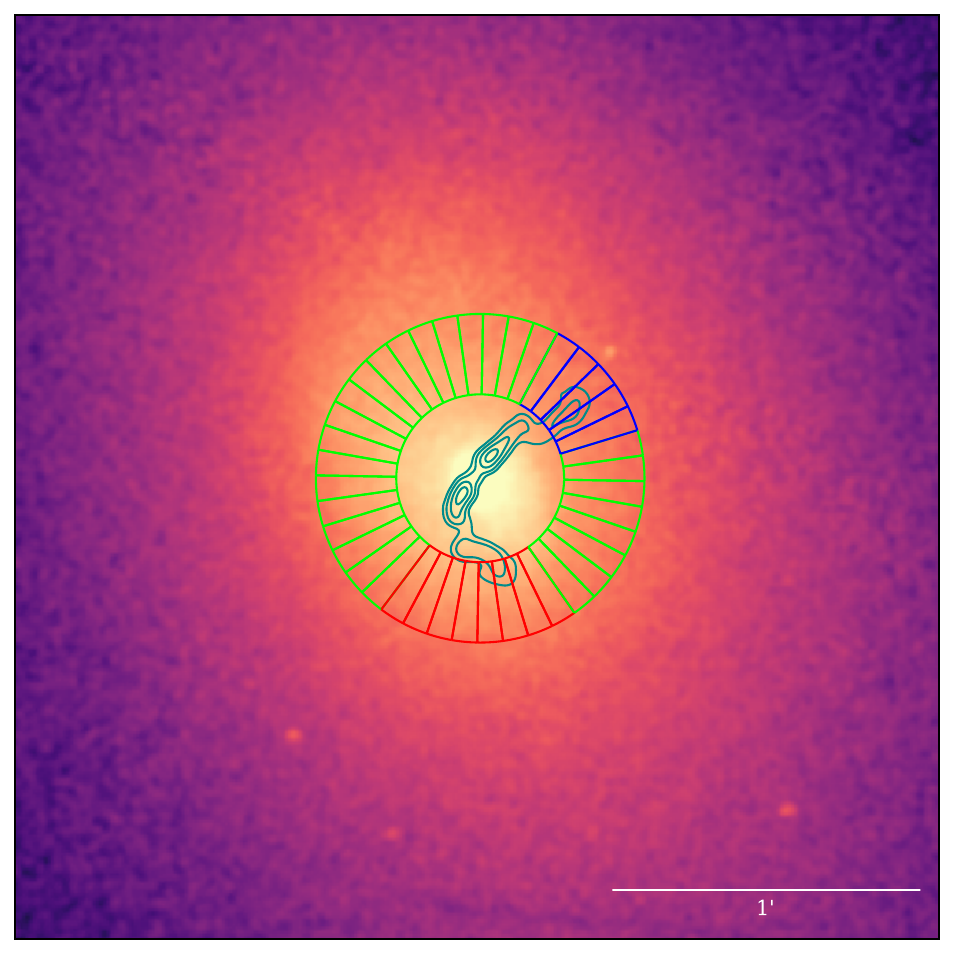} &
      \includegraphics[width=0.5\linewidth]{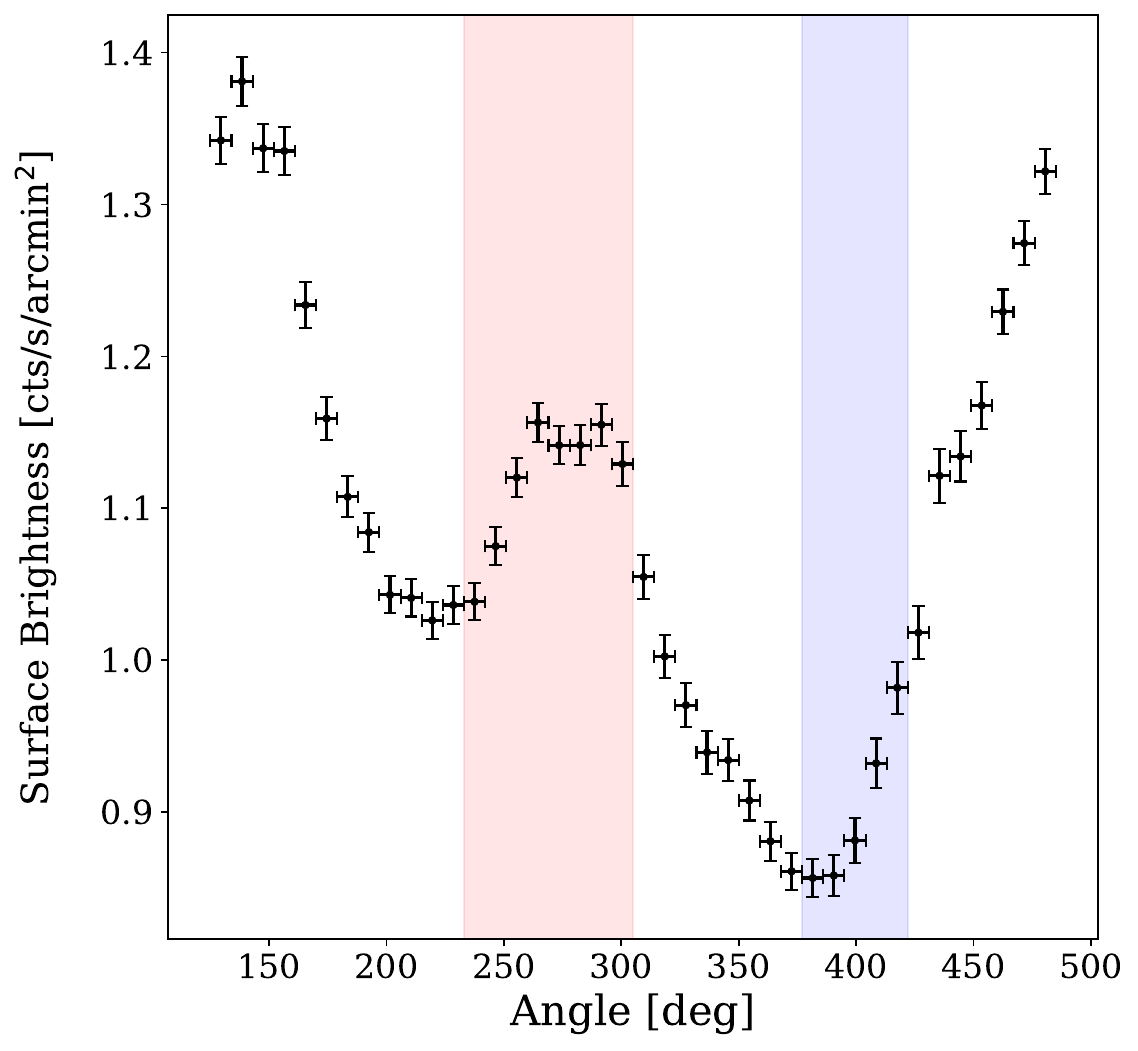} \\
\end{tabular}
    \caption{X-ray image (\emph{left}) showing 1490 MHz radio emission (cyan) and the annular region (green), used to extract the azimuthal surface brightness profile (\emph{right}) around the central WAT source. The red regions highlighted in the left panel correspond to the red-shaded bins in the right panel (and the same for the blue regions). Although there is some decrease in emission coincident with the northern lobe, it is difficult to clearly detect the presence of AGN cavities in this region since it is complicated by the emission from the sloshing spiral.}
    \label{fig:wataz}
\end{figure*}

As a check, we examined the azimuthal profiles of surface brightness at the radius of the radio lobes. We defined a circular annulus encompassing the region around the radio lobes, with an inner radius of $r_i = 26$ kpc (to avoid the jet regions) and an outer radius of $r_o = 47$ kpc. The extracted azimuthal surface brightness profiles for both the northern and southern regions are shown in Fig.~\ref{fig:wataz}. The southern lobe, spanning the angular bins between 233$^\circ$ and 305$^\circ$ (indicated by the red regions in Fig.~\ref{fig:wataz}), interestingly shows an increase in surface brightness, which could suggest that the WAT lobe is being bent backward by the motion induced by the sloshing spiral. In contrast, the northern lobe, covering angular bins between 17$^\circ$ and 62$^\circ$ (highlighted by the blue regions in Fig.~\ref{fig:wataz}), exhibits decreased emission; however, this depression is not isolated to the lobe region and appears to be part of a more extended feature influenced by the sloshing spiral. This added complexity underscores the challenges in distinguishing between cavity signatures and structural features inherent to the cool spiral.

Recent work by \citet{Timmerman2022} applied a hybrid X-ray–radio method to estimate AGN cavity powers in galaxy clusters. For A2029, they utilize high-resolution ($\sim 0.3\arcsec$) observations from the International Low Frequency Array \citep[LOFAR;][]{VanHaarlem2013} telescope (ILT) and low-resolution imaging from the LOFAR Two-Metre Sky Survey \citep[LoTSS;][]{Shimwell2017, Shimwell2019, Shimwell2022}. While the authors report estimates of cavity volumes and power, they do not identify definitive X-ray cavities coincident with the radio lobes, noting that the complex spiral morphology of the ICM complicates the clear identification of such features.

We found in \S\ref{sec:spec} a mass deposition rate within the central 40\arcsec\ (equivalent to 58 kpc at the redshift of A2029) of $\dot{M} =$\fe{349}{20}{25} M$_{\odot}$ yr$^{-1}$ when the low temperature of the cooling flow model is free. However, we find $\dot{M} = $ \fe{17}{2}{2} when the low temperature is fixed to 0.0808 keV and the high temperature is tied to an \texttt{APEC} component. The latter is the more relevant quantity to compare to the star formation rate of the cD galaxy --- \cite{Hicks2010} measure a SFR of only $0.03-0.8$ M$_{\odot}$ yr$^{-1}$ and \cite{Fraser-McKelvie2014} estimate \fe{0.80}{0.17}{0.04} M$_{\odot}$ yr$^{-1}$. The large discrepancy in the results between the free-$kT_{low}$ mass deposition rate and the SFR, coupled with the much smaller fixed-$kT_{low}$ mass deposition rate, indicates that in A2029, most gas does not cool to sub-keV temperatures, consistent with a hidden or partial cooling flow scenario \citep{Fabian2022}.

To see if the central AGN is providing the heating necessary to offset the observed cooling, we compared the energy injection rate, $E/t$, from the radio source to the luminosity of the cooling gas, $L_{\mathrm{cool}}$. For A2029, \cite{PM2013} estimated an energy injection rate\footnote{Calculated as $E=4PV$, where $P$ and $V$ are the cavity pressure and volume, respectively, averaged over a typical AGN repetition rate of $t=5\times10^7$ yr.} of $E/t = 1\times10^{44}$ erg s$^{-1}$. Additionally, \cite{Timmerman2022} used low-resolution LOFAR observations to estimate a total cavity power (i.e., the AGN injection rate) of $\sim 3 \times 10^{44}$ erg s$^{-1}$; however the authors assumed radial propagation of the radio lobes, while it's clear that for A2029, the lobes are bent relative to the jet axis. 

For comparison, the luminosity of the cooling gas, $L_{\mathrm{cool}}$, can be calculated assuming steady-state, isobaric cooling, e.g. 
\begin{equation}
    L_{\mathrm{cool}} = \frac{5}{2}\frac{kT}{\mu m_p}\dot{M},
\end{equation}
where $\mu$ is the mean atomic mass of gas particles in the cluster, taken as 0.6, and $m_p$ is the proton mass. Taking $\dot{M} = 784$ M$_{\odot}$ yr$^{-1}$ and $kT = 11.82$ keV, from our cooling flow fit to the inner 116\arcsec (see \S~\ref{sec:spec} and Tab.~\ref{tab:fittests}), we find $L_{\mathrm{cool}} = 2.3\times10^{45}$ ergs/s.

The differences between the energy injection rate and cooling rate suggest that the energy injected by the AGN ($\sim 1-3 \times 10^{44}$ erg s$^{-1}$) is not enough to offset the observed cooling rate ($2.3 \times10^{45}$ erg s$^{-1}$). Thus, it is likely that the large-scale sloshing is playing a role in balancing ICM cooling in the region enclosing the radio source, contained within the cD galaxy.

\subsection{Comparison to Simulation \label{sec:sims}}

We compare our observations to simulations of galaxy cluster mergers in order to get a better picture of the past merger history that would result in some of the features we have identified in A2029 so far. We examined the idealized binary merger simulations from the Galaxy Cluster Merger Catalog\footnote{\href{https://gcmc.hub.yt}{https://gcmc.hub.yt}} \citep{ZuHone2018} and found that our observations were qualitatively reproduced by a simulation of a 1:10 mass ratio merger, with an impact parameter of 500 kpc, originally presented in \cite{Zuhone2011a}. 

\begin{figure*}
    \centering
    \includegraphics[width=\linewidth]{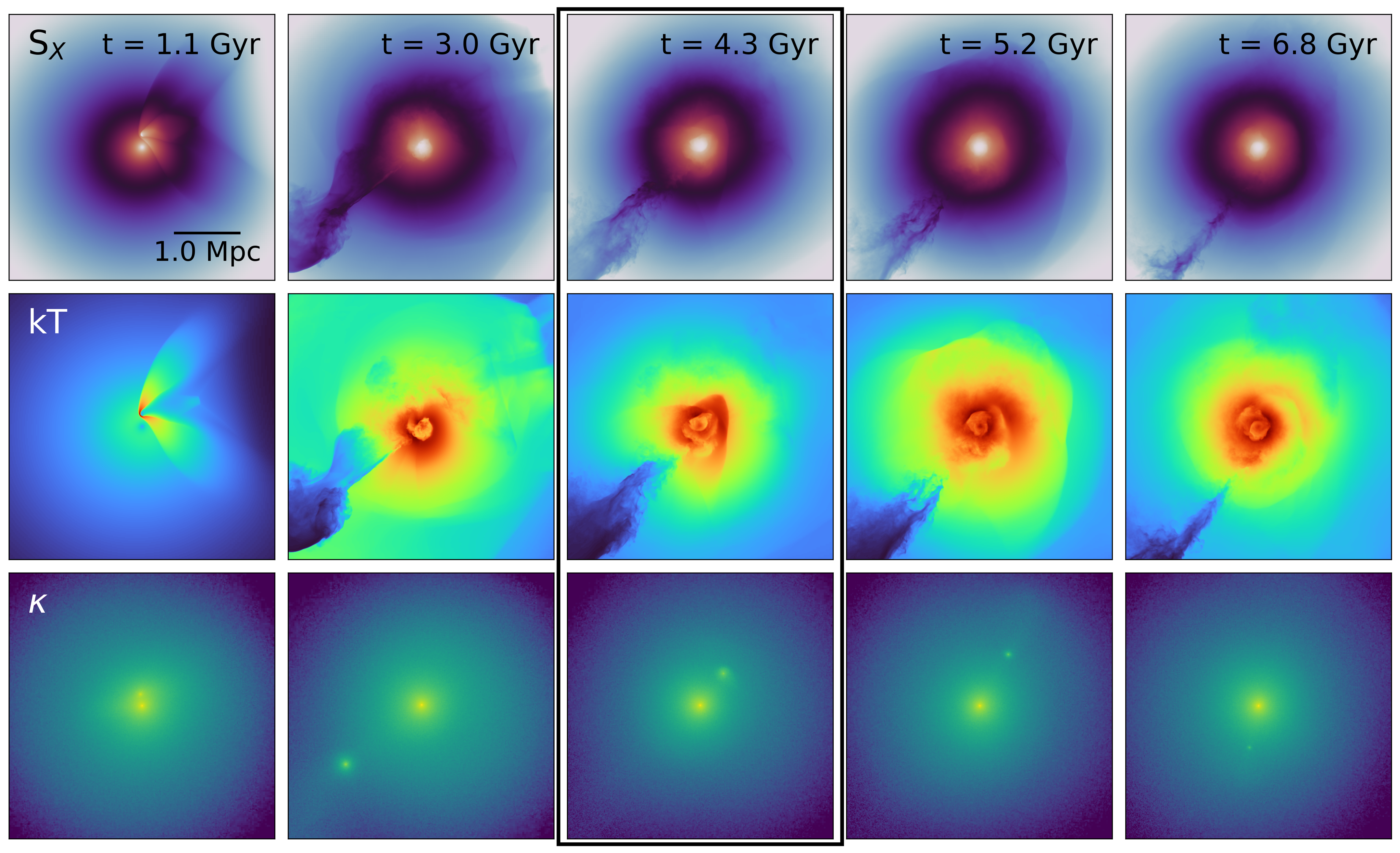}
    \caption{Snapshots of the X-ray emissivity (top row), projected temperature map (middle row), and projected surface density (bottom row) of the 1:10 mass ratio and $b=500$ kpc merger simulation of \cite{Zuhone2011a}. In the temperature maps, cooler gas is represented by bluer hues while hotter gas is represented by redder hues. The timestamps shown in the top left of each panel refer to the time since the simulation started, $t_0$. The epoch which matches the observations of A2029 is in the center-most panels, at 4.3 Gyr. Each image is $\sim4$ Mpc $\times$ 4 Mpc. }
    \label{fig:simpanels}
\end{figure*}

Figure~\ref{fig:simpanels} shows snapshots of the X-ray emissivity and projected temperature map for this 1:10 mass ratio merger model at different timestamps, where $t_0 = 0$ Gyr is the start of the simulation. Each image shows a $\sim$ 8 Mpc $\times$ 8 Mpc area. In this 1:10 mass ratio merger, the perturbing subcluster initially passed by the main cluster just north of the cluster core, traveling towards the east (first column in Fig.~\ref{fig:simpanels}). As the subcluster continued heading east, the initial core passage displaced the core ICM from the central potential, setting off sloshing. Since the perturbing subcluster is on a bound orbit, it re-approached the main cluster from the east, now heading west (second column in Fig.~\ref{fig:simpanels}). As the subcluster progresses through the second core passage, which occurs $\sim 3$ Gyr after the initial core passage, it leaves behind a ``splash'' of cold gas to the SE while driving a shock as it moves back through the cluster ICM (third through fifth columns in Fig.~\ref{fig:simpanels}).

\begin{figure*}
    \centering
    \includegraphics[width=\linewidth]{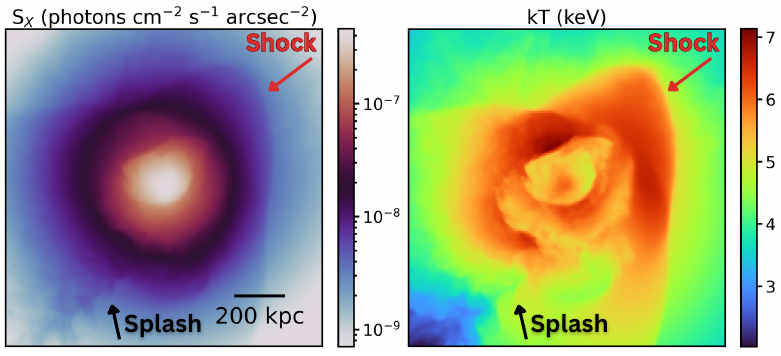}
\caption{Simulated X-ray emissivity (\emph{left}) and projected temperature (\emph{right}) at $4.3$ Gyr after the start of a 1:10 mass ratio cluster-cluster merger, with an impact parameter of $b=500$ kpc. This is the same epoch shown in the middle panels of Fig.~\ref{fig:simpanels}, now zoomed into the inner 1.45 Mpc of the simulated merger. The identified ``splash'' feature (black arrow) is seen in the SE as a trail of cold gas left in the wake of the returning perturber. A shock front (red arrow) can be seen to the NW of the core.}
    \label{fig:singlesim}
\end{figure*}

The epoch\footnote{\url{https://gcmc.hub.yt/fiducial/1to10_b0.5/0215.html}} which matches the observations of A2029 the closest is $\sim0.2$ Gyr after the second core passage of the perturbing subsystem (center column in  Fig.~\ref{fig:simpanels}) which is 4.3 Gyr after the start of the merger simulation. Figure~\ref{fig:singlesim} shows this epoch's X-ray emissivity and projected temperature. At this time, the sloshing gas of the main cluster has developed the expected spiral-like structure, while the cold ``splash'' (i.e., the wake of gas left behind the subsystem's returning passage; highlighted by the black arrow in Fig.~\ref{fig:singlesim}) remains to the SE. The perturbing system drives a shock front that is located $\sim0.5$ Mpc from the main cluster core. The shock can be seen to the NW of the core (highlighted by the red arrow in Fig.~\ref{fig:singlesim}), consistent with our observations of A2029.

While this simulation provides a useful comparison to A2029, it should be noted that it was not originally intended to reproduce the features in A2029. A simulation more tailored to match the physics of A2029 would be needed to better reproduce the observed sloshing spiral, splash, and shock features.

\subsection{Optical Structure \label{sec:optical}}

\begin{figure*}
    \centering
    \begin{tabular}{ccc}
      \includegraphics[width=0.3\linewidth]{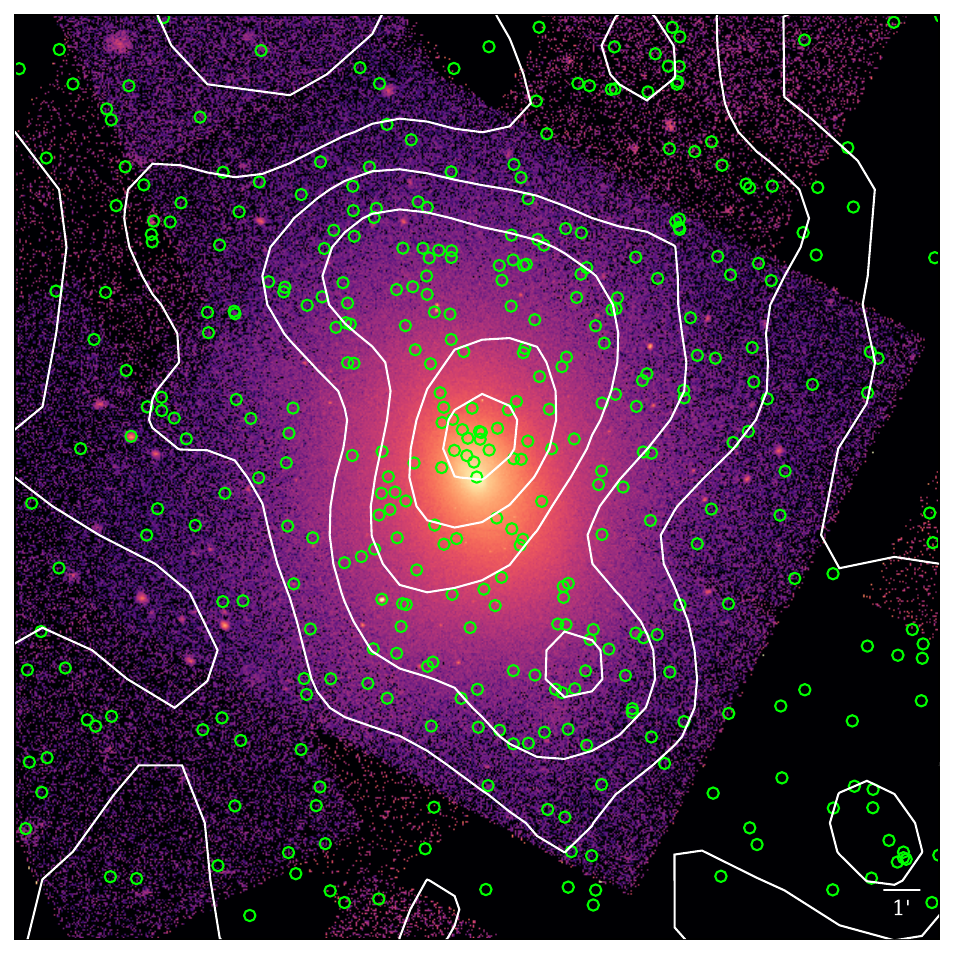} &
      \includegraphics[width=0.3\textwidth]{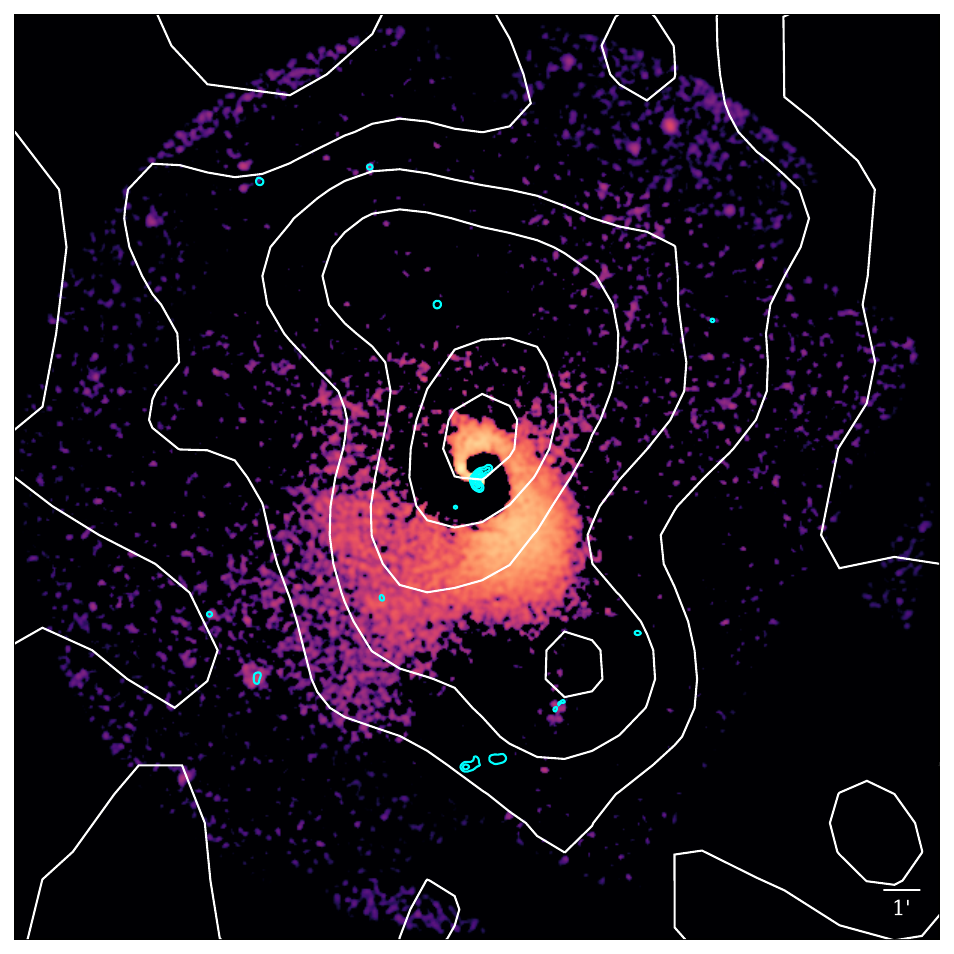} & 
      \includegraphics[width=0.35\textwidth]{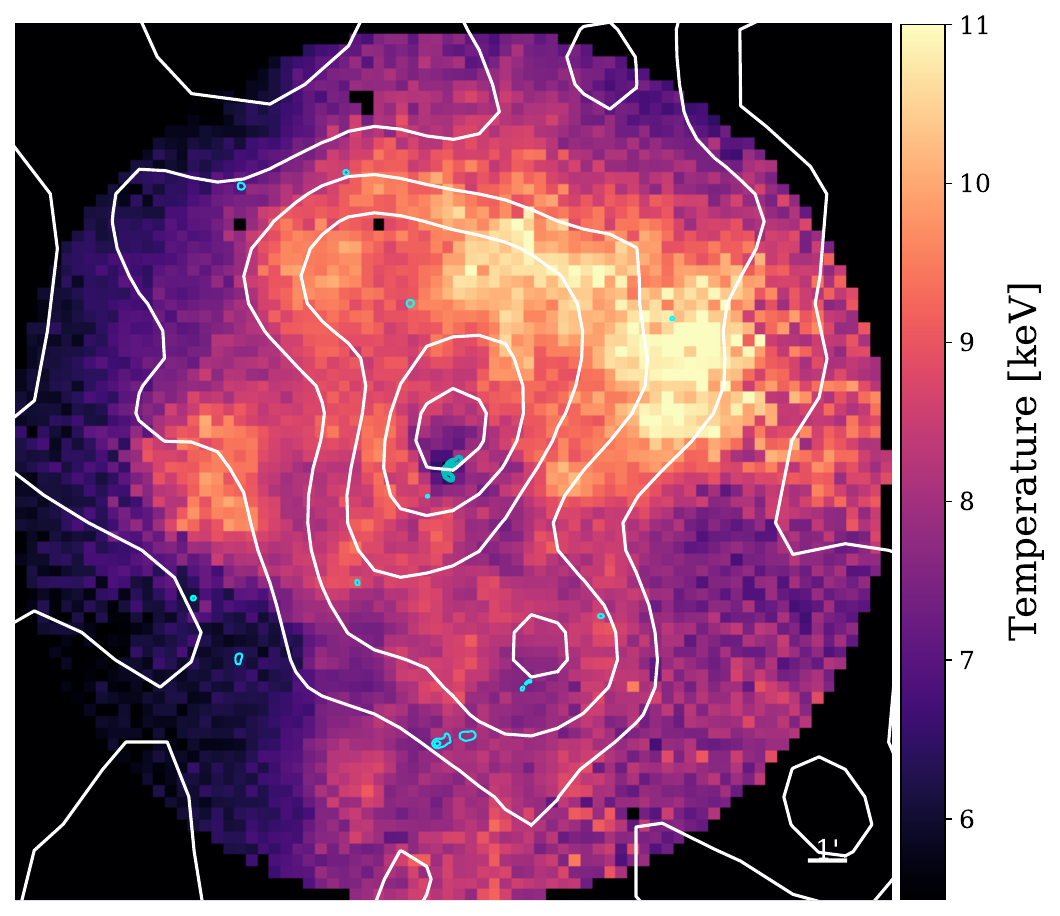}\\
\end{tabular}
    \caption{X-ray image (left), residual emission image (center), and temperature map (right) of A2029. Overlaid are contours of 1.4 GHz radio emission (blue). In the X-ray image, green circles denote spectroscopic members of A2029 \citep{Sohn2019} which are used to generate contours of number density, shown overlaid in white in all three panels. \label{fig:ndensity}}
\end{figure*}

\cite{Sohn2019} identify spectroscopic members of A2029, which we show overlaid on the X-ray image in the left panel of Fig.~\ref{fig:ndensity} in green circles. We create contours (in steps of 0.28 galaxies arcmin$^{-1}$) of galaxy surface density, following \cite{Sohn2019}, which we show as white contours in all panels of Fig.~\ref{fig:ndensity}. 

The distribution of member galaxies in A2029 shows elongations along the NW-SE axis as well as along the NE-SW axis, resulting in a `Z'-shape distribution. This asymmetric distribution suggests large-scale interactions significantly shaped the cluster's current structure. 

In addition to member galaxies, \cite{Sohn2019} show that there are two infalling subsystems, consisting of Abell 2033 and a newly identified group, the Southern Infalling Group (SIG), based on weak lensing and X-ray observations, and phase-space analysis of the member galaxies. Both systems lie outside the Chandra FoV, which is roughly the size of the cluster radius $R_{200} = 1.91$ Mpc \citep{Sohn2019}.

In the X-ray and residual emission images, we identified a region of excess emission located $\sim 570$ kpc SW of the cluster core (highlighted with green arrows in Figs.~\ref{fig:beta2d}; see \S\ref{sec:beta2d}). In \S\ref{sec:azprofs}, we determined that this emission was not likely to be associated with the main cluster and could instead suggest the presence of a foreground or background group or cluster, with its own intragroup or intracluster medium that we are detecting. The thumbnail image of the radio source, shown in Fig.~\ref{fig:optical} panel (b), reveals the presence of numerous red sources clustered together. \cite{Sohn2019} measures a spectroscopic redshift of $z_{spec} = 0.552$, attributed to the largest and brightest source of the group shown in Fig.~\ref{fig:optical} panel (b). The concurrence with radio emission indicates the presence of an AGN within the group of red galaxies. In addition, this group coincides with the region of excess X-ray emission seen to the SW of the cluster (green arrow in Fig.~\ref{fig:beta2d}). All this together supports the picture of a background group or cluster, with its own intragroup or intracluster medium, that lies at a redshift $z=0.55$.

The southernmost double-lobe radio source, located 680 kpc nearly due south of the cluster center, is at a redshift of $z_{spec}=0.549$ \citep{Sohn2019}, indicating it could be associated with the SW radio source, which lies at similar redshift. The thumbnail of the source, shown in panel (c) of Fig.~\ref{fig:optical}, shows two sources near the expected location of the radio source, with the one identified as a point source during the \verb|wavdetect| run (see \S\ref{sec:imaging}) indicated by a red circle. This source is likely the AGN driving the two radio lobes seen in 1.4 GHz. 

Based on the simulations presented in \S\ref{sec:sims}, we would expect the location of the perturbing system to be located to the NW of the cluster center, driving the potential shock front identified in \S\ref{sec:shock}. While there appears to be no distinct group of sources apparent in the distribution of individual sources (left panel of Fig.~\ref{fig:ndensity}) that suggest the presence of a subcluster, there is an elongation in the galaxy number density contours towards the shock region (right panel in Fig.~\ref{fig:ndensity}).

Overall, the close alignment between the optical and X-ray asymmetries, combined with the identification of multiple line-of-sight groups, paints a picture of A2029 as a dynamically evolving system, despite its well known relaxed nature. 

% \begin{figure*}
%     \centering
%     \includegraphics[width=0.6\textwidth]{a2029_cartoon.png}
%     \caption{Cartoon graphic of A2029 substructure showing the same features highlighted in Fig.\ \ref{fig:beta2d}.}
%     \label{fig:cartoon}
% \end{figure*}

\section{Summary and Conclusions}

We present deep (485 ksec) Chandra X-ray observations of the galaxy cluster A2029, hosting one of the most extended, continuous sloshing spiral observed to date. The X-ray imaging shows a generally regular elliptical distribution of the ICM, elongated NNE-SSW. Using beta-model fitting we identify several features of note in the underlying ICM substructure, including a concave bay-like depression to the south, a “splash” of extended emission to the SE (possibly tracing a wake of gas left after the perturber's second core passage), a localized excess in the SW, which we found to be associated with a background structure, and a potential merger shock in the outskirts to the NW.

We find a global, best-fit projected temperature of $kT =$ \fe{7.14}{0.02}{0.03} keV and abundance of $Z = $ \fe{0.75}{0.01}{0.01} $Z_{\odot}$ within the inner 116\arcsec [$\sim170$ kpc]. This is slightly lower temperature than previous studies (e.g., \cite{PM2013}) but consistent within the combined statistical and systematic uncertainties. A two-temperature model provides a significantly improved fit, with components at $kT_{low} = $ \fe{5.63}{0.10}{0.11} keV and $kT_{high} =$ \fe{12.90}{0.18}{0.33} keV, yielding much tighter constraints on the hot temperature component than earlier studies \citep{Clarke2004, PM2013}. Fitting a cooling flow model returned best-fit temperatures of $kT_{low} =$ \fe{3.71}{0.12}{0.12} keV and $kT_{high} = $ \fe{11.82}{0.30}{0.31} keV, abundance $Z = $ \fe{0.77}{0.01}{0.01} $Z_\odot$, and mass deposition rate $\dot{M} = $ \fe{784}{40}{36} M$_\odot$ yr$^{-1}$. This is in agreement with previous studies that have reported similarly high cooling flow rates \citep{Sarazin1992, Peres1998, PM2013}, despite the relatively low levels of star formation in the central cD galaxy compared to expectations from unimpeded cooling \citep{McNamara2007, McDonald2010, Hicks2010}. The low temperature component is consistent with the latest XRISM findings which report a 3.42 keV temperature component detected within the inner $\sim218$ kpc [2.5\arcmin] \citep{Sarkar2025}. However, when keeping the low temperature fixed to 0.0808 keV (the limit in \texttt{XSPEC}) and fitting a cooling flow model, we recover a much lower mass deposition rate of \fe{51}{3}{3} M$_\odot$ yr$^{-1}$, in line with the results of \cite{Clarke2004}. If we instead fit for the cooling within a smaller radius of 40\arcsec [58 kpc] encompassing the cD galaxy, we find \fe{17}{2}{2} M$_{\odot}$ yr$^{-1}$ which, again, is still higher than the observed SFR of the cD galaxy. 

Spectral maps of A2029 show that the sloshing spiral excess coincides with cooler, metal-rich gas --- hallmarks of sloshing-induced uplift from the cluster core. A notable ``splash'' feature in the SE traces a stream of progressively cooler gas, reaching down to $\sim5$ keV. An isolated region of excess X-ray emission farther southwest corresponds to $\sim$7 keV gas. To the NW, a broad, hot ($\sim$11 keV) Mpc-scale region may indicate the presence of shock-heated gas. 

To trace sloshing-induced substructure in A2029's ICM, we compared radial surface brightness profiles extracted from four pairs of diametrically opposed wedge-shaped sectors. These profiles reveal a criss-crossing pattern of alternating excesses at different radii, consistent with cooler, denser gas being displaced outward along a sloshing spiral. Each pair shows distinct transitions in brightness dominance, and when mapped spatially, the radii of these excesses outline a clear spiral pattern extending outward from the cluster core. Notably, the radial profiles show that the spiral excess appears more extended to the N than is immediately apparent in the raw or residual X-ray images.

In the NW sector of A2029, we identified a sharp surface brightness edge at a radius of $\sim$ 687 kpc, consistent with a weak shock front. The broken power-law fit to the surface brightness profile in the NW region of A2029 reveals a clear density discontinuity, consistent with a weak shock at $r\sim 660$ kpc, with a corresponding Mach number, $\mathcal{M} =$ \fe{1.12}{0.04}{0.04}. The projected temperature profile in this sector reveals a pronounced peak at smaller radii, $\sim 330\arcsec$, which we interpreted as a result of projection effects caused by the viewing geometry of the shock. 

The residual emission image revealed the presence of a concave, bay-like feature along the southern edge of the sloshing spiral. This bay feature could be (1) the trough of a large-scale Kelvin–Helmholtz billow rolling up along the cold front, (2) the inner rim of a ghost radio bubble whose contrast is muted by projection, or (3) a projected indentation that simply appears where the outer sloshing arm overlaps the stripped wake of the infalling subcluster on its second passage. Comparisons with cluster merger simulations favor the third scenario. 

While there is no clear evidence of previously undiscovered radio bubbles in the ICM, we are unable to rule out the possibility that the southern bay feature could be associated with the edge of a ghost bubble. The absence of detected cavities associated with the 1.4 GHz emission from the AGN could be the result of projection effects such that potential bubbles appear partially filled in along the line-of-sight or oriented in a way that reduces the contrast with the sloshing spiral regions. In the central 40\arcsec (i.e., encompassing the radio emission, but contained within the cD galaxy), we find that the estimated AGN injection rate ($\sim 1-3\times 10^{44}$ erg s$^{-1}$) is $\sim 8-23\times$ lower than the observed cooling luminosity ($L_{\mathrm{cool}} = 2.3\times 10^{45}$ erg s$^{-1}$.) Our results suggest that while AGN feedback contributes significantly to heating the ICM, it may not fully offset the observed cooling rate; thus, it is likely that the large-scale sloshing is playing a role in balancing ICM cooling towards the cluster center. 

The bending of the central WAT radio source in A2029 is likely a consequence of large-scale bulk motions in the ICM rather than the peculiar motion of the BCG itself. The azimuthal surface brightness profile showed increased brightness in the region of the southern lobe, which could suggest that the WAT lobe is being bent backward by the motion induced by the sloshing spiral. In addition, the northern lobe exhibits a slight kink, where it appears to be bent northwards, following the direction of sloshing. 

Comparison to the simulations of \cite{ZuHone2018} suggests a scenario of a 1:10 mass ratio off-axis merger, with an impact parameter of 500 kpc. In this simulation, a sloshing spiral is set off after the initial core passage of the perturbing subsystem. Upon the second core passage, the subsystem drives a shock front and leaves behind a wake of material, similar to the ``splash'' observed in A2029. The particular epoch that matched the A2029 observations corresponds to $t=4.30$ Gyr after the start of the merger simulation (i.e., $t=0$). The agreement between simulation and observation supports a scenario in which A2029 is currently experiencing the late-time aftermath of a minor merger, with sloshing motions, a NW shock front, and a displaced splash-like structure to the SE. These features reinforce the view that A2029 appears to have experienced a minor merger  $\sim$4 Gyr ago. Even this object, which has been described as one of the most relaxed clusters in the universe \citep{Dressler1979, Buote1996, Dullo2017}, shows evidence of its merger history after deep observations and detailed study.
 %is dynamically active at large radii.

The optical member distribution of A2029 reveals a complex and asymmetric structure, with elongations along both the NE–SW and NW–SE axes forming a distinctive ‘Z’-shaped pattern. While no distinct subcluster is visible in the NW, a modest enhancement in the galaxy density contours toward this direction aligns with the location of a potential shock front, consistent with minor merger simulations. 

Overall, our results present a coherent picture of A2029 as a dynamically evolving system, shaped by the long-lasting aftermath of a minor, off-axis merger. The large-scale sloshing spiral, splash feature, and mild merger-driven shock are all signatures of ongoing dynamical activity despite the overall relaxed appearance of the cluster. Comparison to simulations shows that the observed X-ray features are also found in a 1:10 mass ratio off-axis merger, providing further insight into A2029's merger history. In addition, A2029 offers another example where the surrounding cluster environment can impact the central AGN through the bending of the radio lobes. Taken together, the evidence suggests that A2029 is still settling from past interactions --- showing that even the most relaxed-looking clusters can be hiding a rich history of dynamical activity. %Like a cosmic tornado, A2029 embodies the intense, spiraling chaos that can persist long after a merger's initial impact. 

\begin{acknowledgments}
{\footnotesize
C.B.W. would like to thank Bill Forman and Al Marscher for their helpful comments on this work. 

Support for this work was provided by the National Aeronautics and Space Administration, through Chandra Award Number GO2-23123X. C.B.W was supported in part by the Smithsonian Astrophysical Observatory (SAO) Predoctoral Fellowship Program, through XMM contract 80NSSC22K0568 and Chandra contract GO0-21113X. 

Support for JAZ and SWR was provided by the Chandra X-ray Observatory Center, which is operated by the Smithsonian Astrophysical Observatory for and on behalf of NASA under contract NAS8-03060.

Basic research in Radio Astronomy at the U.S. Naval Research Laboratory is supported by 6.1 Base funding.

This research employs a list of Chandra datasets, obtained by the Chandra X-ray Observatory, contained in~\dataset[DOI: 10.25574/cdc.468]{https://doi.org/10.25574/cdc.468}.

The Legacy Surveys consist of three individual and complementary projects: the Dark Energy Camera Legacy Survey (DECaLS; Proposal ID \#2014B-0404; PIs: David Schlegel and Arjun Dey), the Beijing-Arizona Sky Survey (BASS; NOAO Prop. ID \#2015A-0801; PIs: Zhou Xu and Xiaohui Fan), and the Mayall z-band Legacy Survey (MzLS; Prop. ID \#2016A-0453; PI: Arjun Dey). DECaLS, BASS and MzLS together include data obtained, respectively, at the Blanco telescope, Cerro Tololo Inter-American Observatory, NSF’s NOIRLab; the Bok telescope, Steward Observatory, University of Arizona; and the Mayall telescope, Kitt Peak National Observatory, NOIRLab. Pipeline processing and analyses of the data were supported by NOIRLab and the Lawrence Berkeley National Laboratory (LBNL). The Legacy Surveys project is honored to be permitted to conduct astronomical research on Iolkam Du’ag (Kitt Peak), a mountain with particular significance to the Tohono O’odham Nation.

NOIRLab is operated by the Association of Universities for Research in Astronomy (AURA) under a cooperative agreement with the National Science Foundation. LBNL is managed by the Regents of the University of California under contract to the U.S. Department of Energy.

This project used data obtained with the Dark Energy Camera (DECam), which was constructed by the Dark Energy Survey (DES) collaboration. Funding for the DES Projects has been provided by the U.S. Department of Energy, the U.S. National Science Foundation, the Ministry of Science and Education of Spain, the Science and Technology Facilities Council of the United Kingdom, the Higher Education Funding Council for England, the National Center for Supercomputing Applications at the University of Illinois at Urbana-Champaign, the Kavli Institute of Cosmological Physics at the University of Chicago, Center for Cosmology and Astro-Particle Physics at the Ohio State University, the Mitchell Institute for Fundamental Physics and Astronomy at Texas A\&M University, Financiadora de Estudos e Projetos, Fundacao Carlos Chagas Filho de Amparo, Financiadora de Estudos e Projetos, Fundacao Carlos Chagas Filho de Amparo a Pesquisa do Estado do Rio de Janeiro, Conselho Nacional de Desenvolvimento Cientifico e Tecnologico and the Ministerio da Ciencia, Tecnologia e Inovacao, the Deutsche Forschungsgemeinschaft and the Collaborating Institutions in the Dark Energy Survey. The Collaborating Institutions are Argonne National Laboratory, the University of California at Santa Cruz, the University of Cambridge, Centro de Investigaciones Energeticas, Medioambientales y Tecnologicas-Madrid, the University of Chicago, University College London, the DES-Brazil Consortium, the University of Edinburgh, the Eidgenossische Technische Hochschule (ETH) Zurich, Fermi National Accelerator Laboratory, the University of Illinois at Urbana-Champaign, the Institut de Ciencies de l’Espai (IEEC/CSIC), the Institut de Fisica d’Altes Energies, Lawrence Berkeley National Laboratory, the Ludwig Maximilians Universitat Munchen and the associated Excellence Cluster Universe, the University of Michigan, NSF’s NOIRLab, the University of Nottingham, the Ohio State University, the University of Pennsylvania, the University of Portsmouth, SLAC National Accelerator Laboratory, Stanford University, the University of Sussex, and Texas A\&M University.

BASS is a key project of the Telescope Access Program (TAP), which has been funded by the National Astronomical Observatories of China, the Chinese Academy of Sciences (the Strategic Priority Research Program ''The Emergence of Cosmological Structures'' Grant \#XDB09000000), and the Special Fund for Astronomy from the Ministry of Finance. The BASS is also supported by the External Cooperation Program of Chinese Academy of Sciences (Grant \#114A11KYSB20160057), and Chinese National Natural Science Foundation (Grant \#12120101003, \#11433005).

The Legacy Survey team makes use of data products from the Near-Earth Object Wide-field Infrared Survey Explorer (NEOWISE), which is a project of the Jet Propulsion Laboratory/California Institute of Technology. NEOWISE is funded by the National Aeronautics and Space Administration.

The Legacy Surveys imaging of the DESI footprint is supported by the Director, Office of Science, Office of High Energy Physics of the U.S. Department of Energy under Contract No. DE-AC02-05CH1123, by the National Energy Research Scientific Computing Center, a DOE Office of Science User Facility under the same contract; and by the U.S. National Science Foundation, Division of Astronomical Sciences under Contract No. AST-0950945 to NOAO.

This work made use of data from the Galaxy Cluster Merger Catalog (http://gcmc.hub.yt).

This work was carried out at Boston University which stands on the ancestral lands of The Wampanoag and The Massachusett People. }
\end{acknowledgments}

\software{CIAO (Fruscione et al. 2006),
XSPEC (Arnaud 1996),
Sherpa (Siemiginowska et al. 2024),
pyproffit (Eckert et al. 2020),
SciPy  (Virtanen et al. 2020)
}

\bibliography{A2029.bib}
\bibliographystyle{aasjournal}

\appendix 
\section{Correction for ACIS-I Calibration Uncertainties} \label{sec:cal}

\subsection{Focal Plane Temperature \& Gain Effects on Spectral Fit Parameters \label{sec:a1}}

\begin{figure*}
    \centering
    \includegraphics[width=0.7\linewidth]{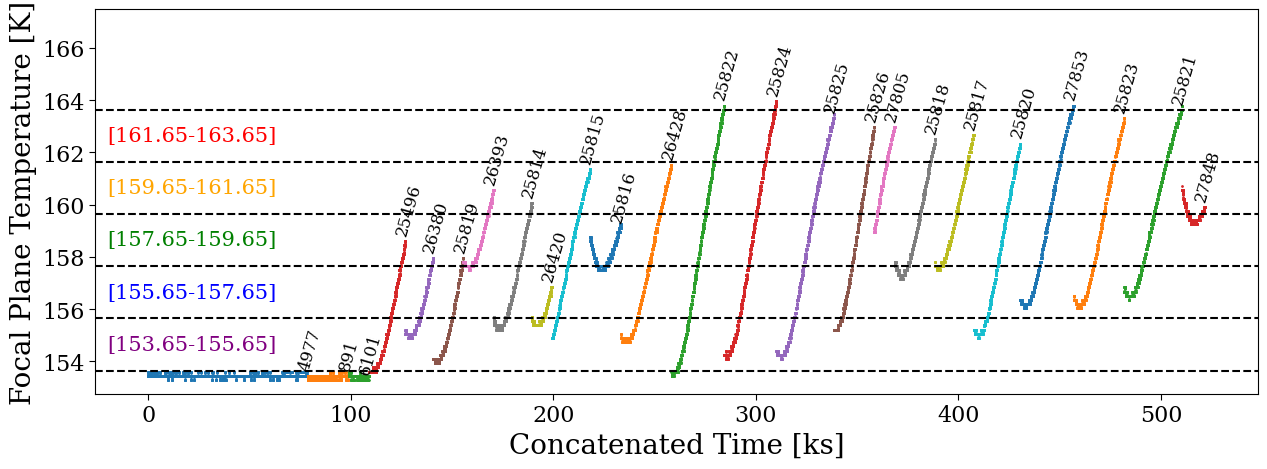}
    \caption{Focal plane temperatures recorded during observation for each ObsID listed in Table \ref{tab:obsinfo}. }
    \label{fig:fptemps}
\end{figure*}

The new set of Chandra observations presented here spans a broader range of focal plane (FP) temperatures during observation (see Fig.~\ref{fig:fptemps}) than seen in the older, pre-2005 data. We investigated the potential dependence of the spectral fits on the FP temperature at the time of the observation. To do this, we defined five FP temperature bins, each with a width of 2~K (denoted by dashed lines and colored labels in Fig.~\ref{fig:fptemps}). The lowest bin is defined to avoid overlapping with the pre-2005 set. Each ObsID in the new set was filtered to include only those events that were recorded during times when the FP temperature remained within a given FP bin range. We used \verb|specextract| to extract the spectrum and temperature-dependent response files.

Spectra were extracted from an annular region, covering a roughly isothermal area, centered on the peak of the cluster emission (as found by the 2D $\beta$-model fitting in \S\ref{sec:beta2d}), defined with outer radius $r_{\text{outer}} = 3\farcm5$ and inner radius equal to the cool core radius plus 15\arcsec, $r_{\text{inner}} = 78\farcs65$ (using the reported cool core radius of A2029 \citep{Sarazin1992} and similar methods of \cite{Schellenberger2015}). For each FP bin range (i.e., considering only the data within each set of dashed lines in Fig.~\ref{fig:fptemps}), the extracted spectra were fit with an absorbed \texttt{APEC} model \citep{Smith2001}, keeping the hydrogen column density --- $n_H$ --- and redshift --- $z$ --- frozen (see \S\ref{sec:spec}). Since the observations were all taken around the same time of year each year and the temperature dependent gain calibration is expected to vary from year to year, the new data set was divided into two groups: one for 2022 observations, and one for 2023 observations --- i.e, fit parameters were tied within each year group, but not across year groups. For each year group, within a given FP bin range, we obtained the temperature (kT), abundance (Z), and normalization with and without including a fit on the gain slope and offset parameters. Spectra of the pre-2005 set were extracted from the same region but without applying any additional filtering. The pre-2005 set was fit as a group with no gain fitting model in Xspec.

\begin{figure*}
    \centering
    \includegraphics[width=\linewidth]{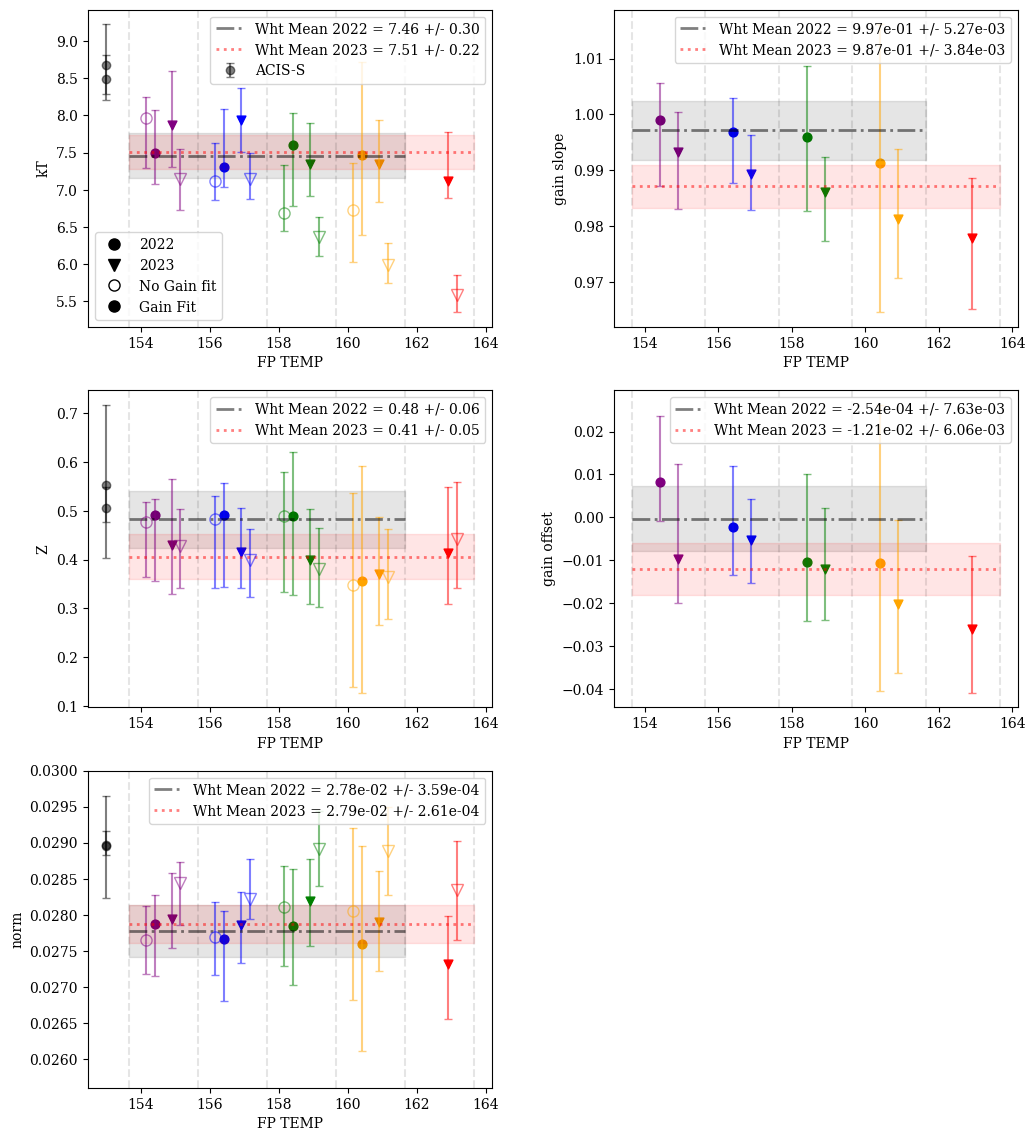}
    \caption{Spectral fit parameters as a function of observed focal plane (FP) temperature range. The dashed lines and marker colors correspond to the horizontal lines and colored labels of Fig.~\ref{fig:fptemps}. Since each point represents fitted data spanning the given FP range, to reduce overlap of the points and aid in visibility of the individual error bars, we include small offsets along the x-axis. The 2022 and 2023 ObsID sets are denoted by circles and inverted triangles, respectively. Results including gain fitting are shown as filled markers and fits without gain corrections are indicated by unfilled markers. The weighted (``Wht'') mean and 3$\sigma$ confidence range of each parameter are calculated for each year group across all temperature bins and shown in each panel (gray for 2022 and red for 2023).}
    \label{fig:fpbinfits}
\end{figure*}

Figure~\ref{fig:fpbinfits} shows the results of these fits, with the marker colors corresponding to the colored labels of Fig.~\ref{fig:fptemps}. Error bars are the 3$\sigma$ confidence ranges. The 2022 and 2023 ObsID sets are denoted by circles and inverted triangles, respectively. Results including gain fitting are shown as filled markers while fits without gain fitting are represented by the unfilled markers. The weighted mean and 3$\sigma$ confidence range of each parameter were calculated for each year group across all temperature bins, and shown in each panel of Fig.~\ref{fig:fpbinfits} (gray for 2022 and red for 2023). We find that there is indeed a dependence on the FP temperature of the best-fit \texttt{APEC} temperature but also a disagreement with the results from the pre-2005 data. We determined that this was the effect of a previously unidentified gain calibration issue that particularly affects high-S/N ACIS-I observations of relatively hot clusters, as we have here. Fig.~\ref{fig:fpbinfits} shows that including a correction on the gain parameters improves the fits for the 2023 group and gives results that are consistent with those from the 2022 group; however, the results from both groups remain inconsistent with those from the old pre-2005 data.

\subsection{Assessing the Significance of Focal Plane Temperature Filtering \& Gain Effects \label{sec:calcombine}}

 To test whether doing a temperature-dependent gain correction significantly improves the fit results for a given ObsID, we compared the fits on the new data with and without the FP filtering described in \S\ref{sec:a1}. To do this, we created a combined spectrum of all the FP-filtered spectra for a given ObsID using the CIAO routine \verb|combine_spec|, which also creates a combined (and now temperature dependent) response file.  

The spectra are grouped by year --- i.e., the ObsIDs within the same year are fit simultaneously --- and fit with the same absorbed \texttt{APEC} model as before and include a fit of the gain parameters. We also fitted the combined FP-filtered new data with the fit parameters for the two year groups tied together, but the gain parameters frozen to the results found when fitting the year groups separately. The results of these fits are listed under `Combined Temperature Dependent Responses' in Table~\ref{tab:xspecfits}, where the uncertainties are $3\sigma$ confidence levels. We performed the same fits on new data that did not have any FP filtering applied. The results of these fits are listed under `No Filtering' in Table~\ref{tab:xspecfits}.

Comparing the fits between the combined FP-filtered spectra (i.e., now considering all data within the top-most and bottom-most dashed lines in Fig.~\ref{fig:fptemps}) and the unfiltered spectra, we find an apparent trend in the fitted gain parameters with FP temperature. However, within each observational year group, the fitted parameters remain consistent within the 3$\sigma$ errors, indicating that any FP dependent effects are small relative to the statistical uncertainties (see results of Table~\ref{tab:xspecfits}), even in this large extraction region with tens of thousands of counts. The dominance of statistical errors in this worst-case scenario suggests that FP dependent gain uncertainties do not significantly impact our spectral measurements. Therefore, to simplify the analysis and avoid introducing unnecessary complexity into the fitting procedure, we proceed without dividing the data into FP temperature dependent data groups for the analysis presented in the main body of this work.

We also find that the best-fit gain parameters for the 2022 data are consistent with a slope of $\sim$ 1 and an offset of $\sim$ 0, indicating that no extra gain correction is needed for this data. However, for the 2023 data, we find that a gain slope of \fe{0.9875}{0.0021}{0.0021} and an offset of \fe{-0.0155}{0.0027}{0.0026} are needed to improve the fits and bring the 2022 and 2023 fitted parameters into agreement. 

\subsection{Determining Calibration Corrections Based On Historic Observations}

We fit the pre-2005 data --- consisting of two ACIS-S (ObsID 4977 and 891) and one ACIS-I (ObsID 6101) observations --- with the same absorbed \texttt{APEC} model as before, and these results are listed in Tab.~\ref{tab:xspecfits}, with 3$\sigma$ errors. These values are what we consider the ``correct'' values, as the fitted data are not subject to the calibration issues we find in the newer data. The fitted values we find for the pre-2005 data are also consistent with previously recorded ICM temperatures. \cite{Schellenberger2015} find a temperature of \fe{8.75}{0.11}{0.11}, where the errors are 1$\sigma$ confidence, and the data was fit in the 0.7-7 keV range. 

\begin{figure*}
    \centering
    \includegraphics[width=\linewidth]{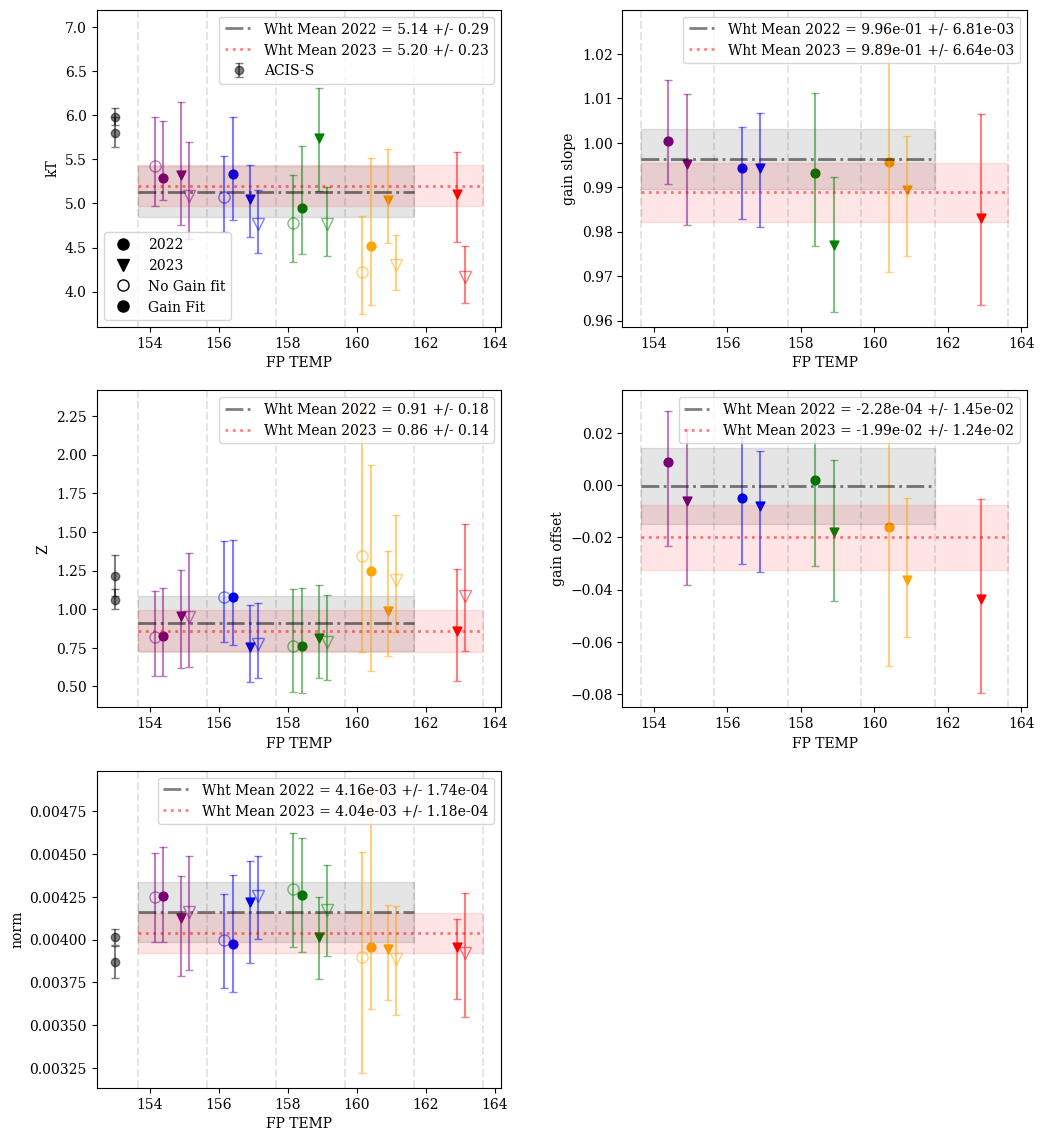}
    \caption{Spectral fit parameters as a function of observed focal plane (FP) temperature range for innermost 15\arcsec\ of the cool core region. The dashed lines and marker colors correspond to the horizontal lines and colored labels of Fig.~\ref{fig:fptemps}. Since each point represents fitted data spanning the given FP range, to reduce overlap of the points and aid in visibility of the individual error bars, we include small offsets along the x-axis. The 2022 and 2023 ObsID sets are denoted by circles and inverted triangles, respectively. Results including gain fitting are shown as filled markers and fits without gain corrections are indicated by unfilled markers. The weighted (``Wht'') mean and 3$\sigma$ confidence range of each parameter are calculated for each year group across all temperature bins and shown in each panel (gray for 2022 and red for 2023).}
    \label{fig:fpbinfits2}
\end{figure*}

We estimate the constant offset between the new and pre-2005 spectral fits by comparing the results of the `No Filtering' and the `Pre-2005 Data' results listed in Tab.~\ref{tab:xspecfits}. We find that the temperature is offset by a constant factor of 1.13 while the abundance is offset by a constant factor of 1.28. We also performed a similar comparison in a relatively cool region of the cluster and found consistent scaling factors (see Fig.~\ref{fig:fpbinfits2} and Tab.~\ref{tab:xspecfits2}), suggesting that these offsets are stable across the range of temperatures observed in A2029. Thus, when fitting spectra extracted from the new observations, we apply scaling corrections of 1.13 to the fitted temperatures and 1.28 to the fitted abundances to bring them into agreement with the old ACIS-S observations. These scale factors are applied consistently throughout all spectral analyses in this work. 

\subsection{Caveats}

While these scaling factors bring the measurements from the pre-2005, 2022, and 2023 observations into agreement, we note that our results based on using these values should be treated with caution. The correction factors were derived from relatively large, high–signal-to-noise extraction regions, and it is possible that they vary on smaller spatial scales due to position-dependent calibration effects across the ACIS-I array (although, the statistical errors will be larger in smaller regions). Without a full recalibration of the instrument response, our approach represents an interim solution that ensures internal consistency, but residual position-dependent or chip-dependent systematics may remain. For this reason, we conservatively recommend that a systematic uncertainty of 13\% in temperature and 28\% in abundance should be considered in any absolute measurements.

\begin{deluxetable}{ccccc}
\tablecaption{XSPEC fits to an annular isothermal region in A2029. Error ranges are 3$\sigma$ confidence ranges.\label{tab:xspecfits}}
\tablehead{
  \colhead{Group} &
  \colhead{$kT$ (keV)} &
  \colhead{$Z$ ($Z_{\odot}$)} &
  \colhead{Gain slope} &
  \colhead{Gain offset}
}
\startdata
Pre-2005 Data\tablenotemark{a} & \fe{8.51}{0.32}{0.15} & \fe{0.51}{0.04}{0.03} & \nodata & \nodata \\\hline
\multicolumn{5}{c}{Combined Temperature Dependent Responses\tablenotemark{b}} \\\hline
2022\tablenotemark{c}       & \fe{7.56}{0.14}{0.34} & \fe{0.40}{0.03}{0.04} & \fe{0.9991}{0.0019}{0.0027} & \fe{-0.0049}{0.0040}{0.0030}  \\
2023\tablenotemark{c}       & \fe{7.47}{0.18}{0.13} & \fe{0.40}{0.03}{0.02} & \fe{0.9863}{0.0028}{0.0020} & \fe{-0.0151}{0.0026}{0.0034}  \\
2022+2023\tablenotemark{d}  & \fe{7.54}{0.09}{0.19} & \fe{0.41}{0.02}{0.02} & \nodata & \nodata \\\hline
\multicolumn{5}{c}{No Filtering\tablenotemark{b}} \\ \hline
2022\tablenotemark{c}       & \fe{7.46}{0.23}{0.18} & \fe{0.39}{0.04}{0.02} & \fe{0.9989}{0.0021}{0.0021} & \fe{-0.0028}{0.0037}{0.0030}  \\
2023\tablenotemark{c}       & \fe{7.51}{0.17}{0.14} & \fe{0.39}{0.03}{0.02} & \fe{0.9875}{0.0013}{0.0019} & \fe{-0.0155}{0.0026}{0.0027}  \\
2022+2023\tablenotemark{d}  & \fe{7.54}{0.08}{0.19} & \fe{0.40}{0.02}{0.03} & \nodata & \nodata  \\
\enddata
\tablenotetext{a}{Consisting of the two ACIS-S (ObsIDs 4977 and 891) and one ACIS-I (ObsID 6101) observations}
\tablenotetext{b}{From combining separate FP-filtered spectra for each ObsID (see Appendix \ref{sec:calcombine})}
\tablenotetext{c}{Fit parameters are tied for ObsIDs within this group}
\tablenotetext{d}{Both with gain parameters set to values found when fitting separately}
\end{deluxetable}

\begin{deluxetable}{ccccc}
\tablecaption{XSPEC fits to the innermost cool core region for A2029. Error ranges are 3$\sigma$ confidence ranges.\label{tab:xspecfits2}}
\tablehead{
  \colhead{Group} &
  \colhead{$kT$ (keV)} &
  \colhead{$Z$ ($Z_{\odot}$)} &
  \colhead{Gain slope} &
  \colhead{Gain offset}
}
\startdata
Pre-2005 Data\tablenotemark{a} & \fe{5.99}{0.27}{0.23} & \fe{1.10}{0.18}{0.16} & \nodata & \nodata  \\\hline
\multicolumn{5}{c}{Combined Temperature Dependent Responses\tablenotemark{b}} \\\hline
2022\tablenotemark{c}  & \fe{5.11}{0.48}{0.27} & \fe{0.93}{0.18}{0.18} & \fe{0.9967}{0.0085}{0.0062} & \fe{-0.0016}{0.0116}{0.0157} \\
2023\tablenotemark{c}  & \fe{5.19}{0.26}{0.26} & \fe{0.84}{0.15}{0.13} & \fe{0.9896}{0.0046}{0.0068} & \fe{-0.0257}{0.0143}{0.0053}  \\
2022+2023\tablenotemark{d}  & \fe{5.16}{0.17}{0.16} & \fe{0.87}{0.11}{0.10} & \nodata & \nodata \\\hline
\multicolumn{5}{c}{No Filtering\tablenotemark{b}} \\\hline
2022\tablenotemark{c}  & \fe{5.13}{0.31}{0.25} & \fe{0.89}{0.18}{0.16} & \fe{0.9982}{0.0074}{0.0059} & \fe{-0.0019}{0.0114}{0.0158} \\
2023\tablenotemark{c}  & \fe{5.24}{0.28}{0.25} & \fe{0.82}{0.07}{0.12} & \fe{0.9907}{0.0059}{0.0071} & \fe{-0.0268}{0.0134}{0.0108} \\ 
2022+2023\tablenotemark{d}  & \fe{5.19}{0.17}{0.15} & \fe{0.85}{0.10}{0.10} & \nodata & \nodata 
\enddata
\tablenotetext{a}{Consisting of the two ACIS-S (ObsIDs 4977 and 891) and one ACIS-I (ObsID 6101) observations}
\tablenotetext{b}{From combining separate FP-filtered spectra for each ObsID (see Appendix \ref{sec:calcombine})}
\tablenotetext{c}{Fit parameters are tied for ObsIDs within this group}
\tablenotetext{d}{Both with gain parameters set to values found when fitting separately}
\end{deluxetable}

\end{document}